\documentclass[acmsmall]{acmart}
  
\usepackage{lineno,hyperref}
 
\usepackage{amsmath,amssymb,amsfonts}
\usepackage{graphicx}

\usepackage{float}
\usepackage{textcomp}
\usepackage{xcolor}
\usepackage{multirow}
\usepackage{array}
\usepackage{colortbl}
\usepackage{verbatim}
\usepackage{url}
\usepackage{tcolorbox}
\usepackage[ruled,linesnumbered,lined,noend]{algorithm2e}
\usepackage{algorithmic} 
\usepackage{booktabs}
\usepackage{bm}
\usepackage{listings}
\usepackage{subfigure}
\usepackage{makecell}
\usepackage{fancyhdr}
\usepackage{cases}
\usepackage{lineno,hyperref} 
\usepackage{textcomp}
\usepackage{tabularx}  
\usepackage{enumitem}

\usepackage{cases}
\usepackage{textcomp}

\definecolor{dkgreen}{rgb}{0,0.6,0}
\definecolor{gray}{rgb}{0.5,0.5,0.5}
\definecolor{mauve}{rgb}{0.58,0,0.82}
\usepackage{todonotes}

\usepackage{amsmath}
\usepackage{multirow}
\usepackage{cases}
\usepackage{textcomp}
\usepackage{wrapfig,lipsum}

\newcommand{\yg}[1]{\textcolor{black}{#1}}
\newcommand{\tl}[1]{\textcolor{black}{#1}}

\def\code#1{\texttt{#1}}

\definecolor{light-gray}{gray}{0.95}   

\newcommand{\tool}{\textsf{RADAR}} 
\newcommand{\pd}{\textsf{FD}}
\newcommand{\pdsig}{\textsf{FD}$^{\rm Sig}$}

\AtBeginDocument{%
  \providecommand\BibTeX{{%
    \normalfont B\kern-0.5em{\scshape i\kern-0.25em b}\kern-0.8em\TeX}}}

\setcopyright{acmcopyright}
\copyrightyear{}
\acmYear{}
\acmDOI{}

\acmVolume{0}
\acmNumber{0}
\acmArticle{0}
\acmMonth{0}

\begin{document}

\title{How Important are Good Method Names in Neural Code Generation? A Model Robustness Perspective}

\author{Guang Yang}
\email{novelyg@outlook.com}
\author{Yu Zhou}
\email{zhouyu@nuaa.edu.cn}
\author{Wenhua Yang}
\email{ywh@nuaa.edu.cn}

\affiliation{%
  \institution{Nanjing University of Aeronautics and Astronautics}
 \city{Nanjing}
  \country{China}
}

\author{Tao Yue}
\email{tao@simula.no}
\affiliation{%
	\institution{Simula Research Laboratory}
	\country{Norway}
}

\author{Xiang Chen}
\email{xchencs@ntu.edu.cn}
\affiliation{%
	\institution{Nantong University}
	\city{Nantong}
	\country{China}
}

\author{Taolue Chen}
\email{t.chen@bbk.ac.uk}
\authornote{Corresponding author.}
\affiliation{%
 \institution{Birkbeck, University of London}
 \city{London}
 \country{UK}}

\renewcommand{\shortauthors}{G. Yang, et al.}

\begin{abstract}
Pre-trained code generation models (PCGMs) have been widely applied in neural code generation which can generate executable code from functional descriptions in natural languages, possibly together with signatures. Despite substantial performance improvement of PCGMs, the role of method names in neural code generation has not been thoroughly investigated. In this paper, we study and demonstrate the potential of benefiting from method names to enhance the performance of PCGMs, from a model robustness perspective. Specifically, we propose a novel approach, named {\tool} (neu\underline{RA}l co\underline{D}e gener\underline{A}tor \underline{R}obustifier).  
{\tool} consists of two components: {\tool}-Attack and {\tool}-Defense. The former attacks a PCGM 
by generating adversarial method names as part of the input, which are semantic and visual similar to the original input, but may trick the PCGM to generate completely unrelated code snippets. As a countermeasure to such attacks,
{\tool}-Defense synthesizes a new method name from the functional description and supplies it to the PCGM. 
Evaluation results show that {\tool}-Attack can reduce the CodeBLEU of generated code by 19.72\% to 38.74\% in three state-of-the-art PCGMs (i.e., CodeGPT, PLBART, and CodeT5) \yg{in the fine-tuning code generation task, and reduce the Pass@1 of generated code by 32.28\% to  44.42\% in three state-of-the-art PCGMs (i.e., Replit, CodeGen, and CodeT5+) in the zero-shot code generation task}.  
Moreover, {\tool}-Defense is able to 
reinstate the performance of PCGMs with synthesized method names.
These results highlight the importance of good method names in neural code generation and implicate the benefits of studying model robustness in software engineering. 
\end{abstract}



\maketitle
\section{Introduction}

\noindent\textit{Context.} Neural code generation generally refers to the task of generating executable code from functional descriptions in natural language using neural networks and it has the potential to reduce the development pressure on programmers. While early studies on automatic code generation mainly focus on domain-specific programming languages (e.g., card game code~\cite{ling2016latent}, Bash~\cite{lin2018nl2bash}, and regular expressions~\cite{locascio2016neural}), recent neural code generation for common programming languages takes the inspiration from the impressive achievements of pre-trained deep learning models in natural language processing, and has attracted a lot of attention recently~\cite{svyatkovskiy2020intellicode, lu2021codexglue, chen2021evaluating, clement2020pymt5, ahmad2021unified, phan2021cotext, wang2021codet5, chakraborty2022natgen}. 

In literature, neural code generation typically focuses on method-level code generation, i.e., generating a method body by taking mainly two types of input: (1) functional description of the intended code only \cite{lu2021codexglue,phan2021cotext,ahmad2021unified,wang2021codet5}, henceforth denoted by \pd; or (2) both the functional description and the method signature (i.e., the combination of the method name and the parameter list \cite{clement2020pymt5, hao2022AixBench, chen2022codet, christopoulou2022pangu}), henceforth denoted by FD$^{\rm Sig}$. 
\yg{
Furthermore, we categorize the existing benchmarks into two groups based on their data size and the availability of test cases, i.e., fine-tuning code generation tasks and zero-shot code generation tasks.
For example, we classify Human-Eval~\cite{chen2021evaluating} as a zero-shot code generation task due to its limited dataset size (164 data items), which includes test cases. This dataset is insufficient to adequately fine-tune the model.
In contrast, CONCODE~\cite{iyer2018mapping} is classified as a fine-tuning code generation task. It consists of numerous data items without accompanying test cases, thereby providing an extensive dataset for fine-tuning the model.}
%

\noindent\textit{Motivation.}
Evidence from the literature has shown that taking signature information as input can largely boost the performance of neural code generation, i.e., generating more syntactically and semantically correct code~\cite{liguori2021evil, liguori2022can}.
For example, the BLEU score of the PyMT5 model was nearly doubled by taking signature information as input~\cite{clement2020pymt5}. Our experiment results (Section~\ref{sec:results}) also confirmed this observation. 
However, a natural, scientifically intriguing question of engineering importance is: 
\tl{what contribution does the additional signature information make so the {\pdsig} approaches become more effective?} 
Clearly, a thorough investigation of this question would be very useful in further improving the performance of neural code generation. 
Considering that not every code method contains the parameter list, we \tl{prioritize our research on} 
the method names in the signature.
In this paper, we study \tl{the impact of method names} through the lens of \emph{robustness} of the pre-trained deep learning models. 
%

Robustness refers to the ability of a model to cope with erroneous inputs and errors that occurred during its execution~\cite{isc1990ieee}.
In particular, in deep learning, by adding minor perturbations to the benign inputs of a neural network model, one can generate adversarial examples, which may spoof the model, thereby causing significant derivations in the model output. A vast amount of attention has been paid to studying the robustness of deep learning models, typically in domains such as image classifications, computer vision, and natural language processing~\cite{carlini2017towards, goswami2018unravelling, wang2021robot}, where adversarial examples often pose both safety and security concerns. We remark that, though, in the context of neural code generation applications, adversarial examples of deep learning models may not bring severe safety concerns, which is in stark contrast to other application domains such as autonomous driving~\cite{deng2020analysis}, studying and improving the robustness of pre-trained code generation models (PCGMs), indeed, bring (previously-unexpected) benefits, as we will demonstrate in this paper.


\begin{figure*}[htbp]
	\centering
	\subfigure[Evaluation on the Python dataset]{%
		\includegraphics[width=1\textwidth]{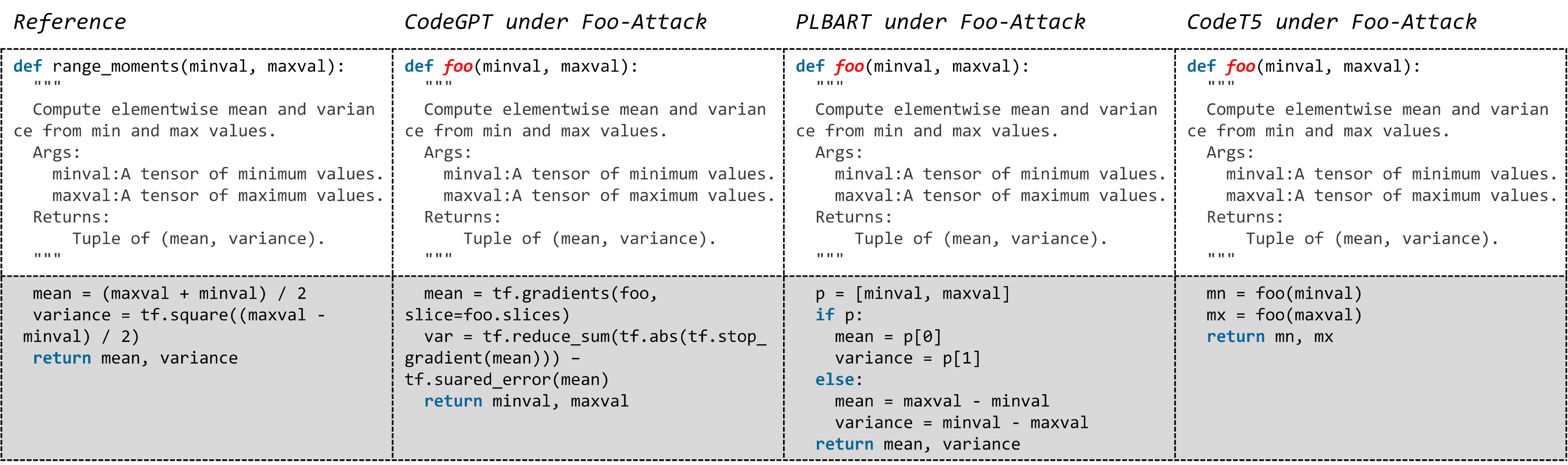}%
	}%
	\hfill
	\subfigure[Evaluation on the Human-Eval dataset]{%
		\includegraphics[width=1\textwidth]{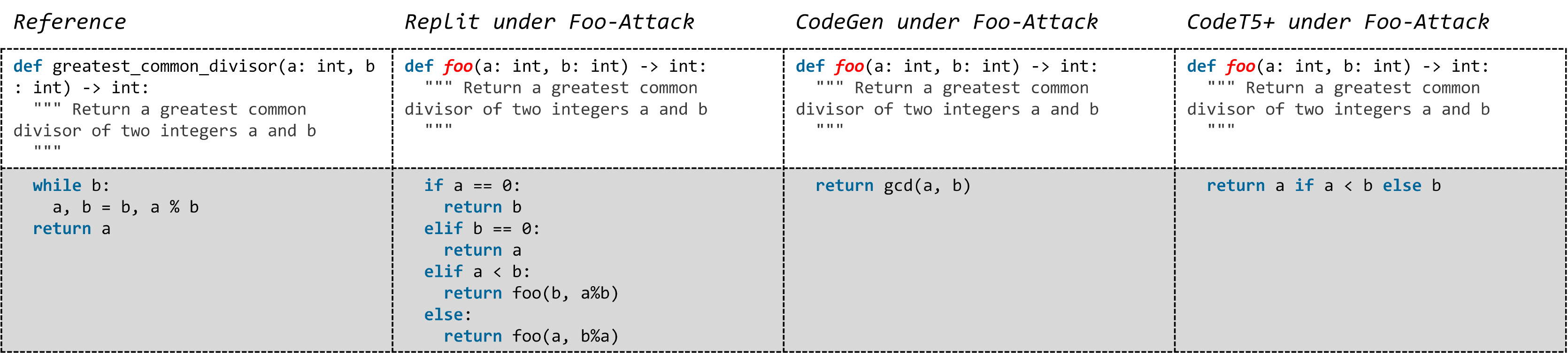}%
	}%
	\caption{The motivating examples illustrating the non-robustness challenge faced by popular PCGMs}
	\label{fig:motivation}
\end{figure*}

However, state-of-the-art PCGMs may \emph{not} be robust.
Fig.~\ref{fig:motivation}(a) presents an example (with the code collected from the PyPi project.\footnote{\url{https://pypi.org/project/fomoro-pyoneer/}}) to illustrate the robustness challenge faced by the three representative PCGMs (i.e., CodeGPT, PLBART, and CodeT5) \yg{in the fine-tuning code generation task}. After fine-tuning, we use the functional description and the signature as the input to each model (code highlighted in light grey in Fig.~\ref{fig:motivation}). The generated code snippets are exactly the same as the reference (the leftmost).
However, if we simply replace the method name \code{range\_moments} with \code{foo} and keep the functional description untouched, all three models generate totally incorrect code (highlighted in the dark grey). 
\yg{
Fig.~\ref{fig:motivation}(b) presents an illustrative example, utilizing code collected from Human-Eval~\cite{chen2021evaluating}, to highlight the challenge of robustness encountered by three representative PCGMs (i.e., Replit~\cite{replit}, CodeGen~\cite{nijkamp2022codegen} and CodeT5+~\cite{wang2023codet5+}) in the zero-shot code generation task.
For each model, we input the functional description and the signature, resulting in generated code snippets that successfully pass the test cases, akin to the reference code shown on the leftmost side. However, when a simple substitution is made by replacing the method name \code{greatest\_common\_divisor} with \code{foo} while retaining the functional description, all three models produce completely incorrect code that fails to pass the test cases (highlighted in the dark grey).
}
Note that \code{foo} is the most commonly used variable name in computer tutorial textbooks. This clearly shows that these models are not robust for the current input. Indeed, as shown in Section~\ref{sec:results}, poor robustness of PCGMs is commonly seen and greatly impacts their performance. For instance, our attack method can generate meaningful (adversarial) and natural method names that could reduce the CodeBLEU score of the generated code by 19.72\%--38.74\% in CodeGPT~\cite{lu2021codexglue}, PLBART~\cite{ahmad2021unified} and CodeT5~\cite{wang2021codet5} \yg{in the fine-tuning code generation task}. 
\yg{In the zero-shot code generation task, our attack can reduce the Pass@1 score of the generated code by 32.28\%--44.42\% in Replit~\cite{replit}, CodeGen~\cite{nijkamp2022codegen}, and CodeT5+~\cite{wang2023codet5+}.}
Hence, we conclude that FD$^{\rm Sig}$ approaches, albeit demonstrating a better performance, are fragile (hence less robust) as they heavily rely on the selection of the input method name. 
This is a serious matter, since developers (i.e., users of PCGMs) may select a low-quality name in coding practice (due to inexperience, carelessness, bad habits, or otherwise just a typo), an ill-formed method name might largely degrade the performance of PCGMs, which thus generate unwanted code. 

\begin{figure*}[htbp]
	\centering
	\includegraphics[width=0.8\textwidth]{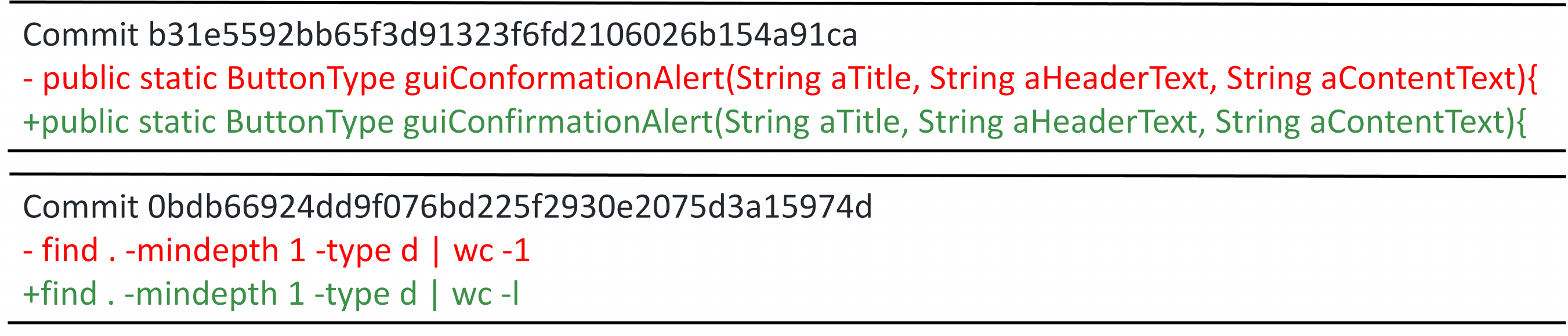}
	\caption{Two typo fixes for code refactoring in Github}
	\label{fig:method-name}
\end{figure*}

In a real-world software development context, it is often the case that developers refactor their code simply due to typos.
The study conducted by Liu et al.~\cite{liu2019learning} shows that an important code refactoring operation is due to simple typos (cf.\ Fig.~\ref{fig:method-name}. For instance, developers spelled `Confirmation' as `Conformation' in a method name or spelled `l' as `1' in bash code).
Meanwhile, a study conducted by Murphy-Hill et al.~\cite{murphy2011we} on activity from over 13,000 Java developers finds that renaming methods was the most commonly used refactoring operation, accounting for 74.8\% of all refactoring operations. This indicates that existing naming guidelines make it difficult for developers, especially novices, to come up with meaningful, concise, and compact method names~\cite{gao2019neural}. 
Moreover, developers might have different naming styles~\cite{caliskan2015anonymizing, hofmeister2017shorter}. It is also likely that a code generation system fails due to different styles in method names. 
Previous works~\cite{wang2021lightweight, xu2019method, ge2021keywords, qu2022method} focus on studying the impact of the method name quality on the readability and maintainability of source code. However, the role of the method name quality for code automation tasks has not been thoroughly investigated.

A possible approach to address the robustness challenge is to synthesize proper method names to replace those provided by developers, by which the performance of {\pdsig} approaches can hopefully be reinstated. Generating high-quality method names is an interesting task in its own right. 

\noindent\textit{Proposed solution.} In this paper, we propose a novel method, along with a tool suite, named {\tool} (neu\underline{RA}l co\underline{D}e gener\underline{A}tor \underline{R}obustifier), of two major components: {\tool}-Attack and {\tool}-Defense. 
Specifically, {\tool}-Attack imitates the undesirable behavior (just like typos) of developers mentioned above and then generates 
natural, visually, and semantically similar method names.
They serve as adversarial examples to reveal the robustness problem of PCGMs, but 
can also be considered as a tool to assess the robustness of PCGMs.
{\tool}-Defense, on the other hand, aims to reinstate the performance of PCGMs. One way is via adversarial training whereby we adapt the ACCENT approach \cite{zhou2021adversarial}, leveraging the generated adversarial examples to retrain a model.
The other is to sanitize the input whereby we propose a passive and lightweight defense method, which synthesizes meaningful and concise method names based on the given functional descriptions. 
These method names are inputted into the PCGMs together with the functional descriptions and other signature information (e.g., parameter lists). 

To evaluate the effectiveness of {\tool}, we consider six state-of-the-art, large-scale PCGMs (i.e., CodeGPT, PLBART, and CodeT5 \yg{in the fine-tuning code generation task and CodeGen, CodeT5+, and Replit in the zero-shot code generation task}). Experiment results show that {\tool}-Attack is effective in attacking these PCGMs, and {\tool}-Defense can improve their robustness and thus reinstate their performance 
by generating higher-quality method names. 
For instance, the CodeT5 model has a CodeBLEU value of 46.09 when not being attacked on the Java dataset, which drops to 31.58 under {\tool}-Attack. Using the method names synthesized by {\tool}-Defense, the CodeBLEU value is back to 46.11.


\smallskip

\noindent\emph{Contributions.} 
\begin{itemize}
	\item We devise {\tool}-Attack to attack PCGMs based on functional descriptions and signatures, showing that their performance is susceptible to provided method names. 
	
	\item We propose a defense method {\tool}-Defense to recover the performance of the attacked PCGMs. 
	
	\item As a byproduct, we provide novel approaches to automatically synthesize method names, which are meaningful in various contexts such as software refactoring. 
	
\end{itemize}
%
%
%


\smallskip

\noindent\emph{Key findings.} Based on our empirical study, we conclude that good names play a crucial role in neural code generation, and they can be synthesized from the functional description with a well-designed approach. In other words, functional description + parameter list + {\tool}-Defense would provide a strong performance boost for state-of-the-art PCGMs. To the best of our knowledge, this represents one of the first works on studying the robustness of neural code generation models via adversarial examples. 
More importantly, 
at the methodological level, this paper promotes, with solid evidence, the importance of studying the robustness of deep learning models in neural code generation and even software engineering in general, where they are playing an increasingly important role. 

To facilitate reproducibility and further research, source code, benchmarks, and experimental data
are released at \url{https://github.com/NTDXYG/RADAR}.
%

%
 

\smallskip
\noindent\emph{Structure.} The rest of the paper is organized as follows. Section~\ref{sec:ground} presents the related work. Section~\ref{sec:method} describes the framework and key approaches in {\tool}. Section~\ref{sec:setup} provides the experiment results and their analysis.  
Section~\ref{sec:discuss} discusses the quantitative study of the effectiveness of {\tool} and the potential threats to the validity of our empirical study. 
Section~\ref{sec:conclusion} concludes this paper and discusses future work.

\section{Related Work}
\label{sec:ground}


\subsection{Neural Code Generation}

Previous studies on code generation mainly
focus on domain-specific languages~\cite{ling2016latent, lin2018nl2bash, locascio2016neural}. 
Studies on code generation for general programming languages~\cite{mou2015end, see2017get} use sequence-to-sequence models, and they formalize code generation as text sequence generation based on the hypothesis of code naturalness~\cite{hindle2016naturalness, allamanis2018survey}. Some studies~\cite{yin2017syntactic, sun2020treegen} use tree-based models, by capturing the grammar of the natural language as a priori-knowledge to generate complex programs. Other studies~\cite{hayati2018retrieval, hashimoto2018retrieve} use retrieval-enhanced models, i.e., benefiting from information retrieval to compensate for the lack of ability of neural networks to memorize large and complex structures.

\yg{
In recent years, researchers have gradually utilized pre-trained models for neural code generation tasks, which can be classified into two types based on benchmark requirements: fine-tuning code generation tasks and zero-shot code generation tasks.
Fine-tuning code generation tasks are typically applied to benchmarks that lack test cases, such as CONCODE~\cite{iyer2018mapping} and CoNaLa~\cite{yin2018learning}. These benchmarks are divided into training, validation and test sets, with pre-trained models (often with parameter numbers less than a billion) fine-tuned on the training set to be adapted  to the specific task.
For example, models like CodeGPT~\cite{lu2021codexglue}, PLBART~\cite{ahmad2021unified}, and CodeT5~cite{wang2021codet5} leverage the GPT, BART, and T5 architectures of language models pre-trained on code corpora. Extensive evaluations on the CONCODE benchmark have demonstrated their robust code generation capabilities.
Moreover, models such as PyMT5~\cite{clement2020pymt5}, CoTexT~\cite{phan2021cotext}, and NatGen~\cite{chakraborty2022natgen} have also exhibited promising performance on code generation tasks, depending on the specific pre-training tasks. However, these models are more suitable for fine-tuning code generation tasks, as their parameter numbers are not large enough to demonstrate emergent capabilities in zero-shot scenarios.}

\yg{With the development of neural networks, Hestness et al.~\cite{hestness2017deep} point out that the performance of Transformer-based models improved in a predictable way as the amount of computation or the size of the network increased, and is called ``scaling laws''~\cite{ kaplan2020scaling}. When the model scales to a certain level, the phenomenon of ``emergent capacity" ~\cite{weiemergent} can occur. 
Building upon this understanding, researchers have increasingly employed large language models with over a billion parameters for zero-shot code generation tasks. These models have demonstrated substantial enhancements in the performance of code generation benchmarks, aligning with the aforementioned theory.
}

\yg{
The zero-shot code generation task is typically applied to benchmarks that include test cases but often have limited data size due to the costly manual construction of test cases.
In this context, Chen et al~\cite{chen2021evaluating} first introduced and evaluated the capabilities of Codex, which is pre-trained on GitHub code with 12 billion model parameter.
Subsequently, Li et al.~\cite{li2022competition} proposed AlphaCode with 1.1 billion parameters, and Chowdhery et al.~\cite{chowdhery2022palm} introduced PaLM-Coder, 
with 540 billion parameters. These models were evaluated for their performance on HumanEval.
However, all of these models are of closed-source.
For the open-source models, Fried et al.~\cite{fried2022incoder} proposed InCoder, which is trained for program synthesis (via left-to-right generation) and editing (via masking and infilling).
Nijkamp et al.~\cite{nijkamp2022codegen, nijkamp2023codegen2} proposed CodeGen and CodeGen2, which are large language models for code with multi-turn program synthesis.
Zheng et al.~\cite{zheng2023codegeex} proposed CodeGeeX, a multilingual model with 13 billion parameters for code generation. CodeGeeX is pre-trained on 850 billion tokens of 23 programming languages.
Li et al.~\cite{li2023starcoder} proposed StarCoder, a 15.5 billion parameter model with an 8K context length, infilling capabilities, and fast large-batch inference enabled by multi-query attention. 
In addition, the Replit company proposed replit-code-v1-3b model~\cite{replit}, which is trained on a subset of the Stack Dedup v1.2 dataset, and the training mixture includes 20 different languages.
Differing from the aforementioned decoder-only model, Wang et al.~\cite{wang2023codet5+} introduced CodeT5+, a family of encoder-decoder LLMs for code-related tasks.
}

\yg{In contrast to the previous studies, our primary objective is to evaluate the influence of method names on neural code generation from the perspective of model robustness.
We have observed a significant improvement in the performance of neural code generation when incorporating signature information as input. This observation has motivated us to further investigate the impact of method names, an essential component of signatures, on the code generation process.
By examining the relationship between method names and code generation, we  gain insights into the overall robustness and effectiveness of neural models in generating high-quality code.
To achieve this objective, we have conducted empirical investigations on both fine-tuning code generation tasks and zero-shot code generation tasks.}


\subsection{Adversarial Attack on Code-related Models}

Adversarial attacks on code can be divided into two categories: white-box adversarial attacks and black-box adversarial attacks.
\yg{
These attack methods  differ primarily in their underlying assumptions.
In the case of white-box attacks, the attacker assumes access to the internal structure of the victim models and their training parameters. For instance, Yefet et al.~\cite{yefet2020adversarial} proposed the white-box attack method DAMP, which leverages gradient information from the victim model to manipulate variables in the code.
However, white-box attacks are often less practical in real-world scenarios. This is because victim models are typically deployed remotely, and obtaining  model's internal  details can be challenging or even impossible.
}

\yg{
In contrast to white-box attacks, black-box attacks assume that the attacker has no knowledge of the internal details of the victim models and can only interact with the model through its output.}
For instance, Applis et al.~\cite{applis2021assessing} proposed LAMPION, a method that evaluates the robustness of the CodeBERT model by generating new code snippets that are equivalent to the original test set.  
Zhang et al.~\cite{zhang2020generating} proposed MHM, which utilizes Metropolis-Hastings sampling-based identifier renaming to perform code obfuscation.  
Tian et al.~\cite{tian2021generating} proposed QMDP, a Q-learning-based Markov decision process, which enables semantically equivalent transformations on the structure of source code.
\yg{Rabin et al.~\cite{rabin2021generalizability} employed variable renaming to evaluate the generalizability of neural program analyzers for the task of method name prediction.
Liguori et al.~\cite{liguori2022can} explored the use of unseen synonyms and missing information to evaluate line-based code generation tasks. 
Zeng et al.~\cite{zeng2022extensive} employed a wide range of NLP-based adversarial attack methods to evaluate pre-trained models and discovered that random attack methods can outperform carefully designed adversarial attack methods in most cases.}

In recent research, there has been a growing focus on addressing the naturalness aspect of adversarial examples.  
\yg{Yang et al.~\cite{yang2022natural} proposed a naturalness-aware attack called ALERT, which takes into account the natural semantics of generated examples. ALERT generates multiple natural candidates using the GraphCodeBERT model and the mask language model task in the CodeBERT model. It then calculates the cosine similarity to filter out natural and similar adversarial samples.}
Zhou et al.~\cite{zhou2021adversarial} proposed ACCENT, an identifier substitution approach for crafting adversarial code snippets in source code summarization. ACCENT aims to generate code snippets that are syntactically correct and semantically similar to the original code snippet. 
\yg{Zhang et al.~\cite{zhang2022towards} introduced CARROT, an optimization-based attack technique that assesses and improves the robustness of deep program processing models.
Wang et al.~\cite{wang2022recode} presented ReCode, a tool that provides over 30 transformations specifically designed for code generation. These transformations cover various aspects such as document strings, function and variable names, code syntax, and code formatting. Notably, six of these transformations are dedicated to modifying function names.}

\yg{
Moreover, due to the extensive search space of adversarial examples, numerous attack methods utilize optimization algorithms to enhance the efficiency of searching and thus improve the attack performance. In the field of natural language processing, commonly employed optimization algorithms include greedy algorithms~\cite{yang2020greedy}, genetic algorithms~\cite{alzantot2018generating}, and particle swarm optimization algorithms~\cite{zang2020word}. These optimization algorithms are also widely applied in adversarial attack methods for code-related tasks.}

\yg{Different from the previous studies, we mainly focus on black-box attacks specifically targeting code generation. 
This choice is driven by the practical consideration that, in real-world scenarios, neither users nor attackers have access to the internal structure of PCGMs.
Furthermore, our approach encompasses not only the generation of semantically equivalent adversarial examples but also considers aspects such as typos and visual similarity. By broadening the scope of our attack methodology, we aim to explore various types of adversarial examples.
To enhance the efficiency of attacking PCGMs, we employ genetic algorithms, which play a vital role in optimizing the search process and improving the effectiveness of our attacks.}

\subsection{Adversarial Defense on Code-related Models}
\label{sec:related-defense}
Current studies on adversarial defense for code-related tasks mainly focus on active defense. 
\yg{Bielik et al.~\cite{bielik2020adversarial} attempted adversarial defense with the assistance of gradient-based adversarial training method~\cite{goodfellow2014explaining}. They observed that relying solely on gradient-based adversarial training can provide insights into the model's robustness but may also lead to a decline in performance on the original task.
} 
Zhang et al.~\cite{zhang2020generating} and Yang et al.~\cite{yang2022natural} proposed the adversarial training method, which uses adversarial examples for data augmentation to re-train the model. 
However, this approach is highly dependent on the quality of adversarial examples. 
Zhou et al.~\cite{zhou2021adversarial} and Zhang et al.~\cite{zhang2020training} proposed a lightweight adversarial training method named mask training algorithm, which reduces the model's dependence on the non-robust features since any perturbations on these features may cause a large-scale change in the output. 

\yg{In contrast to the previous studies, our defense method proposes a unique passive defense approach that efficiently reinstates the performance of PCGMs. This defense method holds particular value in scenarios where PCGMs cannot undergo fine-tuning, such as zero-shot code generation tasks.
By employing this passive defense method, our objective is to substantially enhance the robustness of PCGMs, ensuring their effectiveness even in challenging zero-shot code generation scenarios.}

\section{Approach} \label{sec:method}

\begin{figure*}[htbp]
\vspace{-0.3cm}
	\centering
	\includegraphics[width=1\textwidth]{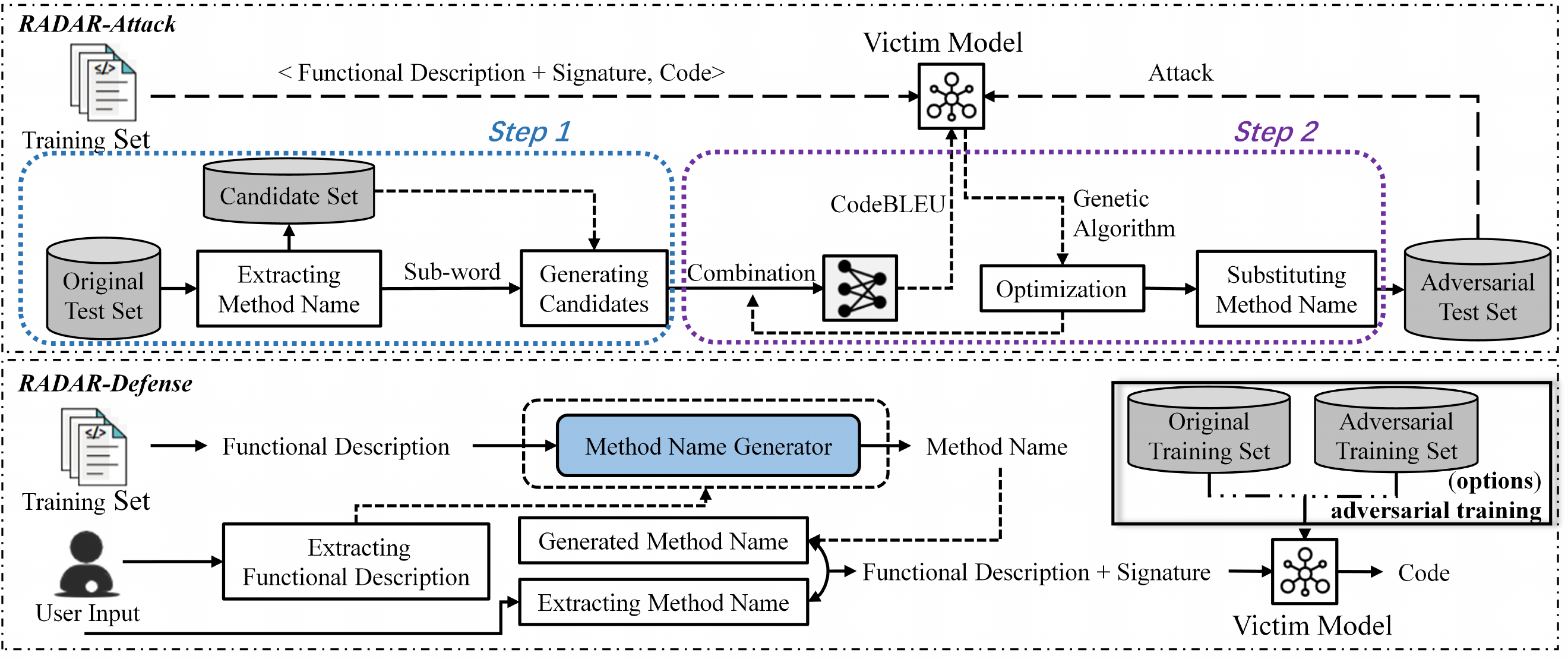}
	\caption{The Framework of {\tool}}
	\label{fig:approach}
\end{figure*}

We show an overview of {\tool} in Fig.~\ref{fig:approach} and
{\tool} includes two main parts: {\tool}-Attack and {\tool}-Defense.
In particular, {\tool}-Attack proposes a black-box, gradient-free optimization attack algorithm and {\tool}-Defense proposes a passive defense method based on retrieval-enhanced prompt learning for passive defense. 


\subsection{{\tool}-Attack}
\label{sec:attack}

\begin{algorithm}[htbp]
	\small
	\caption{Adversarial Example Generation Algorithm} \label{alg:Framwork} 
	\KwIn{
	Pre-trained Code Generation Model $\mathcal{F}$;\\
	Code Generation	DataSet $D$, where $(\boldsymbol{x},\boldsymbol{y})\in D$\\}
	\KwOut{
		Adversarial DataSet $D_{adv}$;}
	Initialize: Candidate Method Name Set $V \gets \emptyset$, Adversarial DataSet $D_{adv}\gets \emptyset$\;
	\For{each $(\boldsymbol{x}, \boldsymbol{y})\in D$} {
	    Extract the method name in $\boldsymbol{x}$\;
		$V \gets V \cup$ \{$M \mid$ $M = <m_{1}, \dots, m_{n}>$ to represent the sequence of sub-words from the method name\};
	}
	Training Method Name Embedding $Embed$ via $V$\;
	\For {each $(\boldsymbol{x}, \boldsymbol{y})\in D$} {
	    Extract the method name set $M$ in $\boldsymbol{x}$\;
	    Adversarial method name set $M^{\prime} \gets \emptyset$\;
        \For{each $\boldsymbol{m}\in M$} {
            $M^{\prime} \gets L_m$ based on semantic and visual similarity via $Embed$ in $V$\;
        }
        $t \gets 0$\;
        Initial population generation $\mathcal{P}^{t}$\;
        \While{$t \leq max\_iterations$} {
    		Calculate fitness value\;
            Selection\;
            Crossover\;
            \If{mutation\_prob $\ge$ random\_prob}{
    			Mutation\;
    		}
            $M^{\prime} \gets$ minimize evaluate fitness of $\mathcal{P}^{t}$\;
            \If{Minimum fitness value is not updated in $n$ iters}{
                Early stop\;
            }
            $\mathcal{P}^{t + 1} \gets buildNewGeneration(\mathcal{P}^{t}) $\;
            $t \gets t + 1$\;
        }
        $D_{adv} \gets$ $D_{adv} \cup \{(\boldsymbol{x}.replace(M, M^{\prime}), \boldsymbol{y})\}$\;
	}
    \Return $D_{adv}$\; 
\end{algorithm}


\yg{
In the fine-tuning code generation task, we commence by fine-tuning pre-trained code generation models using a provided dataset. This process yields a model $\mathcal{F}$, which maps each pair $\mathbf{x}$ consisting of functional description and signature to code $\boldsymbol{y}=\mathcal{F}(\mathbf{x})$.
In the zero-shot code generation task, we directly load the weights of the pre-trained model, resulting in the model $\mathcal{F}$.}
For attacking model $\mathcal{F}$, our goal is to generate an adversarial example $\boldsymbol{x_{adv}}$ for a given $\boldsymbol{x}$, which is visually and semantically similar to $\boldsymbol{x}$, but minimizes the $\mathrm{CodeBLEU}$ score between $\boldsymbol{y}$ and $\mathcal{F}(\boldsymbol{x_{adv}})$. 
Recall that $\mathrm{CodeBLEU}$ is a widely recognized automatic evaluation metric of code generation, which subsumes BLEU in the $n$-gram match and injects code syntax via abstract syntax trees (AST) and code semantics via data flow analysis. In the absence of test cases, CodeBLEU offers a sensible surrogate for automated evaluation.
\yg{
Given the expense of manual test case construction and the absence of corresponding test cases in most datasets, we have utilized CodeBLEU as the optimization objective function for both fine-tuning code generation tasks and zero-shot code generation tasks.}
Meanwhile, there is a correlation between the metrics, as seen in Table~\ref{tab:RQ1-Python}, Table~\ref{tab:RQ1-Java} and Table~\ref{tab:RQ2}, when the CodeBLEU value increases the BLEU metric also increases, so to some extent neither the choice of CodeBLEU nor BLEU has much influence on the selection of the adversarial example.
Formally, we aim to solve the following problem:
\begin{equation}
\label{equ:1}
\boldsymbol{x_{adv}}=\underset{\boldsymbol{x^{\prime}}}{\arg \min } \boldsymbol{CodeBLEU}\left(\boldsymbol{y}, \mathcal{F} (\boldsymbol{x^{\prime}} )\right)
\end{equation}
Note that we only consider part of the input $\boldsymbol{x}$ when generating adversarial examples; we only modify the method name in $\boldsymbol{x}$ (i.e., part of the signature), as parameters are optional for the signature. 
%
We assume that the attacker is unaware of the model architecture, parameters, and training data, and can only interact with the model through its output. Therefore, instead of utilizing the gradient-based optimization, we adopt a gradient-free optimization attacking approach, based on a genetic algorithm (GA) as shown in Algorithm~\ref{alg:Framwork}.

In Algorithm~\ref{alg:Framwork}, {\tool}-Attack first extracts method names from all the signatures in the dataset and then tokenize each method name according to the method naming convention (e.g., the camel case or the snake case) to build a set of sub-words.
{\tool}-Attack then creates a candidate set for each sub-word. The candidates are selected based on their \textbf{visual similarity} (to model typos) and \textbf{semantics similarity} (to model programmers' preferences of the use of English words).
Finally, {\tool}-Attack generates adversarial examples for method names by considering various combinations. It uses GA to generate the best replacement for the original method name by minimizing the CodeBLEU value~\cite{ren2020codebleu}.
We now elucidate these two main steps, i.e., \textit{Step 1} candidates generation (the \textcolor{cyan}{blue} box in Fig.~\ref{fig:approach}) and \textit{Step 2} optimization with GA (the \textcolor{violet}{purple} box in Fig.~\ref{fig:approach}). 

\subsubsection{Step 1. Candidates Generation} 

The first step aims to generate high-quality candidate adversarial examples that have high visual and semantic similarity with the original words.
According to previous studies~\cite{rawlinson2007significance, li2019textbugger}, the text semantic is likely to be retained or deduced after the user changes a few characters. Therefore, we make small-scale changes to the original words for human comprehension, which can help to generate visual similar candidates.
Moreover, as method names often contain a variety of domain-specific acronyms, jargon, and their combinations, they are frequently outside the vocabulary of the word embedding model in the general domain.
In this study, based on our previous work~\cite{zhou2021adversarial}, we first train a general word2vec~\cite{mikolov2013distributed} model based on the Wiki dataset and then continue to train it for a corpus of method names (Line 2-5 in Algorithm~\ref{alg:Framwork}). Finally, we select the $top$ $5$ nearest candidate sub-words for each sub-word in the method name based on the cosine similarity.

Based on these observations, we propose four operators to generate candidate samples (Line 9-10 in Algorithm~\ref{alg:Framwork}): 

\begin{itemize}
\item \textbf{Delete Operator:}  Randomly delete a character of the sub-word. 
\item \textbf{Swap Operator:} Randomly swap two adjacent letters in the word. 
\item \textbf{Replace-vis Operator:} 
Replace characters with visually similar characters (e.g., replacing “l” with “1”, “O” with “0”) or special coding styles words (e.g., replacing “2” with “to”, “4” with “for”).
\item \textbf{Replace-sem Operator:} Replace a sub-word in the method name with its most semantic similar Top5 candidate sub-words in a high-dimensional vector space.
\end{itemize}

\begin{figure}[htbp]
	\centering
	\includegraphics[width=0.7\textwidth]{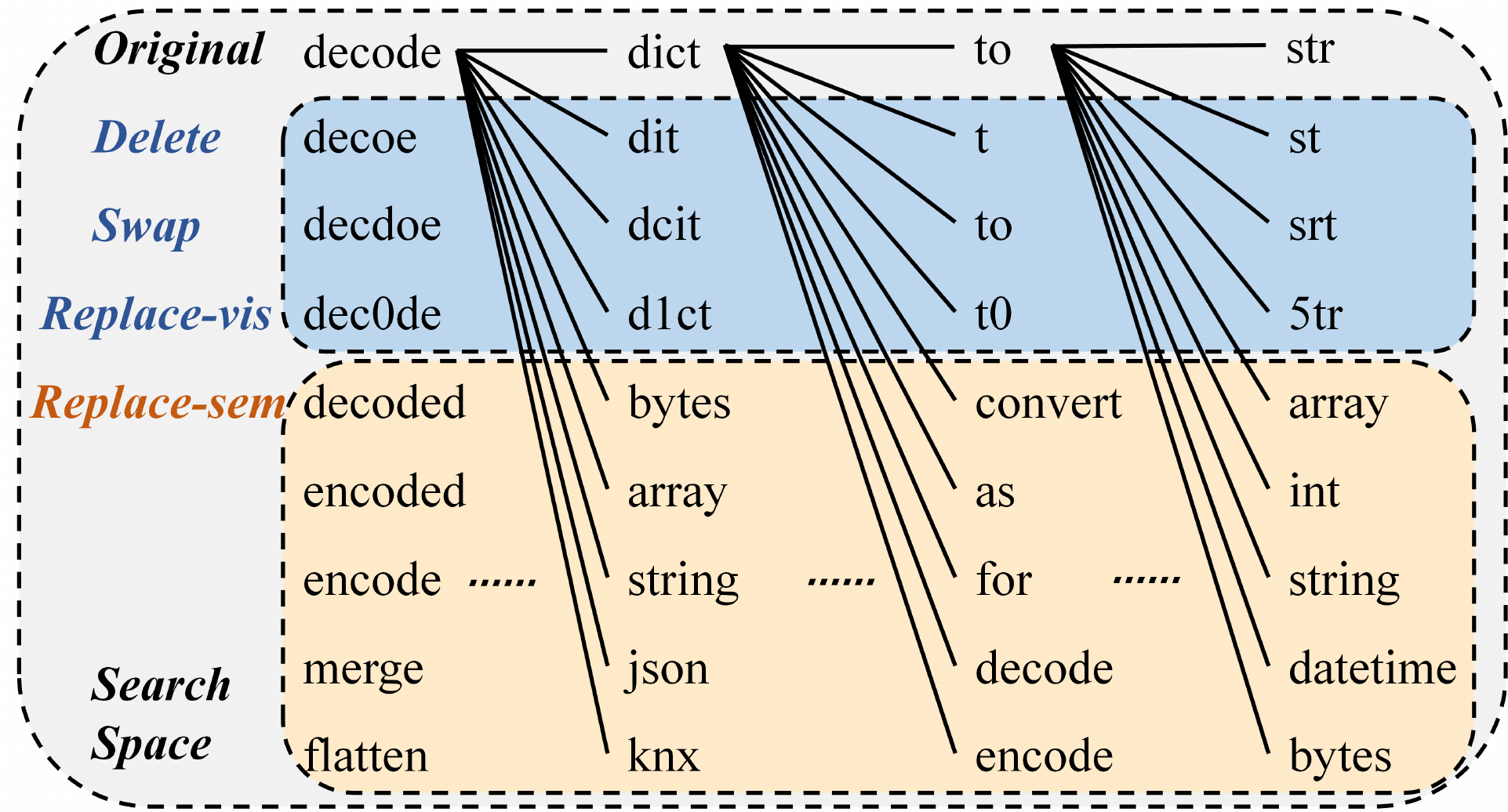}
	\caption{An Example of permutations of candidate sub-words}
	\vspace{-1mm}
	\label{fig:attack_example}
\end{figure}

Notice the first two operators are designed to model that developers type carelessly. The \textbf{Replace-vis} operator is designed to model the novice behaviors (e.g., copy the code from course materials to their program tasks).
%
An example in Fig.~\ref{fig:attack_example} illustrates the four operators. Method name \code{decode\_dict\_to\_str} can be divided into four sub-words (i.e., decode, dict, to, and str). Each operator generates multiple candidate sub-words, which form the discrete search space of the original sub-words.

\subsubsection{Step 2. Optimization with GA}
This step aims to find the most effective adversarial examples in the discrete search space that can successfully fool the victim model, with GA.
%
Let $\boldsymbol{M} = \langle m_{1}, \dots, m_{n}\rangle$ be the sequence of sub-words from the method name. 
The discrete search space can be represented as $\boldsymbol{M^{k}} = \{\langle m^{k}_{1},  \dots, m^{k}_{n}\rangle \mid m^{k}_{i} \in \mathbb{V}\left(m_{i}\right)\}$, 
where $k$ denotes the number of the generated candidate sub-words, 
$\mathbb{V}\left(m_{i}\right)$ is 
the set of candidates of $m_i$.

By Equation~\ref{equ:1}, the fitness function of {\tool}-Attack can be formalized as 
\[
\boldsymbol{y_{goal}}=\underset{\boldsymbol{M^{\prime}}}{\arg\min} \boldsymbol{CodeBLEU}\left(\boldsymbol{y}, \mathcal{F}\left(\boldsymbol{x}.replace(\boldsymbol{M}, \boldsymbol{M^{\prime}})\right)\right)
\]
where $\boldsymbol{M^{\prime}}$ represents the set of solutions with $n$ variables (i.e., the number of sub-words). Values of each variable are in the range $[0,k]$, where $k$ denotes the number of candidates.


We denote the initial population as the initial generation $\mathcal{P}^{0}$ (Line 12 in Algorithm~\ref{alg:Framwork}). The size of the population is denoted as $size\_population$. To get a new generation (i.e., transiting from $\mathcal{P}^{t}$ to $\mathcal{P}^{t+1}$), the operations of selection, crossover (with $crossover\_prob$), and mutation (with $mutation\_prob$) are performed (Line 14-18 in Algorithm~\ref{alg:Framwork}). The termination condition is the maximum number of generations,  which is denoted as $max\_iterations$.
To improve the computational efficiency of GA, we refer to the early-stop strategy used by Garcia et al.~\cite{garcia2006alternative}. The evolution ends when the average fitness of the population does not improve above a certain threshold in the last $n$ generations 
(Line 20-21 in Algorithm~\ref{alg:Framwork}).
To avoid experimental bias due to the randomness of GA, we repeat  the run 30 times, taking the average values as the final result. 

\begin{figure}[htbp]
	\centering
    \vspace{-1mm}
	\includegraphics[width=1\textwidth]{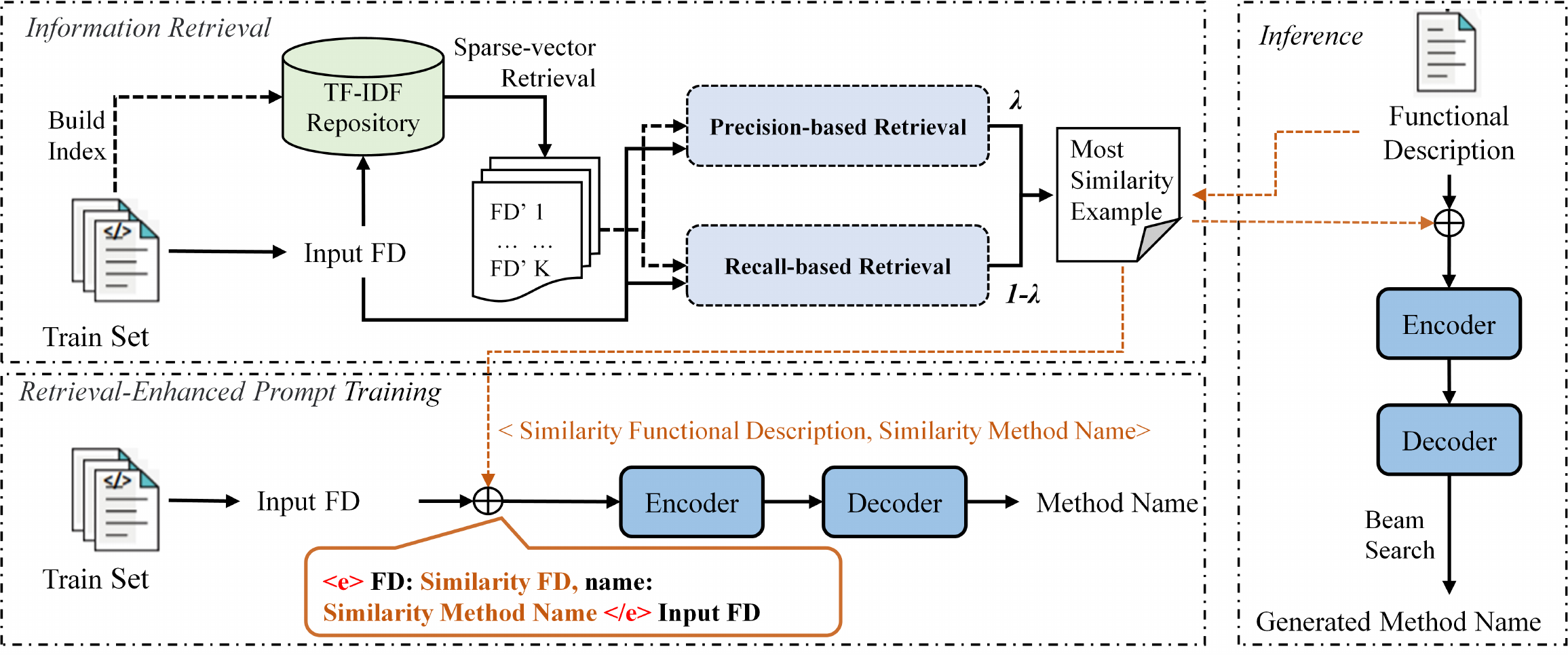}
	\caption{Overview of {\tool}-Defense}
	\label{fig:defense}
\end{figure}

\subsection{{\tool}-Defense}
\label{sec:defense}

{\tool}-Defense can adapt the adversarial training approach which leverages generated adversarial examples to retrain a model in the fine-tuning code generation task, but this is optional, as we mentioned in {\tool}-Attack the model black box assumption, we expect {\tool}-Defense is able to reinstate its performance without retraining PCGMs.
Thus {\tool}-Defense's main purpose is to synthesize a new method name for a given functional description to replace the original method name in the signature. As shown in Fig.~\ref{fig:defense}, {\tool}-Defense mainly consists of two steps: 1) generating the most similar example via information retrieval, and 2) training the model with the augmented function description via prompt training.


In general, we treat the training set as a corpus, 
from which a list of key-value pairs ($\mathcal{T}=\{(c_{i}, m_{i})\}$) can be constructed, with $c_{i}$ and $m_{i}$ denoting the functional description 
and the method name, respectively. 
Given a functional description $c$, the retrieval model aims to retrieve the most relevant example $z = (c_{r}, m_{r})$ from the corpus. To achieve this, we first retrieve top-$K$ similar functional descriptions from the corpus based on the standard TF-IDF due to low computational cost, out of which we further retrieve the most similar functional description based on lexical similarity.

First, we adopt standard TF-IDF\cite{aizawa2003information} and cosine distance; each functional description $c$ is associated with the semantic sparse-vector $\text{TF-IDF}(c) \in \mathbb{R}^{D}$, where $D$ denotes the total number of words in the corpus, and the similarity is defined as the 
cosine distance:
\[ 
semantic(a, b) = \frac{\operatorname{TF-IDF}(a) \cdot \operatorname{TF-IDF}(b)}{\|\operatorname{TF-IDF}(a)\| \cdot\|\operatorname{TF-IDF}(b)\|}
\]
%
%
%
%
Second, for lexical similarity, we utilize precision-based and recall-based retrieval methods. In our study, we use two evaluation metrics, (i.e., BLEU~\cite{papineni2002bleu} and ROUGE~\cite{lin2004rouge}), which measure the similarity based on precision and recall, respectively. 

For the given functional descriptions $a$ and $b$, lexical similarity can be computed as: 
\[ 
lexical(a, b) = \lambda\operatorname{BLEU}(a, b) + \left(1-\lambda\right)\operatorname{ROUGE}(a, b), 
\]
where $\lambda$ is a hyper-parameter for allowing the flexible control of precision and recall in information fusion. 


%
In the next step, we resort to a retrieval-enhanced prompt training approach. This approach is based on the observations from the previous studies~\cite{gu2018search, cai2019retrieval, parvez2021retrieval, li2022survey} that by giving a model access to external memory via information retrieval techniques, more information can be obtained in the model generation process and the uncertainty can be reduced. 

Recall that for the given functional description $c$, we obtain the most relevant sample $z = \left(c_{r}, m_{r}\right)$ in the first step. %
We augment $c$ to form a retrieval-enhanced functional description $c'$. 
\[
c^{\prime} = \langle\mathbf{e}\rangle \textbf{FD:} c_{r}, \textbf{name:} m_{r} \langle\mathbf{/e}\rangle \oplus c
\]
where, 
$z$ is tagged and concatenated with $c$, such that the model can learn the most similar functional description and method name information. 

Our model is based 
on UniXcoder~\cite{guo2022unixcoder}, a unified cross-modal pre-trained model 
which can support both code-related understanding and generation tasks based on \textit{Transformer}~\cite{vaswani2017attention}, and utilizes mask attention matrices with prefix adapters to control the access to context for each token.

For the input $c^{\prime}$, our model first tokenizes it to obtain an input sequence $\left\{\mathbf{c^{\prime}}_{i}\right\}_{i=1}^{|c^{\prime}|}$. We utilize UniXcoder to encode the $c^{\prime}$ and decode it to synthesize the method name. Note that the parameters of the encoder and decoder in UniXcoder are shared. The final decoder's output of the UniXcoder $\mathbf{H}^{t}$ is sent to a fully connected neural network. This network can pass a $\operatorname{softmax}$ layer to predict the probability of the next token, which can be defined as follows.
\[ 
p\left(m_{t+1} \mid m_{1}, \cdots, m_{t} \right)=\operatorname{softmax}\left(\mathbf{H}^{t} \mathbf{W}+\mathbf{b}\right)
\]

In model training, we use the Incomplete-Trust (In-trust)~\cite{huang2021named} loss function, viz.,
\[  
	\mathcal{L}_{\rm In-trust}(\theta)= \alpha \mathcal{L}_{\rm CE} (\theta) + \beta \mathcal{L}_{\rm DCE} (\theta )
\]
where $\mathcal{L}_{\rm CE}(\theta)=-\sum_{i=1}^{|m|} q \log p$
and $\mathcal{L}_{\rm DCE} (\theta)=-\sum_{i=1}^{|m|} p \log (\delta p + \left(1-\delta\right) q )$. Here $\mathcal{L}_{\rm CE}$ represents the Cross-Entropy function which is not noise-tolerant but benefits the convergence of the model, $\mathcal{L}_{\rm DCE}$ represents the robust Distrust-Cross-Entropy and can effectively prevent the model from overfitting noisy samples; $p$ denotes the model's prediction distribution and $q$ denotes the trust label distribution.
\section{Evaluation}
\label{sec:setup} 
We aim to evaluate the effectiveness of our approach by answering the following three research questions (RQs).

\begin{description}
\item[\textbf{RQ1}] How effective is {\tool}-Attack in degrading the performance of FD$^{\rm Sig}$ by attacking method names?
	
\item[\textbf{RQ2}] How effective is {\tool}-Defense in reinstating the performance of FD$^{\rm Sig}$?
	
\item[\textbf{RQ3}] How effective is {\tool}-Defense in terms of the method name generation?
	
	
\end{description}


\subsection{Experiment Design}
\subsubsection{Dataset}
\label{sec:dataset}

\yg{
In the fine-tuning code generation task, widely used open-source datasets include CONCODE~\cite{iyer2018mapping} for the Java language, and Django~\cite{oda2015learning}, CoNaLa~\cite{yin2018learning}, and Juice~\cite{agashe2019juice} for the Python language.}

\begin{figure*}[htbp]
\vspace{-0.3cm}
	\centering
	\includegraphics[width=1\textwidth]{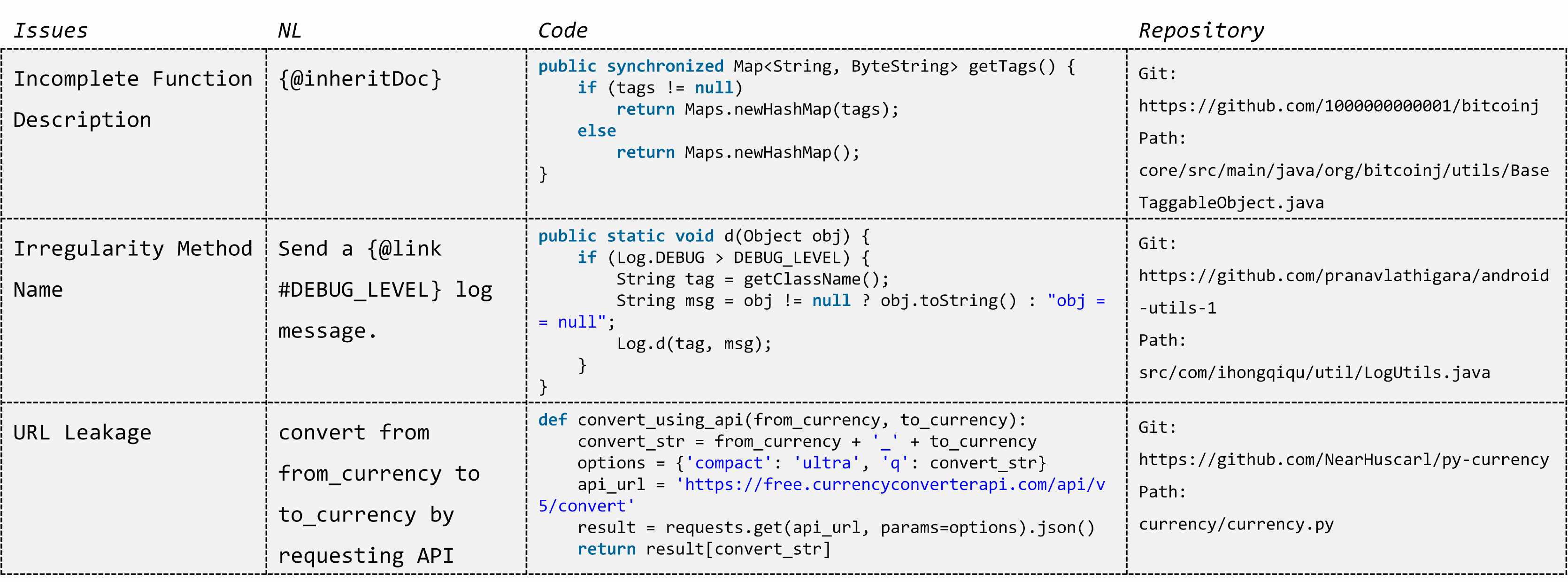}
	\caption{Irregularity issues in the common fine-tuning code generation dataset}
	\label{fig:dataset-issues}
\end{figure*}

\yg{
In our research, we have uncovered irregularity issues within specific datasets that can impact the quality and reliability of the data. These issues are illustrated in Fig.~\ref{fig:dataset-issues}, and we provide a detailed description of each problem.
For example, in the original CONCODE dataset, we have observed instances of incomplete function descriptions and irregular method names. These inconsistencies pose challenges and hinder the advancement of code generation tasks. To support our findings, we present specific examples and indicate their sources within the dataset.
Similarly, in the CodeSearchNet dataset~\cite{husain2019codesearchnet}, we have identified instances of URL leakage issues. These issues contribute to the presence of low-quality data, further limiting the progress in code generation tasks. To illustrate these concerns, we provide concrete examples along with relevant references.
The presence of irregularity issues and low-quality data in these datasets emphasizes the significance of addressing data quality concerns in code generation research.}

To evaluate our approaches in the fine-tuning code generation task, we need to construct new high-quality datasets to avoid these issues and biases, which include functional descriptions, signatures, and their corresponding code. 
To ensure the quality of our newly constructed datasets, we designed six heuristic rules to filter out noisy data items by following previous study~\cite{iyer2018mapping}. 


\begin{description}
\item[H1] The code needs to be parsed through the AST tool to ensure that the syntax is correct.

\item[H2]  The number of sub-words of the method name is no less than 2, and the length of each sub-word is no  more than 16.

\item[H3] The length of the functional description should be no more than 50 and no less than 4.

\item[H4]  The length of the code should be no more than 256.

\item[H5]  Remove annotation information, exception code, and URL information from the code.

\item[H6]  Unify method names in Java data to hump naming rules and unify method names in Python data to snake naming rules.
\end{description}

Our Java dataset is collected from the raw CONCODE~\cite{iyer2018mapping} dataset, which is 
from Java projects on GitHub, and our python dataset is collected from the raw PyTorrent~\cite{bahrami2021pytorrent} dataset, which is 
from Python package libraries on PyPI and Anaconda. 
%

\yg{In the context of the zero-shot code generation task, several popular open-source datasets are available. For the Java language, the Aix-bench dataset~\cite{hao2022AixBench} is commonly utilized. For the Python language, widely evaluated datasets include Human-Eval~\cite{chen2021evaluating}, MBPP~\cite{austin2021program}, and GSM8K-Python~\cite{chowdhery2022palm}. Among these datasets, Human-Eval is particularly prominent.
However, we have observed that the functional descriptions in the Human-Eval dataset contain test case prompts that include method names. To mitigate the potential impact of these method names on the code generated by the model, we adopt an approach of removing the test case prompts from the functional descriptions.
By eliminating the prompts related to the test cases, our aim is to minimize potential bias or influence that the method names in the prompts may have on the code generation process.}

Descriptive statistics of our datasets, including their length distributions of functional description (FD), signature (Sig), method name (MD), and Code, are provided in Table~\ref{tab:statistics}. 
Following the previous work~\cite{iyer2018mapping}, we randomly select 100,000 examples for training, 2,000 examples for validation, and 2,000 examples for testing in the fine-tuning code generation task.
\yg{For the zero-shot code generation task, the Human-Eval dataset consists of a total of 164 test data samples.}

\begin{table}[htbp]
\caption{Descriptive statistics of the datasets when tokenized by BPE algorithm}
\label{tab:statistics}
\begin{center}
	\begin{tabular}{ccccccc} 
		 \toprule 
		    \textbf{FD} & Avg & Mode & Median & $<16$ & $<32$ & $<64$ \\ 
		    Java & 14.25 & 8 & 11 & $69.52\%$ & $93.52\%$ & $99.99\%$ \\ 
		    Python & 17.88 & 8 & 13 & $58.45\%$ & $82.86\%$ & $99.85\%$ \\
		    \midrule 
		    \textbf{Sig} & Avg & Mode & Median & $<8$ & $<16$ & $<32$ \\ 
		    Java & 8.49 & 7 & 7 & $58.44\%$ & $93.94\%$ & $99.85\%$ \\ 
		    Python & 7.78 & 6 & 6 & $55.48\%$ & $96.92\%$ & $99.98\%$ \\
		    \midrule 
		    \textbf{MD} & Avg & Mode & Median & $<4$ & $<8$ & $<16$ \\ 
		    Java & 2.85 & 2 & 3 & $79.36\%$ & $99.58\%$ & $99.99\%$ \\ 
		    Python & 2.74 & 2 & 3 & $83.58\%$ & $99.92\%$ & $100\%$ \\
		    \midrule 
		    \textbf{Code} & Avg & Mode & Median & $<64$ & $<128$ & $<256$ \\ 
		    Java & 40.46 & 28 & 38 & $88.86\%$ & $99.99\%$ & $100\%$ \\ 
		    Python & 69.44 & 42 & 63 & $50.38\%$ & $92.54\%$ & $100\%$ \\
		 \bottomrule 
		\end{tabular}
	\end{center}
\end{table}

\subsubsection{Victim Models}
\label{victim}
The victim models (i.e., the target models under adversarial attacks) are based on large-scale pre-trained language models for source code, which can represent state-of-the-art research for the code generation task.

\yg{
In the context of the fine-tuning code generation task, we selected CodeGPT, PLBART, and CodeT5 as our models. These models have parameter sizes ranging from 100 million to 300 million.}

\begin{itemize}[leftmargin=*]
	\item \textbf{CodeGPT}~\cite{lu2021codexglue} is a Transformer-based decoder-only model inspired by GPT~\cite{radford2019language}, following similar pre-training tasks of GPT including the causal language model.
	
	\item \textbf{PLBART}~\cite{ahmad2021unified} is a Transformer-based encoder-decoder model inspired by BART~\cite{lewis2020bart}, following similar pre-training tasks of BART, including token masking, token deletion, and token infilling.
	
	\item \textbf{CodeT5}~\cite{wang2021codet5} is a Transformer-based encoder-decoder model inspired by T5\cite{raffel2020exploring}. It proposes a novel identifier-aware pre-training task to leverage code-specific structural information.
\end{itemize}

\yg{In the context of the zero-shot code generation task, we selected Replit, CodeGen, and CodeT5+ with the best performance within 3 billion parameters, based on the evaluation results of Gunasekar et al.~\cite{gunasekar2023textbooks} and Wang et al.~\cite{wang2023codet5+}}

\begin{itemize}[leftmargin=*]
	\item \yg{\textbf{Replit}~\cite{lu2021codexglue} is a Transformer-based decoder-only model~\cite{radford2019language}, which uses Flash Attention~\cite{dao2022flashattention} for efficient training and inference, and incorporates AliBi positional embeddings~\cite{press2021train} to handle variable context length during inference.}
	
	\item \yg{\textbf{CodeGen}~\cite{ahmad2021unified} is a Transformer-based decoder-only model, which adopts rotary position embedding for improving the ability to handle long documents, and uses JAX~\cite{bradbury2018jax} for training the model.}
	
	\item \yg{\textbf{CodeT5+}~\cite{wang2021codet5} is a Transformer-based encoder-decoder model, which employs a ``shallow encoder and deep decoder” architecture~\cite{li2022competition}, both encoder and decoder are initialized from pretrained checkpoints and connected by cross-attention layers.}
\end{itemize}
\subsubsection{Baselines}
As for baselines, we select six attack methods to generate adversarial examples, 
one 
defense method to improve the robustness of PCGMs, as well as eight method name generation methods, which are described below.

\noindent\textbf{Baselines for adversarial attack and defense.} 
In terms of the baselines for the adversarial attack, we select Foo-Attack, Random-Attack, ALERT-Attack, Genetic-Attack, ReCODE-Attack, and ACCENT-Attack. 

\begin{itemize}[leftmargin=*]
\item \yg{\textbf{Foo-Attack} is the attack method we introduced in the motivation, involving the replacement of all method names with the term ``foo''.}

\item \yg{\textbf{Random-Attack} is a method proposed by Zeng et al.~\cite{zeng2022extensive} that involves randomly substituting method names. In their empirical study, Random-Attack demonstrates improved attack effectiveness compared to existing NLP-based adversarial attack algorithms.}

\item \yg{\textbf{ALERT-Attack}  is a method proposed by Yang et al.~\cite{yang2022natural}. It utilizes CodeBERT and GraphCodeBERT to generate natural candidates and employs a combination of greedy search and genetic algorithm for optimization.}

\item \yg{\textbf{Genetic-Attack} is a method proposed by Alzantot et al.~\cite{alzantot2018generating}. It utilizes Glove and GoogleLM to generate candidates and employs a genetic algorithm for optimization.
}

\item \yg{\textbf{ReCODE-Attack} is a method proposed by Wang et al.~\cite{wang2022recode}. It utilizes rule-based transformations to generate candidates and employs a greedy search for optimization.}

\item \yg{\textbf{ACCENT-Attack} is a method proposed by Zhou et al.~\cite{zhou2021adversarial}. It first selects several of the most important tokens and then employs word2vec to generate candidates.}
\end{itemize}

\yg{
When addressing adversarial defense, several common defense methods can be employed, such as gradient-based adversarial training, data augmentation, and mask training (proposed by ACCENT-Defense). It is important to note that gradient-based adversarial training may lead to a decline in model performance, while the effectiveness of data augmentation relies on the quality of the adversarial samples.
Among these defense methods, ACCENT-Defense stands out as a lightweight mask learning approach based on active defense. Its objective is to enhance both the robustness and performance of the model. Given its effectiveness and relevance to our research, we consider ACCENT-Defense as the primary baseline for our study.}

\noindent\textbf{Baselines for method name generation.}
We consider eight 
name generation methods, which are classified into three groups: information-retrieval (including BM25\cite{robertson2009probabilistic}, NNGen~\cite{liu2018neural}, and CCGIR~\cite{yang2022ccgir}), deep-learning (including RNN-Att-Copy~\cite{gao2019neural}, CodeBERT~\cite{feng2020codebert}, and UniXcoder~\cite{guo2022unixcoder}), and retrieval-enhanced methods (including Rencos~\cite{zhang2020retrieval} and REINA~\cite{wang2022training}).

These methods are widely used in method name generation, text summarization, and code summarization. In this study, we train them with functional descriptions as the input and method names as the output, as per the individual model.

\begin{table}[htbp]
	\caption{Hyper-parameters settings of {\tool}}
	\vspace{-1mm}
	\begin{center}
		\begin{tabular}{ccc}
			\toprule
			\textbf{Category} & \textbf{Hyper-parameter Name} & \textbf{Hyper-parameter Value} \\ \midrule
			\multirow{4}{*}{{\tool}-Attack} 
			& size\_population   & 20     \\
			& max\_iterations    & 50     \\
			& crossover\_prob & 0.9   \\
			& mutation\_prob & 0.001   \\
			& early\_stop   & 3   \\
			\midrule
			\multirow{10}{*}{{\tool}-Defense} & top-$K$ in Java  & 9  \\
			& $\lambda$ in Java      & 0.6  \\
			& top-$K$ in Python  & 3  \\
			& $\lambda$ in Python      & 0.1  \\
			& max\_source\_length      & 128  \\
			& max\_target\_length      & 24  \\
			& batch\_size      & 64  \\
			& max\_epoch  & 50  \\
			& learning\_rate      & 4e-5  \\
			& early\_stop   & 3   \\
			\bottomrule
		\end{tabular}
	\end{center}
	\vspace{-1mm}
	\label{Hyper-parameters}
\end{table}

\subsubsection{Evaluation Metrics and Hyper-parameters}
\yg{
To assess the effectiveness of adversarial attacks in the fine-tuning code generation task, we consider three evaluation metrics: BLEU~\cite{papineni2002bleu}, CodeBLEU~\cite{ren2020codebleu}, and Attack Success Rate (ASR)~\cite{zhou2021adversarial}. ASR is defined as the percentage of generated adversarial examples that successfully decrease the CodeBLEU score of the generated code.
For the zero-shot code generation task, we employ four evaluation metrics: BLEU, CodeBLEU, Pass@1~\cite{chen2021evaluating}, and Attack Success Rate (ASR). ASR is defined as the percentage of generated adversarial examples that successfully reduce the Pass@1 score of the generated code.
}
For method name generation, we use three evaluation metrics, i.e., Exact Match (EM)~\cite{gao2019neural}, BLEU and Edit Distance (ED)~\cite{gao2019neural}.
These performance measures have been widely used in previous studies for neural code generation and automatic method name generation~\cite{gao2019neural, zhou2021adversarial, wang2021codet5, lu2021codexglue, guo2022unixcoder, feng2020codebert, guo2021graphcodebert}. 
Note that the scores of \emph{BLEU}, \emph{CodeBLEU}, \emph{Pass@1}, \emph{Exact Match}, and \emph{Success rate} are in the range of [0,1]; 
the higher, the better. \emph{Edit Distance} is measured in actual values; the smaller, the better.
		
The hyper-parameters are optimized according to actual performance and the values are summarized in Table~\ref{Hyper-parameters}. 
The first four rows mean the parameters of GA in {\tool}-Attack and the following rows mean the parameters of model training and inference in {\tool}-Defense.
For the implementation of GA, we utilize the  scikit-opt\footnote{\url{https://github.com/guofei9987/scikit-opt}} library. 
For the implementation of {\tool}-Defense, we utilize the Pytorch\footnote{\url{https://pytorch.org/}} and Transformers\footnote{\url{https://github.com/huggingface/transformers}} libraries.

\subsubsection{Experiment Platform}

All the experiments were run on Intel(R) Xeon(R) Silver 4210 CPU and GeForce RTX3090 GPU with 24 GB memory. The operating system is Linux Debian. 


		

\subsection{Experimental Results}\label{sec:results}

\noindent\textbf{RQ1: How effective is {\tool}-Attack in degrading the performance of FD$^{\rm Sig}$ by attacking method names?}

We investigate whether the existing {\pdsig} PCGMs are vulnerable to method name attacks, and in case they are, whether our defense method can reinstate their performance.
\yg{As discussed in Section \ref{victim}, we include three PCGMs, namely CodeGPT, PLBART, and CodeT5, for the fine-tuning code generation task. For the zero-shot code generation task, we consider three PCGMs, namely Replit, CodeGen, and CodeT5+.} 
Here we consider four performance measures (i.e., BLEU, CodeBLEU, Pass@1, Attack Success rate), which have been widely used in previous studies of neural code generation~\cite{svyatkovskiy2020intellicode, lu2021codexglue, chen2021evaluating, clement2020pymt5, ahmad2021unified, phan2021cotext, wang2021codet5, chakraborty2022natgen} and adversarial example generation~\cite{yefet2020adversarial, tian2021generating, liguori2022can, zeng2022extensive, applis2021assessing, yang2022natural, zhou2021adversarial}.

\begin{table*}[]
\caption{Evaluation results of comparing {\tool} and the baselines in terms of adversarial attack in the Java dataset}
\begin{center}
\vspace{-1mm}
\setlength{\tabcolsep}{3mm}{
    \begin{tabular}{c|c|ccc}
        \toprule
        \multicolumn{1}{c|}{Model} & \multicolumn{1}{c|}{Method} & 
        \multicolumn{1}{c}{BLEU} & \multicolumn{1}{c}{CodeBELU} & \multicolumn{1}{c}{ASR} \\
        \midrule
        \multirow{9}{*}{CodeGPT}
        & \pd & 11.56 &  14.78 & \text { -- }  \\
        & \pdsig & 23.18 &  26.33 & \text { -- } \\
        \cline{2-5} 
        & Foo-Attack & 16.95 ($\downarrow$ 26.88\%) & 20.09 ($\downarrow$ 23.70\%) & 55.40\%  \\
        & Random-Attack & 15.52 ($\downarrow$ 33.05\%) & 19.82 ($\downarrow$ 24.72\%) &  58.25\% \\
        & ALERT-Attack & 13.85 ($\downarrow$ 40.25\%) & 17.24 ($\downarrow$ 34.52\%) &  65.52\% \\
        & Genetic-Attack & 14.25 ($\downarrow$ 38.52\%) & 17.88 ($\downarrow$ 32.09\%) &  60.48\% \\
        & ReCODE-Attack & 15.11 ($\downarrow$ 34.81\%) & 18.48 ($\downarrow$ 29.81\%) & 59.58\% \\
        & ACCENT-Attack & 14.31 ($\downarrow$ 38.27\%) & 17.60 ($\downarrow$ 33.16\%) & 61.05\% \\
        & {\tool}-Attack & \textbf{13.02 ($\downarrow$ 43.83\%)} & \textbf{16.13 ($\downarrow$ 38.74\%)} &  \textbf{67.25\%}  \\
        \midrule 
        \multirow{9}{*}{PLBART}
        & \pd & 20.84 & 29.38  & \text {--} \\
        & \pdsig & 35.19 &  43.71  & \text { -- } \\
        \cline{2-5} 
        & Foo-Attack & 27.47 ($\downarrow$ 21.94\%) & 36.32 ($\downarrow$ 16.91\%) &  56.15\% \\
        & Random-Attack & 25.22 ($\downarrow$ 28.33\%) & 33.67 ($\downarrow$ 22.97\%) & 58.85\% \\
        & ALERT-Attack & 23.52 ($\downarrow$ 33.16\%) & 32.62 ($\downarrow$ 25.37\%) &  63.58\% \\
        & Genetic-Attack & 22.85 ($\downarrow$ 35.07\%) & 31.52 ($\downarrow$ 27.89\%) & 67.20\% \\
        & ReCODE-Attack & 24.59 ($\downarrow$ 30.12\%) & 32.98 ($\downarrow$ 24.55\%) & 62.48\% \\
        & ACCENT-Attack & 23.34 ($\downarrow$ 33.67\%) & 32.53 ($\downarrow$ 25.58\%) & 64.40\% \\
        & {\tool}-Attack & \textbf{22.61 ($\downarrow$ 35.75\%)} & \textbf{31.31 ($\downarrow$ 28.37\%)} & \textbf{67.60\%}  \\
        \midrule 
        \multirow{9}{*}{CodeT5}
        & \pd & 20.53 &  30.43 & \text { -- } \\
        & \pdsig & 38.45 &  46.09 & \text { -- } \\
        \cline{2-5} 
        & Foo-Attack & 31.21 ($\downarrow$ 18.83\%) & 37.83 ($\downarrow$ 17.92\%) & 54.15\% \\
        & Random-Attack & 28.74 ($\downarrow$ 25.25\%) & 36.39 ($\downarrow$ 21.05\%) &  59.10\% \\
        & ALERT-Attack & 26.40 ($\downarrow$ 31.34\%) & 34.16 ($\downarrow$ 25.88\%) & 64.88\% \\
        & Genetic-Attack & 25.45 ($\downarrow$ 33.81\%) & 33.66 ($\downarrow$ 26.97\%) & 67.52\% \\
        & ReCODE-Attack & 25.87 ($\downarrow$ 32.72\%) & 33.95 ($\downarrow$ 26.34\%) & 66.21\% \\
        & ACCENT-Attack & 25.81 ($\downarrow$ 32.87\%) & 33.38 ($\downarrow$ 27.58\%) & 66.25\% \\
        & {\tool}-Attack & \textbf{24.48 ($\downarrow$ 36.33\%)} & \textbf{31.58 ($\downarrow$ 31.48\%)} & \textbf{74.65\%}\\
        \bottomrule
    \end{tabular}
}
\end{center}
\label{tab:RQ1-Java}
\end{table*}
	
\begin{table*}[]
\caption{Evaluation results of comparing {\tool} and the baselines in terms of adversarial attack in the Python dataset}
\begin{center}
\vspace{-1mm}
\setlength{\tabcolsep}{3mm}{
    \begin{tabular}{c|c|ccc}
        \toprule
        \multicolumn{1}{c|}{Model} & \multicolumn{1}{c|}{Method} & 
        \multicolumn{1}{c}{BLEU} & \multicolumn{1}{c}{CodeBELU} & \multicolumn{1}{c}{ASR} \\
        \midrule
        \multirow{9}{*}{CodeGPT}
        & \pd & 5.06 &  18.77 & \text { -- }  \\
        & \pdsig & 11.94 &  24.27 & \text { -- }  \\
        \cline{2-5} 
        & Foo-Attack & 9.02 ($\downarrow$ 24.46\%) & 22.10 ($\downarrow$8.94\%) & 56.05\%  \\
        & Random-Attack & 8.11 ($\downarrow$ 32.08\%) & 20.88 ($\downarrow$ 13.97\%) & 56.55\%  \\
        & ALERT-Attack & 7.94 ($\downarrow$ 33.50\%) & 18.47 ($\downarrow$ 23.90\%) &  61.20\% \\
        & Genetic-Attack & 7.48 ($\downarrow$ 37.35\%) & 18.32 ($\downarrow$ 24.52\%) &  60.50\% \\
        & ReCODE-Attack & 7.92 ($\downarrow$ 33.67\%) & 19.12 ($\downarrow$ 21.22\%) & 59.28\% \\
        & ACCENT-Attack & 7.65 ($\downarrow$ 35.93\%) & 18.58 ($\downarrow$ 23.44\%) & 60.00\%  \\
        & {\tool}-Attack & \textbf{7.09 ($\downarrow$ 40.62\%)} & \textbf{17.86 ($\downarrow$ 26.41\%)} & \textbf{63.20\%}  \\
        \midrule 
        \multirow{9}{*}{PLBART}
        & \pd & 7.85 & 20.60 & \text { -- }  \\
        & \pdsig & 19.99 & 30.12 & \text { -- }  \\
        \cline{2-5} 
        & Foo-Attack & 16.93 ($\downarrow$ 15.31\%) & 26.13 ($\downarrow$ 13.25\%) &  56.15\%  \\
        & Random-Attack & 14.39 ($\downarrow$ 28.01\%) & 25.89 ($\downarrow$ 14.04\%) & 57.95\%  \\
        & ALERT-Attack & 14.21 ($\downarrow$ 28.91\%) & 25.24 ($\downarrow$ 16.20\%) &  60.55\% \\
        & Genetic-Attack & 13.68 ($\downarrow$ 31.57\%) & 24.98 ($\downarrow$ 17.07\%) & 63.85\% \\
        & ReCODE-Attack & 14.63 ($\downarrow$ 26.81\%) & 25.85 ($\downarrow$ 14.18\%) & 57.80\% \\
        & ACCENT-Attack & \textbf{13.00 ($\downarrow$ 34.97\%)} & 24.61 ($\downarrow$ 18.29\%) &  62.35\%  \\
        & {\tool}-Attack & 13.31 ($\downarrow$ 33.42\%) & \textbf{24.18 ($\downarrow$ 19.72\%)} & \textbf{65.80\% } \\
        \midrule 
        \multirow{9}{*}{CodeT5}
        & \pd & 5.35 & 19.11 & \text { -- }  \\
        & \pdsig & 21.69 & 33.26 & \text { -- }  \\
        \cline{2-5} 
        & Foo-Attack & 19.37 ($\downarrow$ 10.70\%) & 29.23 ($\downarrow$ 12.12\%) &  53.50\%  \\
        & Random-Attack & 15.11 ($\downarrow$ 30.34\%) & 27.59 ($\downarrow$ 17.05\%) & 58.95\%  \\
        & ALERT-Attack & 14.59 ($\downarrow$ 32.73\%) & 26.53 ($\downarrow$ 20.23\%) &  64.75\% \\
        & Genetic-Attack & 13.84 ($\downarrow$ 36.19\%) & 25.68 ($\downarrow$ 22.79\%) & 69.50\% \\
        & ReCODE-Attack & 14.21 ($\downarrow$ 34.49\%) & 25.94 ($\downarrow$ 22.01\%) & 68.50\% \\
        & ACCENT-Attack & 13.57 ($\downarrow$ 37.44\%) & 25.04 ($\downarrow$ 24.71\%) & 71.00\%  \\
        & {\tool}-Attack & \textbf{13.23 ($\downarrow$ 39.00\%)} & \textbf{24.52 ($\downarrow$ 26.28\%)} & \textbf{72.80\% } \\
        \bottomrule
\end{tabular}
}
\end{center}
\label{tab:RQ1-Python}
\end{table*}

\begin{table*}[]
\caption{\yg{Evaluation results of comparing {\tool} and the baselines in terms of adversarial attack in the Human-Eval dataset}}
\begin{center}
\vspace{-1mm}
\setlength{\tabcolsep}{3mm}{
  \resizebox{1.0\textwidth}{!}{
    \begin{tabular}{c|c|cccc}
        \toprule
        \multicolumn{1}{c|}{Model} & \multicolumn{1}{c|}{Method} & 
        \multicolumn{1}{c}{BLEU} & \multicolumn{1}{c}{CodeBELU} & \multicolumn{1}{c}{Pass@1} & \multicolumn{1}{c}{ASR} \\
        \midrule
        \multirow{9}{*}{Replit}
        & \pd  & \text { -- }  & \text { -- } & \text { -- } & \text { -- }  \\
        & \pdsig & 28.56 & 29.98 & 18.90 & \text { -- } \\
        \cline{2-6} 
      & Foo-Attack & \textbf{25.48 ($\downarrow$ 10.78\%)} & \textbf{27.73 ($\downarrow$ 7.51\%)}  & 15.85 ($\downarrow$  16.14\%) & 29.03\%  \\
      & Random-Attack & 26.26 ($\downarrow$ 8.05\%) & 28.99 ($\downarrow$ 3.30\%)  & 16.46 ($\downarrow$ 12.91\%) & 25.81\%  \\
        & ALERT-Attack & 26.24 ($\downarrow$ 8.12\%) & 29.21 ($\downarrow$ 2.57\%)  & 14.02 ($\downarrow$ 25.82\%) & 32.26\%  \\
        & Genetic-Attack & 26.50 ($\downarrow$ 7.21\%) & 29.14 ($\downarrow$ 2.80\%)  & 15.24 ($\downarrow$ 19.37\%)& 29.03\%  \\
        & ReCODE-Attack & 26.40 ($\downarrow$ 7.56\%) & 28.62 ($\downarrow$ 4.54\%)  & 15.85 ($\downarrow$ 16.14\%) & 25.81\%  \\
      & ACCENT-Attack & 25.90 ($\downarrow$ 9.31\%) & 28.36 ($\downarrow$ 5.40\%)  & 13.41 ($\downarrow$ 29.05\%) & 35.48\%  \\
      & {\tool}-Attack & 25.87 ($\downarrow$ 9.42\%) & 28.27 ($\downarrow$ 5.70\%)  & \textbf{12.80 ($\downarrow$ 32.28\%)} & \textbf{45.16\%}  \\
        \midrule 
        \multirow{9}{*}{CodeGen}
        & \pd  & \text { -- }  & \text { -- } & \text { -- } & \text { -- }  \\
        & \pdsig & 30.18 &  33.01 & 21.34 & \text { -- }  \\
        \cline{2-6} 
        & Foo-Attack & 30.71 ($\uparrow$ 1.76\%) & 32.48 ($\downarrow$ 1.61\%)  & 17.68 ($\downarrow$ 17.15\%) & 25.71\%  \\
        & Random-Attack & 28.12 ($\downarrow$ 6.83\%) & 31.80 ($\downarrow$ 3.67\%)  & 15.24 ($\downarrow$ 28.58\%)& 42.86\%  \\
        & ALERT-Attack & 26.71 ($\downarrow$ 11.50\%) & 29.75 ($\downarrow$ 9.88\%)  & 14.02 ($\downarrow$ 34.30\%) & 45.71\%  \\
        & Genetic-Attack & 28.76 ($\downarrow$ 4.71\%) & 30.89 ($\downarrow$ 6.42\%)  & 13.41 ($\downarrow$ 37.16\%) & 37.14\%  \\
        & ReCODE-Attack & 28.90 ($\downarrow$ 4.24\%) & 30.96 ($\downarrow$ 6.21\%)  & 18.90 ($\downarrow$ 11.43\%) & 20.00\%  \\
      & ACCENT-Attack & 27.70 ($\downarrow$ 8.22\%) & 30.19 ($\downarrow$ 8.54\%)  & 14.02 ($\downarrow$ 34.30\%)& 42.86\%  \\
      & {\tool}-Attack & \textbf{26.51 ($\downarrow$ 12.16\%)} & \textbf{28.44 ($\downarrow$ 13.84\%)}  & \textbf{12.20 ($\downarrow$ 42.83\%)} & \textbf{51.43\%}  \\
        \midrule 
        \multirow{9}{*}{CodeT5+}
        & \pd  & \text { -- }  & \text { -- } & \text { -- } & \text { -- }  \\
        & \pdsig & 27.21 & 30.92 & 21.95 & \text { -- }  \\
        \cline{2-6} 
        & Foo-Attack & 25.75 ($\downarrow$ 5.37\%) & 29.10 ($\downarrow$ 5.89\%)  & 20.73 ($\downarrow$ 5.56\%) & 25.00\%  \\
        & Random-Attack & 25.63 ($\downarrow$ 5.81\%) & 29.31 ($\downarrow$ 5.21\%)  & 16.46 ($\downarrow$ 25.01\%) & 36.11\%  \\
        & ALERT-Attack & 24.18 ($\downarrow$ 11.14\%) & 26.88 ($\downarrow$ 13.07\%)  & 13.41 ($\downarrow$ 38.91\%)& 44.44\%  \\
        & Genetic-Attack & \textbf{23.35 ($\downarrow$ 14.19\%)} & \textbf{26.04 ($\downarrow$ 15.78\%)} & 13.41 ($\downarrow$ 38.91\%)& 44.44\%  \\
        & ReCODE-Attack & 24.89 ($\downarrow$ 8.53\%) & 27.58 ($\downarrow$ 10.80\%)  & 18.29 ($\downarrow$ 16.67\%)& 25.00\%  \\
      & ACCENT-Attack & 24.13 ($\downarrow$ 11.32\%) & 26.63 ($\downarrow$ 13.87\%)  & 14.63 ($\downarrow$ 33.35\%) & 47.22\%  \\
      & {\tool}-Attack & 26.51 ($\downarrow$ 2.57\%) & 28.48 ($\downarrow$ 7.89\%)  & \textbf{12.20 ($\downarrow$ 44.42\%)} & \textbf{50.00\%}  \\
        \bottomrule
\end{tabular}
}
}
\end{center}
\label{tab:RQ1-HumanEval}
\end{table*}

Table~\ref{tab:RQ1-Java} and Table~\ref{tab:RQ1-Python} show the evaluation results of these three victim models before and after the attacks for fine-tuning code generation tasks, respectively. 
The second column gives the used method. Columns 3--5 in Table~\ref{tab:RQ1-Java} show the performance metrics for the Java dataset while columns 3--5 in Table~\ref{tab:RQ1-Python} show the counterparts for the Python dataset. The rows marked by {\pd} and {\pdsig} show the performance of each PCGM when the signature is either excluded or included in the input. The following three rows show how the model performs under different adversarial attacks (i.e., with modified method names). 

From this table, we can first observe that the performance of the code generation with {\pdsig} is consistently better than that with {\pd}, in terms of all the metrics. For instance, for the CodeT5 model, on the Java dataset, in terms of both BLEU and CodeBLEU, the code generation with {\pdsig} performs nearly 1.5 times better than with {\pd}. On the Python dataset, the code generation with {\pdsig} performs nearly four times better than with {\pd} in BLEU performance and nearly twice as well as in CodeBLEU performance.
In short, the code generation with {\pdsig} performs nearly twice as well as with {\pd} in  most cases.



%
Furthermore, we observe that all the PCGMs are vulnerable to adversarial attacks in the fine-tuning code generation task, as their performance decreases largely when the method names are modified. 
However, the impact of adversarial attacks varies across these models. 
Among them, the simplest foo-Attack can cause 9\%-27\% performance degradation in code generation on the test set for all three models. 
In addition, well-designed attacks (such as ACCENT-Attack and {\tool}-Attack) can have a more severe impact on the model performance.

Take the CodeT5 model as an example, {\tool}-Attack degrades its BLEU and CodeBLEU performance on the Java dataset by 36.33\% and 31.58\% respectively, and can successfully attack 74.65\% of the test set samples. On the Python dataset, the CodeT5's BLEU and CodeBLEU performance is degraded by 39.00\% and 26.28\% respectively, and {\tool}-Attack can successfully attack 72.80\% of the test set samples. 


\yg{
Table~\ref{tab:RQ1-HumanEval} presents the evaluation results of three victim models (Replit, CodeGen, and CodeT5+) before and after the attacks in the zero-shot code generation task. Similar to the findings in the fine-tuning code generation task, it is evident that all PCGMs are susceptible to adversarial attacks, resulting in significant performance degradation when method names are modified.
In our experiments, we observed that in certain cases, the model generated incorrect code based on the original prompt but made correct predictions when presented with perturbed prompts, which aligns with the findings of Wang et al.~\cite{wang2022recode}. To accurately evaluate the ASR, we computed the ratio of samples where the model correctly generated code based on the original prompt but made incorrect predictions on perturbed prompts, to the total number of samples where the model correctly generated code based on the original prompt.
Using the CodeGen model as an example, the {\tool}-Attack method leads to a reduction in BLEU and CodeBLEU performance by 12.16\% and 13.84\%, respectively. Moreover, it successfully attacks 51.43\% of the samples in the test set.
These results highlight the vulnerability of PCGMs to adversarial attacks, emphasizing the importance of robust defense mechanisms in code generation tasks.}

\yg{
All the existing attack methods, including our proposed {\tool}-Attack, have a detrimental impact on the performance of Replit, CodeGen, and CodeT5+ PCGMs, particularly in terms of the Pass@1 metric. However, in contrast to the PCGMs used in the fine-tuning code generation task, these models (Replit, CodeGen, and CodeT5+) do not exhibit significant differences in token-level similarity metrics such as BLEU and CodeBLEU.
The lack of substantial differentiation in token-based similarity metrics can be attributed to the gap between these metrics and execution-based metrics. As a result, the impact of {\tool}-Attack on the CodeT5+ model, for example, only leads to a modest degradation of 2.57\% in BLEU and 7.89\% in CodeBLEU. Nonetheless, {\tool}-Attack successfully attacks 50.00\% of the samples in the test set.
These findings highlight the limitations of token-level similarity metrics when assessing the robustness of PCGMs and emphasize the need to consider execution-based metrics for a comprehensive evaluation.
}

\yg{
In general, we have observed that the ASR performance of {\tool}-Attack is optimal across all datasets and victim models. Specifically, on the Java dataset, the ASR performance of {\tool}-Attack is, on average, 4.40\% higher than the second best baseline method. On the Python dataset, the ASR performance of {\tool}-Attack is, on average, 2.96\% higher than the second best baseline method. On the Human-Eval dataset, the ASR performance of {\tool}-Attack is, on average, 17.73\% higher than the second best baseline method.
It is worth mentioning that since the Java dataset and the Python dataset do not support the calculation of the Pass@1 metric, we calculated the ASRs on these two datasets by reducing the CodeBLEU value. However, this method may not be as accurate as the Human-Eval dataset in terms of semantic consistency.
Considering the significant improvement in performance on the Human-Eval dataset, it can be concluded that {\tool}-Attack has a substantial impact on the ASR performance.
}

\begin{tcolorbox}[width=1.0\linewidth, title={Summary for RQ1}]
Existing PCGMs are generally vulnerable to adversarial attacks on method names \yg{both in fine-tuning and zero-shot code generation tasks}, which shows that the quality of the method names in the signature is crucial for PCGMs. In general, {\tool}-Attack is the most effective method in attacking the models. 
\end{tcolorbox}

\noindent\textbf{RQ2: How effective is {\tool}-Defense in reinstating the performance of FD$^{\rm Sig}$?}





\begin{table}[]
 \caption{Evaluation results of comparing {\tool} and the baselines in terms of attack and defense}
 \vspace{-0.3cm}
 \begin{center}
 \setlength{\tabcolsep}{2mm}{
\begin{tabular}{c|c|cc|cc}
  \toprule
\multicolumn{1}{c|}{\multirow{2}{*}{Model}} & \multicolumn{1}{c|}{\multirow{2}{*}{Method}} & \multicolumn{2}{c|}{Java} & \multicolumn{2}{c}{Python}\\
\multicolumn{1}{c|}{} & \multicolumn{1}{c|}{} & BLEU & CodeBLEU & BLEU & CodeBLEU\\
  \midrule
\multirow{3}{*}{CodeGPT}
& \pdsig & \textbf{23.18} &  \textbf{26.33} & 11.94 &  24.27\\
& {\tool}-Attack & 13.02 &  16.13 & 7.09 &  17.86\\
\cline{2-6} 
& ACCENT-Defense & 17.95 & 20.90 & 9.20 & 21.61 \\
& {\tool}-Defense & 22.15  & 25.45 & \textbf{12.54} & \textbf{24.44} \\
\midrule 
\multirow{3}{*}{PLBART}
& \pdsig & 35.19 & \textbf{43.71} & \textbf{19.99} & 30.12 \\
& {\tool}-Attack & 22.61 &  31.31 & 13.31 &  24.18\\
\cline{2-6} 
& ACCENT-Defense & 27.57 & 36.24 & 14.49 & 26.52 \\
& {\tool}-Defense & \textbf{35.84} & 43.61  & 19.64  & \textbf{30.88} \\
\midrule 
\multirow{3}{*}{CodeT5}
& \pdsig & 38.45 & 46.09 & \textbf{21.69} & \textbf{33.26} \\
& {\tool}-Attack & 24.28 & 31.58 & 13.23 & 24.52\\
\cline{2-6} 
& ACCENT-Defense & 30.31 & 37.43 & 16.01 & 27.22 \\
& {\tool}-Defense & \textbf{39.29} & \textbf{46.11} & 21.31 & 32.90 \\
  \bottomrule
\end{tabular}
 }
 \end{center}
 \label{tab:RQ2}
\end{table}

\begin{table}[]
 \caption{Evaluation results of comparing {\tool} and the baselines in terms of attack and defense}
 \vspace{-0.3cm}
 \begin{center}
 \setlength{\tabcolsep}{2mm}{
\begin{tabular}{c|c|ccc}
  \toprule
  Model & Method & BLEU & CodeBLEU & Pass@1\\
  \midrule
\multirow{4}{*}{Replit}
& \pdsig & \textbf{28.56} & 29.98 & \textbf{18.90} \\
& {\tool}-Attack & 25.87 & 28.27 & 12.80 \\
\cline{2-5} 
& ACCENT-Defense & \text { -- } & \text { -- } & \text { -- } \\
& {\tool}-Defense & 28.51 & \textbf{30.21} & 18.29 \\
\midrule 
\multirow{4}{*}{CodeGen}
& \pdsig & \textbf{30.18} & \textbf{33.01} & 21.34 \\
& {\tool}-Attack & 26.51 & 28.44 & 12.20 \\
\cline{2-5} 
& ACCENT-Defense & \text { -- } & \text { -- } & \text { -- } \\
& {\tool}-Defense & 29.95 & 32.99 & \textbf{21.95} \\
\midrule 
\multirow{4}{*}{CodeT5+}
& \pdsig & \textbf{27.21} & \textbf{30.92} & \textbf{21.95} \\
& {\tool}-Attack & 26.51 & 28.48 & 12.20 \\
\cline{2-5} 
& ACCENT-Defense & \text { -- } & \text { -- } & \text { -- } \\
& {\tool}-Defense & 26.94 & 30.04 & 20.12 \\
  \bottomrule
\end{tabular}
 }
 \end{center}
 \label{tab:RQ2-HumanEval}
\end{table}

Table~\ref{tab:RQ2} summarizes evaluation results on the three victim models of the two defense strategies for fine-tuning code generation task. Rows of {\pdsig} and {\tool}-Attack recapitulate the performance of PCGMs when the method name is unattacked or attacked respectively, followed by two rows showing how the model performs under the two different defense strategies.

In terms of defense, we find that the mask training employed in ACCENT-Defense can indeed resist some attack examples, mainly because the mask training masks the attacked method name and lets the model learn the corresponding code generation after the mask. 
Compared to ACCENT-Defense,  {\tool}-Defense is a passive defense method to sanitize the input, and the performance of the defended model is almost the same as that of the original environment (e.g., CodeT5 has a BLEU metric of 21.69 on the Python dataset, and the metric drops to 13.23 after being attacked by {\tool}-Attack, but after {\tool}-Defense the metric reinstates to 21.31) 
Moreover, we are surprised to observe that some models can slightly improve their code generation performance after defending the method names in the signature. 
For instance, CodeT5's performance measured in BLEU and CodeBLEU is improved by 61.82\% and 46.01\% respectively, by {\tool}-Defense on the Java dataset, when compared with that of the attacked model. ACCENT-Defense, on the other hand, only improved 24.84\% of the BLEU performance and 18.52\% of the CodeBLEU performance.
These results show that the defense of {\tool}-Defense is superior. 
Indeed, {\tool}-Defense even exceeds the performance of the original methods on some combinations (e.g., CodeBLEU in Python using CodeGPT, BLEU in Java and CodeBLEU in Python using PLBART, and both BLEU and CodeBLEU in Java using CodeT5).
It also indicates that the quality of method names in the signature is crucial for the model to generate code.

\yg{
In the zero-shot code generation task, since the PCGMs are not fine-tuned on the HumanEval dataset, an approach based on active defense is not suitable for this scenario.
Table~\ref{tab:RQ2-HumanEval} provides a summary of the evaluation results for the three victim models under our defense method in the zero-shot code generation task. Consistent with the findings from the fine-tuning code generation task, the defended models exhibit performance that is nearly equivalent to the original environment.
Furthermore, we observe that some models can experience slight improvements in their code generation performance after defending the method names in the signatures. For example, CodeGen's Pass@1 metric increases from 21.34 in the original environment to 21.95 in the {\tool}-Defense.
These results highlight the significance and advantages of employing well-chosen method names in neural code generation, both in the fine-tuning and zero-shot code generation tasks.
}

\yg{
In general, we observe that our proposed {\tool}-Defense method is a passive defense approach that ensures both clean performance and robustness of the model without the need for retraining. Therefore, our {\tool}-Defense method provides a viable way that enhances model robustness without sacrificing clean performance. This passive defense approach has certain advantages over active defense methods, especially in scenarios with high costs and limitations in zero-shot scenarios. 
}


\begin{tcolorbox}[width=1.0\linewidth, title={Summary for RQ2}]
{\tool}-Defense, as a passive defense method,  shows better defense performance and is capable of bringing the performance of FD$^{\rm Sig}$ back. As well, it also shows that the quality of the method names in the signature is crucial for PCGMs.
\end{tcolorbox}



\noindent\textbf{RQ3: How effective is our proposed {\tool}-Defense in terms of method name generation?}

Results of RQ1 and RQ2 demonstrate the importance of method names in neural code generation. 
In RQ3, we investigate whether our method can synthesize high-quality method names for programmers.
\yg{Note that for our zero-shot evaluation in the Human-Eval task, we utilize the model trained by {\tool}-Defense on the Python dataset that we collected in Section \ref{sec:dataset}.}

For the baselines with shared code (e.g., NNGen, CCGIR, CodeBERT, UniXcoder, Rencos, and REINA), we directly used their implementation to obtain the optimal values of parameters and trained the models. Otherwise (e.g., BM25 and RNN-Att-Copy), we replicated them according to the description of the original studies.

\begin{table}[htbp]
 \caption{Evaluation results of comparing {\tool}-Defense with the baselines for the Java dataset} 
 \vspace{-0.3cm}
 \label{tab:RQ3-1-java}
 \begin{center}
 \begin{tabular}{c|c|cccc} 
  \toprule 
    \textbf{Type} & \textbf{Method} & \textbf{EM} & \textbf{BLEU} & \textbf{ED} \\
    \midrule 
    \multirow{4}{*}{Information Retrieval} 
    & BM25 & 22.00 & 42.24 & 9.39 \\ 
    & NNGen & 23.65 & 45.93 & 8.93 \\ 
    & CCGIR & 23.50 & 46.97 & 8.71 \\ 
    & \textbf{\tool-IR} & 24.10 & 46.66 & 8.70 \\
     \midrule 
    \multirow{4}{*}{Deep Learning} 
    & RNN-Att-Copy & 22.20 & 47.99 & 8.37 \\ 
    & CodeBERT & 40.95 & 63.76 & 6.13 \\ 
    & UniXcoder & 43.35 & 65.66 & 5.99 \\ 
     \midrule 
    \multirow{3}{*}{IR-Enhanced}
    & Rencos & 27.75 & 53.53 & 7.39 \\ 
    & REINA & 41.00 & 63.51 & 6.39 \\ 
     & \textbf{{\tool}-Defense} & \textbf{47.60} & \textbf{68.86} & \textbf{5.28} \\ 
  \bottomrule 
 \end{tabular}
 \end{center}
\end{table}

\begin{table}[htbp]
 \caption{Evaluation results of comparing {\tool}-Defense with the baselines for the Python dataset} 
 \vspace{-0.3cm}
 \label{tab:RQ3-1-python}
 \begin{center}
 \begin{tabular}{c|c|cccc} 
  \toprule 
    \textbf{Type} & \textbf{Method} & \textbf{EM} & \textbf{BLEU} & \textbf{ED} \\
    \midrule 
    \multirow{4}{*}{Information Retrieval} 
    & BM25 & 14.50 & 31.39 & 10.68 \\ 
    & NNGen & 14.75 & 32.00 & 10.42 \\ 
    & CCGIR & 15.20 & 32.62 & 10.34 \\ 
    & \textbf{\tool-IR} & 15.10 & 34.58 & 9.98 \\ 
     \midrule 
    \multirow{4}{*}{Deep Learning} 
    & RNN-Att-Copy & 11.60 & 37.66 & 9.29 \\ 
    & CodeBERT & 25.35 & 50.18 & 7.58 \\ 
    & UniXcoder & 27.40 & 52.46 & 7.67 \\ 
     \midrule 
    \multirow{3}{*}{IR-Enhanced}
    & Rencos & 17.55 & 39.63 & 9.12 \\ 
    & REINA & 25.35 & 49.98 & 7.93 \\ 
     & \textbf{{\tool}-Defense} & \textbf{32.60} & \textbf{57.56} & \textbf{6.65} \\ 
  \bottomrule 
 \end{tabular}
 \end{center}
\end{table}

\begin{table}[htbp]
 \caption{Evaluation results of comparing {\tool}-Defense with the baselines for the Human-Eval dataset} 
 \vspace{-0.3cm}
 \label{tab:RQ3-1-Human-Eval}
 \begin{center}
 \begin{tabular}{c|c|cccc} 
  \toprule 
    \textbf{Type} & \textbf{Method} & \textbf{EM} & \textbf{BLEU} & \textbf{ED} \\
    \midrule 
    \multirow{4}{*}{Information Retrieval} 
    & BM25 & 0.61 & 7.90 & 13.42 \\ 
    & NNGen & 0.61 & 4.98 & 12.95 \\ 
    & CCGIR & 0.00 & 4.66 & 12.84 \\ 
    & \textbf{\tool-IR} & 1.22 & 10.05 & 12.43 \\ 
     \midrule 
    \multirow{4}{*}{Deep Learning} 
    & RNN-Att-Copy & 1.22 & 9.71 & 11.07 \\ 
    & CodeBERT & 14.63 & 32.33 & 8.22 \\ 
    & UniXcoder & 29.88 & 46.62 & 7.24 \\ 
     \midrule 
    \multirow{3}{*}{IR-Enhanced}
    & Rencos & 7.58 & 18.45 & 10.14 \\ 
    & REINA & 22.81 & 42.60 & 8.19 \\ 
     & \textbf{{\tool}-Defense} & \textbf{32.93} & \textbf{49.62} & \textbf{6.09} \\ 
  \bottomrule 
 \end{tabular}
 \end{center}
\end{table}
Table~\ref{tab:RQ3-1-java}, Table~\ref{tab:RQ3-1-python}, and Table~\ref{tab:RQ3-1-Human-Eval} show the results of {\tool}-Defense and the baselines for the Java, Python, and Human-Eval datasets respectively. The second column  of the tables shows the considered baselines. Columns 3--5 show the results of the performance metrics. 
%

\begin{table}[htbp]
 \caption{Ablation experiments between three components} 
 \vspace{-0.3cm}
 \label{tab:RQ3-2}
 \begin{center}
 \begin{tabular}{c|ccccccc} 
  \toprule 
\textbf{Dataset} & \textbf{IR} & \textbf{Prompt} & \textbf{In\_trust Loss} & \textbf{EM} & \textbf{BLEU} & \textbf{ED} \\
\midrule 
\multirow{6}{*}{Java} 
 & \text { -- } & \text { -- } & \text { -- } & 43.35 & 65.66 & 5.99 \\ 
 & \text { -- } & \text { -- } & $\checkmark$ & 43.75 & 66.07 & 5.90 \\ 
 & $\checkmark$ & \text { -- } & \text { -- } & 43.45 & 66.04 & 5.83 \\ 
 & $\checkmark$ & \text { -- } & $\checkmark$ & 43.55 & 66.27 & 5.83 \\ 
 & $\checkmark$ & $\checkmark$ & \text { -- } & 47.10 & 67.70 & 5.34 \\ 
 & $\checkmark$ & $\checkmark$ & $\checkmark$ & \textbf{47.60} & \textbf{68.86} & \textbf{5.28} \\ 
 \midrule 
\multirow{6}{*}{Python} 
 & \text { -- } & \text { -- } & \text { -- } & 27.40 & 52.46 & 7.67 \\ 
 & \text { -- } & \text { -- } & $\checkmark$ & 28.30 & 52.77 & 7.52 \\ 
 & $\checkmark$ & \text { -- } & \text { -- } & 27.60 & 53.05 & 7.23 \\ 
 & $\checkmark$ & \text { -- } & $\checkmark$ & 28.40 & 53.69 & 7.33 \\
 & $\checkmark$ & $\checkmark$ & \text { -- } & \textbf{32.60} & 56.74 & 6.76 \\
 & $\checkmark$ & $\checkmark$ & $\checkmark$ & \textbf{32.60} & \textbf{57.56} & \textbf{6.65} \\
 \midrule 
\multirow{6}{*}{Human-Eval} 
 & \text { -- } & \text { -- } & \text { -- } & 29.88 & 46.62 & 7.24 \\ 
 & \text { -- } & \text { -- } & $\checkmark$ & 29.88 & 46.23 & 6.95 \\ 
 & $\checkmark$ & \text { -- } & \text { -- } & 30.58 & 47.85 & 6.88 \\ 
 & $\checkmark$ & \text { -- } & $\checkmark$ & 31.05 & 48.11 & 6.56 \\
 & $\checkmark$ & $\checkmark$ & \text { -- } & 32.76 & 49.11 & 6.27 \\
 & $\checkmark$ & $\checkmark$ & $\checkmark$ & \textbf{32.93} & \textbf{49.62} & \textbf{6.09} \\
  \bottomrule 
 \end{tabular}
 \end{center}
\end{table}


First, when comparing {\tool}-Defense with the information retrieval baselines, we observe that, since CCGIR uses dense vectors for retrieval while both BM25 and NNGen use sparse vectors for retrieval, CCGIR performs slightly better than BM25 and NNGen on both datasets. Then  CodeBERT used by CCGIR for semantic vectorization representation will take more time, and our proposed information retrieval method can achieve better performance in less time, showing that our proposed method's information retrieval part is effective.

Second, when comparing {\tool}-Defense with the deep learning baselines, we find that among all the deep learning baselines, {\tool}-Defense has the best performance.

Last, results of comparing the hybrid baselines with our method show that {\tool}-Defense can largely improve the performance of the methods. More specifically, compared to the best-performing baseline UniXcoder, on the Java dataset, {\tool}-Defense improves the EM, BLEU, and ED performances by 9.80\%, 4.87\%, and 11.85\% respectively; on the Python dataset, {\tool}-Defense improves the EM, BLEU, and ED performances by 18.98\%, 9.72\%, and 12.27\%, respectively; \yg{on the Human-Eval dataset, {\tool}-Defense improves the EM, BLEU, and ED performances by 9.26\%, 6.44\%, and 15.88\%, respectively.}

\begin{figure}[htbp]
	\centering
	\subfigure[Evaluation on the Replit]{%
		\includegraphics[width=0.33\textwidth]{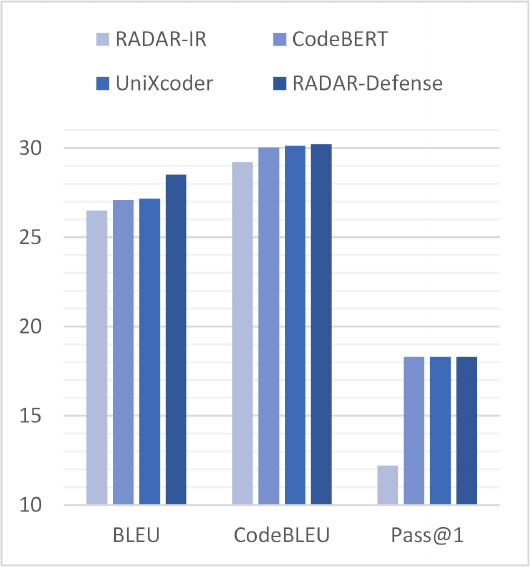}%
	}%
	\hfill
	\subfigure[Evaluation on the CodeGen]{%
		\includegraphics[width=0.33\textwidth]{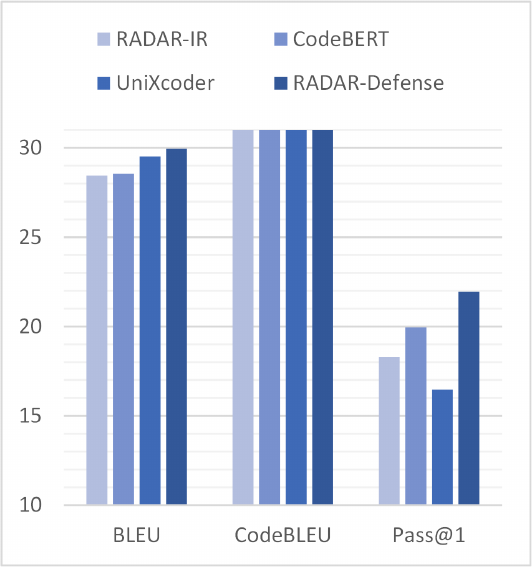}%
	}%
        \hfill
	\subfigure[Evaluation on the CodeT5+]{%
		\includegraphics[width=0.33\textwidth]{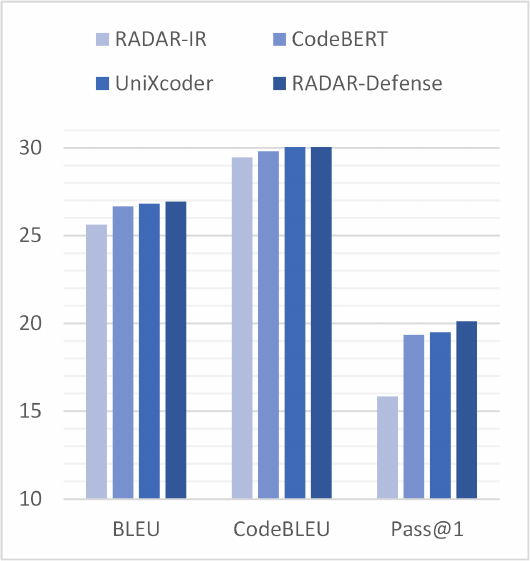}%
	}%
	\caption{The impact of the quality of generated method names on the robustness improvement of PCGMs}
	\label{fig:func_name_quality}
	\vspace{-1mm}
\end{figure}

To further investigate the component setting rationality of our proposed method {\tool}-Defense, we carry out an ablation study. We have considered five variants through permutations between components. The experimental results are given in Table~\ref{tab:RQ3-2} and show that the inclusion of each component is reasonable. The most significant impact on model performance among these three components is our proposed prompt method. With the same settings for the remaining two components, adding the prompt will give {\tool}-Defense a more substantial performance boost. 

\yg{
Furthermore, we conduct an investigation into the impact of data quality on the improvement of robustness. In the zero-shot code generation task, we generate method names using {\tool}-IR, CodeBERT, UniXcoder, and {\tool}-Defense. These methods for generating method names demonstrate increasing performance in the method name generation task.
As depicted in Fig.~\ref{fig:func_name_quality}, 
we observe a correlation between the quality of the generated data and the improvement in robustness. Across all three models, we notice that the BLEU and CodeBLEU metrics improve as the quality of the generated data increases. Moreover, in most cases, the Pass@1 metric also shows improvement as the quality of the generated data increases.
These experimental findings further highlight the importance of utilizing high-quality method names in neural code generation tasks.
}

\yg{
In general, we observe that our proposed {\tool}-Defense method is ability to generate method names that are closer to the golden truth and the method names generated by {\tool}-Defense can improve the accuracy of code generation by PCGMs. The success of {\tool}-Defense can be attributed to the following factors:
(1) the choice of the base model: UniXcoder. UniXcoder has demonstrated the best performance among existing baselines, making it a strong foundation for {\tool}-Defense; 
(2) the retrieval-enhanced prompt learning method and the application of the In\_trust loss, which are reflected in the ablation experiments presented in Table~\ref{tab:RQ3-2}.
}

\begin{tcolorbox}[width=1.0\linewidth, title={Summary for RQ3}]
{\tool}-Defense can achieve better performance than eight state-of-the-art baselines of three different types. 
In our ablation study, the prompt component demonstrates the most influence on the performance of the method.
More importantly, the quality of the method names also impacts the robustness improvement.
\end{tcolorbox} 

\section{Discussion}
\label{sec:discuss} 

\subsection{Qualitative Analysis}

\begin{figure}[htbp]
	\centering
	\subfigure[An example in the Python Dataset]{%
		\includegraphics[width=0.5\textwidth]{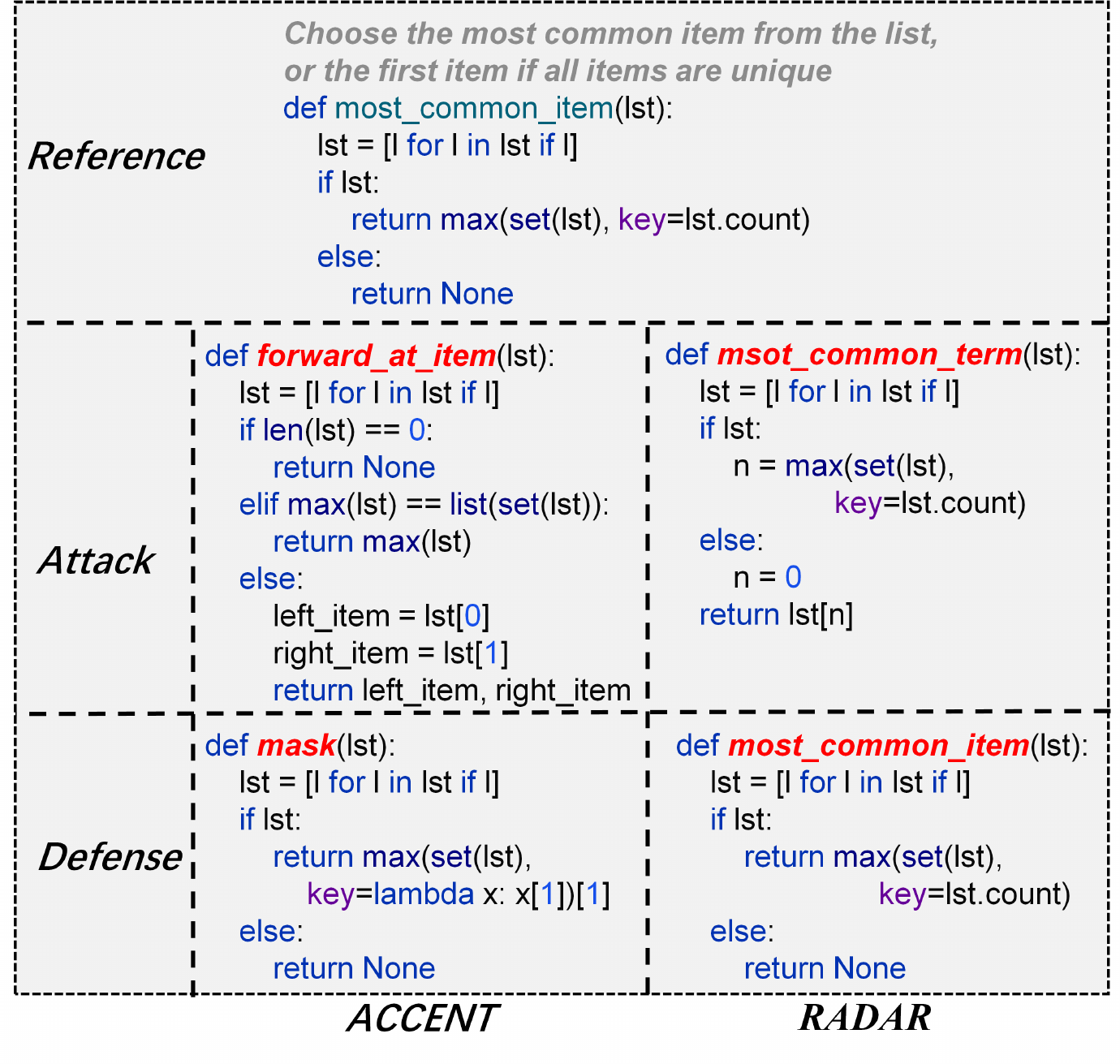}%
	}%
	\hfill
	\subfigure[An example in the Java Dataset]{%
		\includegraphics[width=0.5\textwidth]{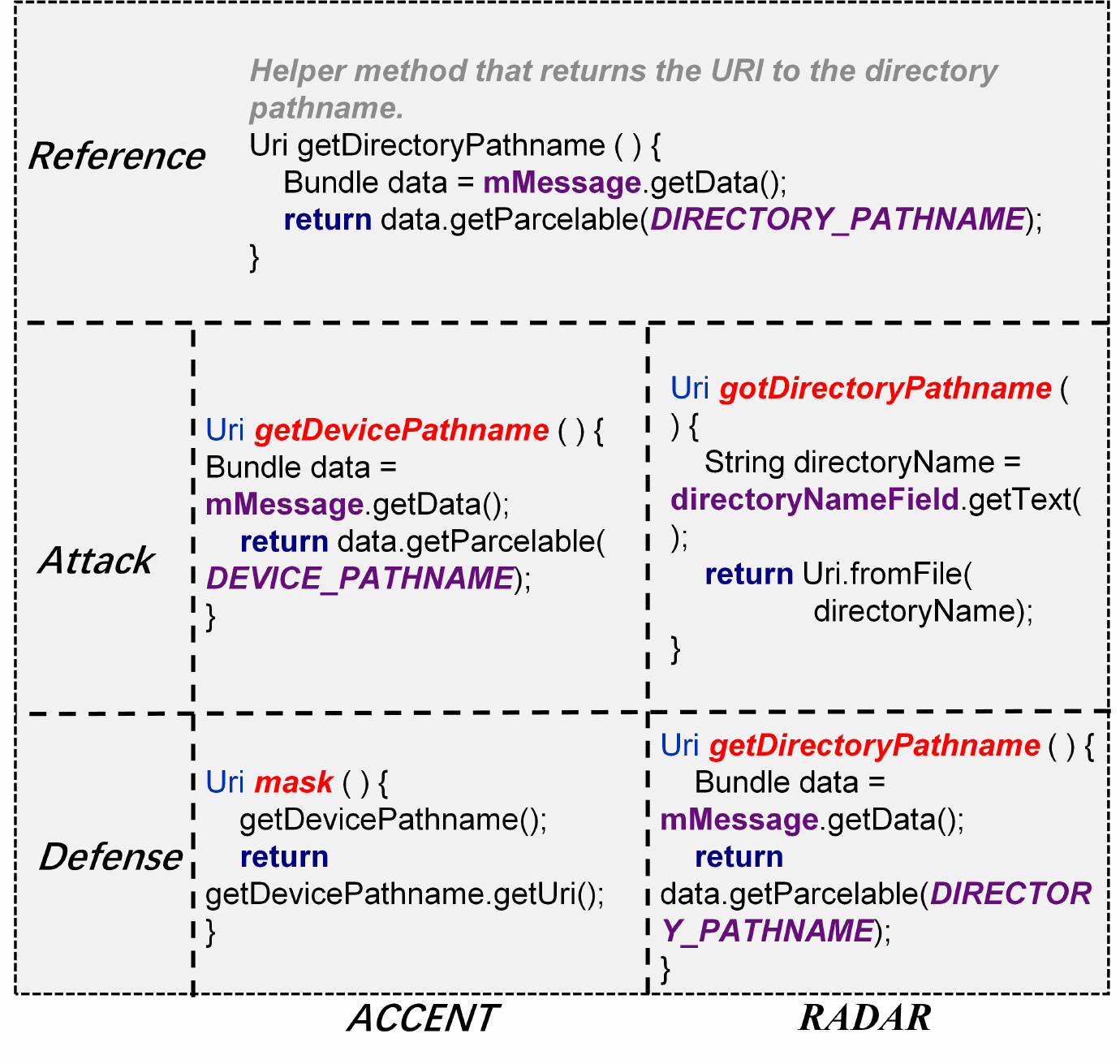}%
	}%
	\caption{Two examples of generated code by CodeT5 when attacked and defended by {\tool} and ACCENT}
	\label{fig:case_study}
	\vspace{-1mm}
\end{figure}

In Section~\ref{sec:results}, we design three RQs to provide a quantitative study of the effectiveness of conducted performance comparisons between 
{\tool} and baselines automatically 
in terms of performance measures. 
However, these performance measures may not truly reflect the semantic similarity~\cite{stapleton2020human}.
To further demonstrate the effectiveness of {\tool}, we conduct qualitative analysis. 

\begin{figure}[htbp]
	\centering
	\subfigure[Heat map before being attacked in Python example code]{%
		\includegraphics[width=0.85\textwidth]{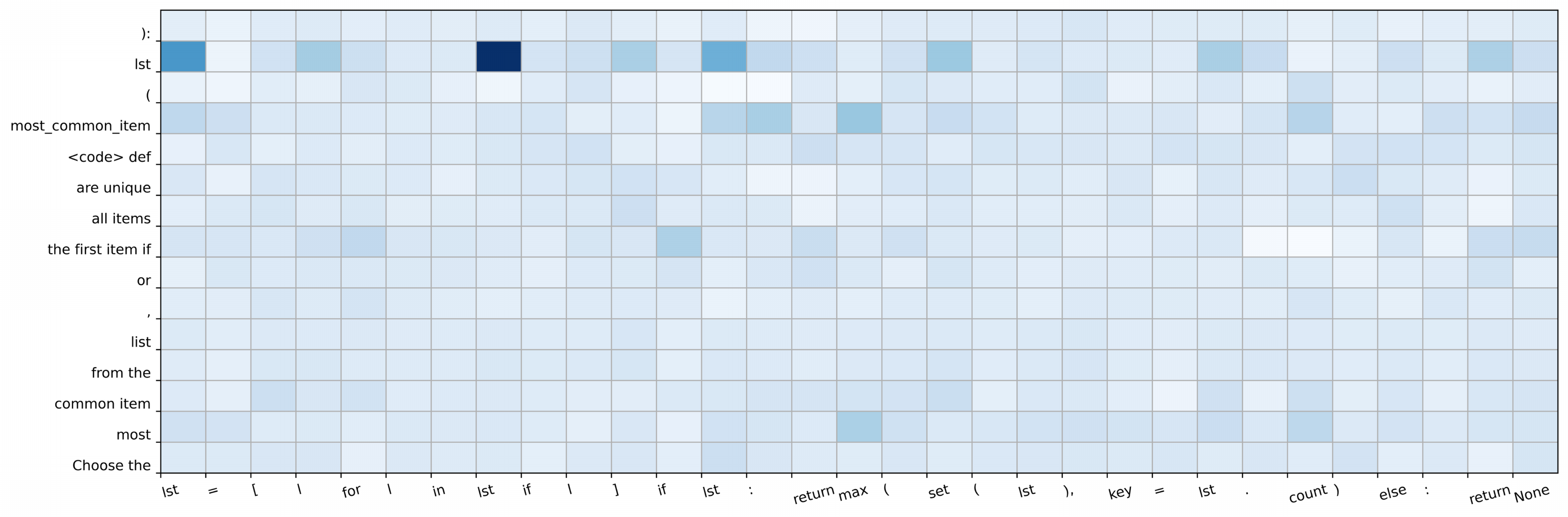}%
	}%
	\hfill
	\subfigure[Heat map after being attacked in Python example code]{%
		\includegraphics[width=0.85\textwidth]{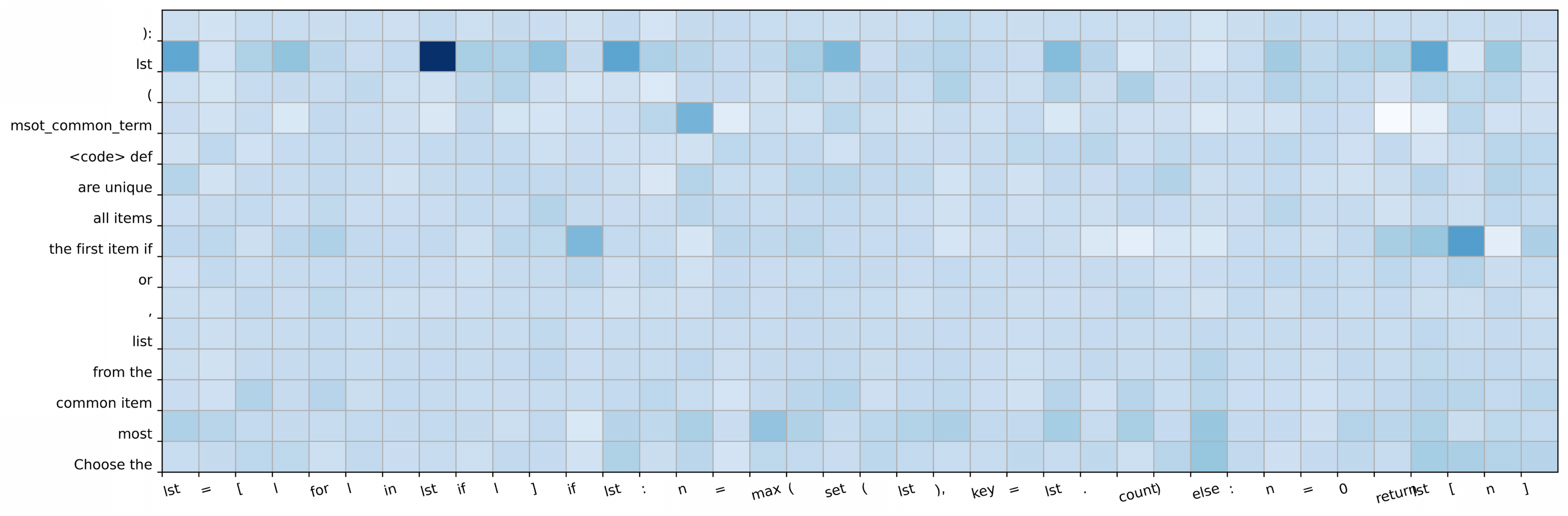}%
	}%
	\hfill
        \subfigure[Heat map before being attacked in Java example code]{%
		\includegraphics[width=0.85\textwidth]{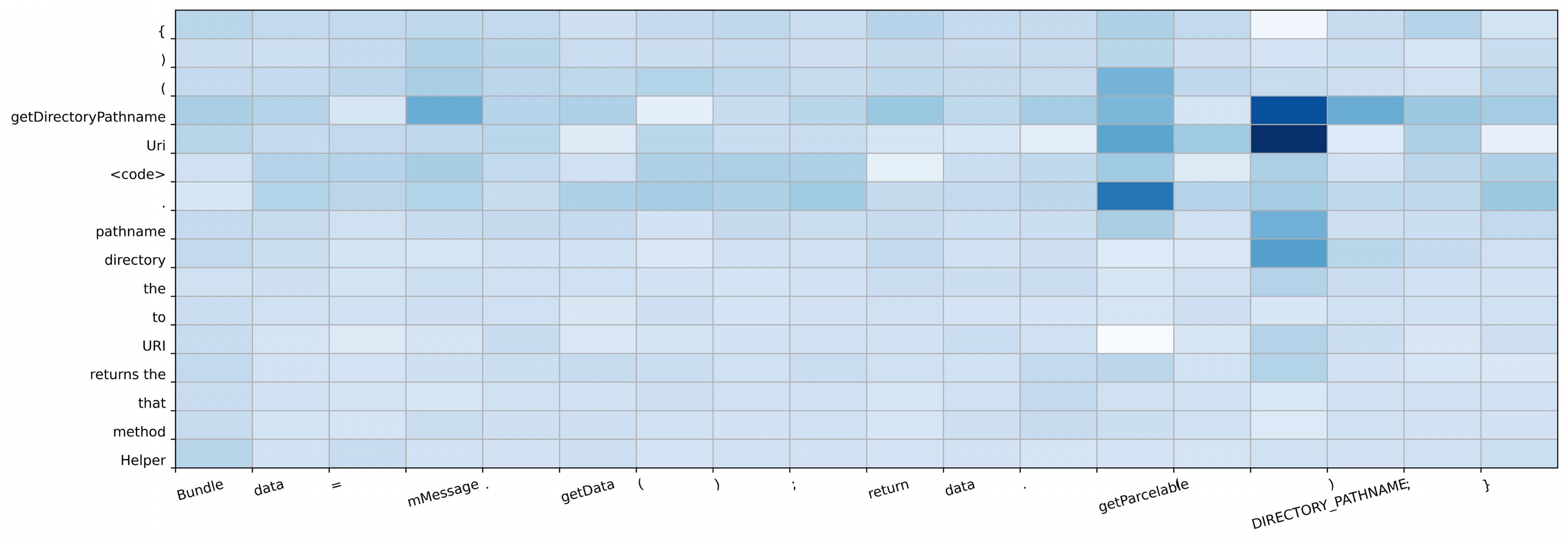}%
	}%
	\hfill
        \subfigure[Heat map after being attacked in Java example code]{%
		\includegraphics[width=0.85\textwidth]{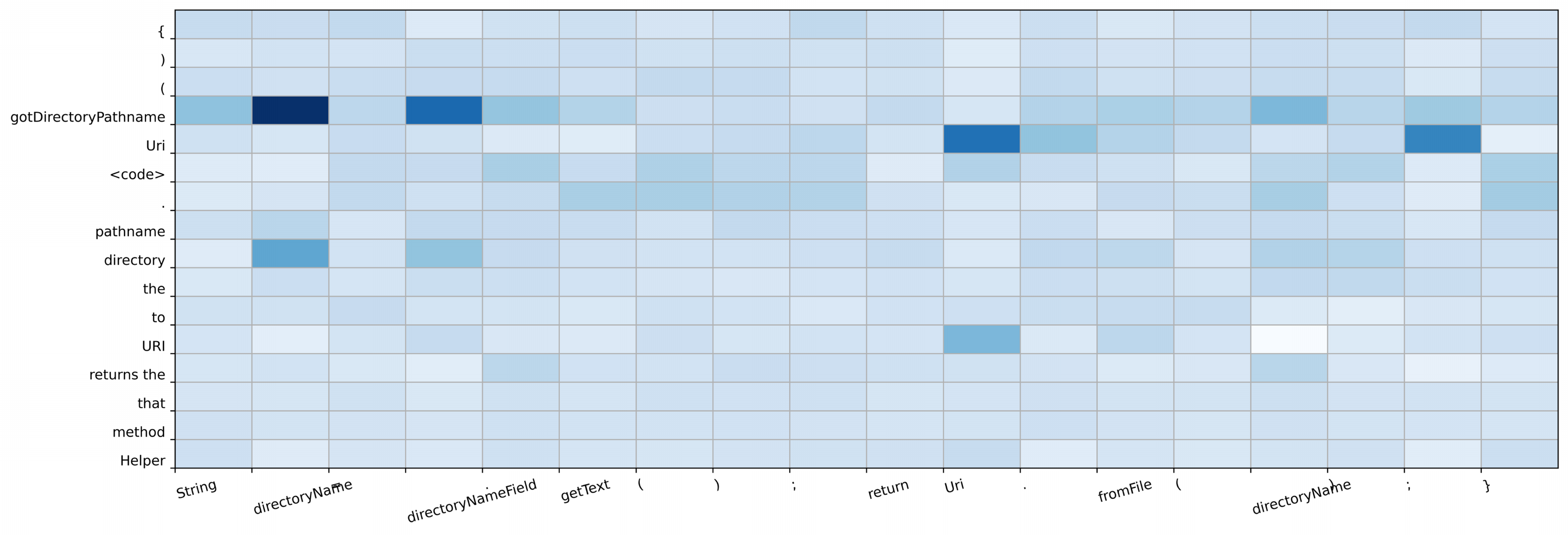}%
	}%
	\caption{Explore the effect of method names on the Python example code generated by CodeT5 before and after being attacked}
	\label{fig:explain}
	\vspace{-1mm}
\end{figure}


\noindent \textbf{Examples in Robustness of Pre-trained Code Generation.}
For the fine-tuning code generation task, we give a Python example based on a real-world project\footnote{\url{https://pypi.org/project/spirit/2.1.1/}} 
and a Java example based on a real-world project\footnote{\url{https://github.com/douglascraigschmidt/POSA-15}}  using the CodeT5 model. 
Fig.~\ref{fig:case_study} shows these two examples of generated code by CodeT5 when attacked and defended by {\tool} and ACCENT. 
The first row gives the functional description, signature, and reference code, where the generated code by CodeT5 is the same as the reference code. The second row shows adversarial examples generated by {\tool}-Attack and ACCENT-Attack while the third row shows the effectiveness of two defensive methods.

From Fig.~\ref{fig:case_study} (a), we can see that the original method name is \code{most\_common\_item}.
The adversarial example \code{forward\_at\_item} generated by ACCENT-Attack is based on semantic similarity, which is not as natural as \code{msot\_common\_term} generated by {\tool}-Attack, in which ``msot'' is generated by the \textbf{Swap} operator and ``term'' is generated by the \textbf{Replace-sem} operator. 

From the Fig.~\ref{fig:case_study}(b), we can see that the original method name is \code{getDirectoryPathname}.
ACCENT-Attack generates \code{getDevicePathname} as the adversarial method name based on semantic similarity, which is arguably  not as natural as \code{gotDirectoryPathname} generated by {\tool}-Attack, in which ``got'' is generated by the \textbf{Replace-sem} operator.

The code generated by {\tool}-Attack in the above two examples can cause functional errors that can lead to the failure of PCGMs. This demonstrates the effectiveness of our {\tool}-Attack, and that the robustness issue in PCGMs needs to be addressed properly.

We also explore the effectiveness of two defensive methods.
ACCENT-Defense replaces the method name with $\langle mask \rangle$ and then feeds it into the mask learned model and generates the corresponding code. In contrast, {\tool}-Defense synthesizes method names based on functional descriptions, replaces them in the adversarial examples, and then generates the corresponding code by the model.  Two examples in Fig.~\ref{fig:case_study} show that {\tool}-Defense is capable of generating the correct method names, and the code generated by CodeT5 after being defended by {\tool}-Defense can be reinstated to what it was before being attacked.

Moreover, in order to explore the effect of method names on the code  generated by CodeT5 before and after being attacked, we visualize and analyze them with the SHAP tool.\footnote{\url{https://github.com/slundberg/shap}}
In contrast to the work  on model interpretation based on attention weight visualization, SHAP is based on game theory, which defines the additive feature attribution method and guarantees a unique solution. Research~\cite{NIPS2017_8a20a862} shows that SHAP is similar to human intuition measurement and more effective.

Fig.~\ref{fig:explain} visualize the Python code and Java code in Fig.~\ref{fig:case_study}, as a way to analyze the effect of method names on the code generated by CodeT5 before and after being attacked.
In Fig.~\ref{fig:explain}(a), before being attacked, the normal method name ``most\_common\_item'' can lead CodeT5 to generate the correct code; from this heat map one can see that it has a greater impact on the tokens ``max'' and ``count'' of the code snippet. However, in Fig.~\ref{fig:explain}(b), under the attack of ``msot\_common\_term'', CodeT5 generates semantically incorrect code, and the heat map shows that this method name only has a large effect on token ``n'' in the code snippet.
Likewise, in Fig.~\ref{fig:explain}(c), before being attacked, the normal method name ``getDirectoryPathname'' can lead CodeT5 to generate the correct code and it has a greater impact on tokens ``getParcelable'' and ``DIRECTORY\_PATHNAME'' of the code snippet. However, in Fig.~\ref{fig:explain}(d), under the attack of  ``gotDirectoryPathname'', CodeT5 outputs incorrect code.

\begin{figure*}[htbp]
\vspace{-0.3cm}
	\centering
	\includegraphics[width=1\textwidth]{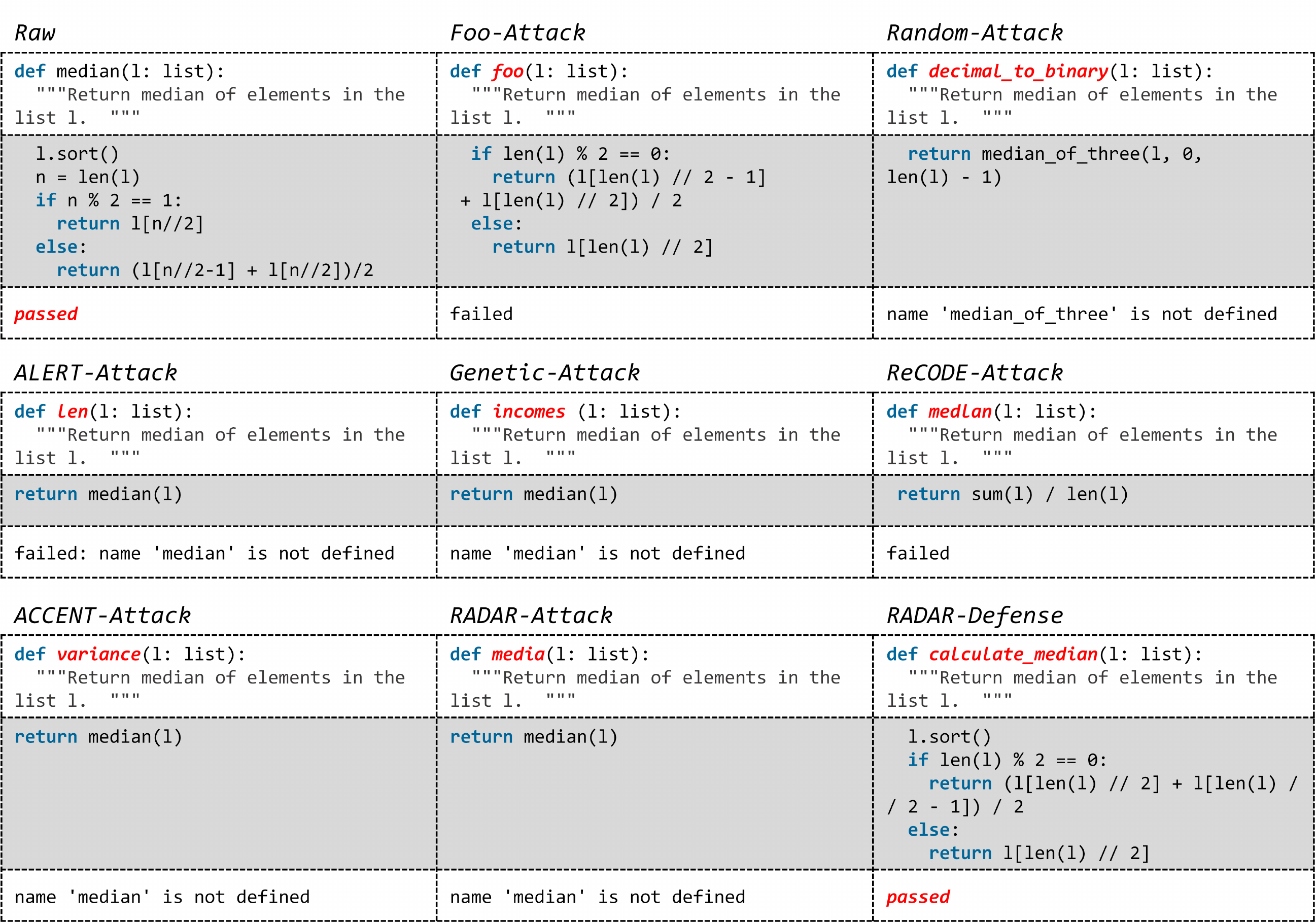}
	\caption{One example of generated code by CodeGen under various attacks as well as the RADAR-Defense in Human-Eval dataset}
	\label{fig:HumanEval}
\end{figure*}

\yg{
In the zero-shot code generation task, we provide an example from the Human-Eval dataset, specifically using the CodeGen model. In Fig.~\ref{fig:HumanEval}, we showcase the generated code snippets under various attacks as well as the RADAR-Defense approach. Additionally, we provide information on whether the generated code can be successfully compiled given the provided test cases.
Initially, the original method name is \code{median}, and CodeGen can generate the correct code when this method name is utilized. However, when the method name is changed to \code{foo} or an adversarial method name generated by different attack methods, the code generated by CodeGen either fails the test cases or contains syntax errors.
In contrast, RADAR-Defense synthesizes the method name \code{calculate\_median} based on functional descriptions, replaces it in the adversarial examples, and subsequently, CodeGen is able to generate the corresponding code that aligns with the desired functionality.}

\smallskip

\begin{table}[htbp]
	\caption{Examples of synthesized method name by {\tool}-Defense and baselines in both Java and Python dataset}
	\vspace{-0.3cm}
	\begin{tabular}{c|p{32em}}
		\toprule
		Case &Example  \\ 
		\midrule
		\multirow{12}{*}{Java} & Parse the string as a websocket request and return the value from WebSocket-Protocol header (See RFC 6455). Return empty string if not found. \\
		\cline{2-2}
		&\textbf{BM25:} getClientWebSocketOrigin \\
		&\textbf{NNGen:} getClientWebSocketOrigin \\
		&\textbf{CCGIR:} getClientWebSocketOrigin \\
		&\textbf{RNN-Att-Copy:} parseValue \\
		&\textbf{CodeBert:} getWebsocketRequest \\
		&\textbf{UniXcoder:} getWebsocketHeader \\
		&\textbf{Rencos:} getClientWebSocketOrigin \\
		&\textbf{REINA:} getProtocol \\
		&\textbf{{\tool}-Defense:} getClientWebSocketProtocol \\
		&\textbf{Human Written:} getClientWebSocketProtocol \\
		\midrule
		\multirow{11}{*}{Python} & 
		Returns an * RGBA * tuple of 4 ints from 0 - 255 \\ 
		\cline{2-2}
		&\textbf{BM25:} to\_rgb\_255 \\
		&\textbf{NNGen:} to\_rgb\_255 \\
		&\textbf{CCGIR:} to\_rgb\_255 \\
		&\textbf{RNN-Att-Copy:} format\_rgba \\
		&\textbf{CodeBert:} to\_rgb\_255 \\
		&\textbf{UniXcoder:} to\_rgb\_255 \\
		&\textbf{Rencos:} to\_rgb\_255 \\
		&\textbf{REINA:} rgba4 \\
		&\textbf{{\tool}-Defense:} to\_rgba\_255 \\
		&\textbf{Human Written:} to\_rgba\_255 \\
  \midrule
		\multirow{12}{*}{Human-Eval} & 
		Check if in given list of numbers, are any two numbers closer to each other than given threshold. \\ 
		\cline{2-2}
		&\textbf{BM25:} are\_rooms\_adjacent \\
		&\textbf{NNGen:} connected\_pair \\
		&\textbf{CCGIR:} connected\_pair \\
		&\textbf{RNN-Att-Copy:} format\_rgba \\
		&\textbf{CodeBert:} is\_numbers \\
		&\textbf{UniXcoder:} are\_adjacent \\
		&\textbf{Rencos:} are\_rooms\_adjacent \\
		&\textbf{REINA:} are\_adjacent \\
		&\textbf{{\tool}-Defense:} is\_closer \\
		&\textbf{Human Written:} has\_close\_elements \\
		\bottomrule
	\end{tabular}
	\label{tab:Method_Name}
\end{table}

\noindent\textbf{Examples in Method Name Generation.}
To further explore the quality of the method names synthesized by {\tool}-Defense, we select three examples from the Java dataset, the Python dataset, and the Human-Eval dataset respectively for analysis in Table~\ref{tab:Method_Name}. 
In these samples, we find {\tool}-Defense can synthesize more-accurate method names than baselines when compared with human-written method names.

\subsection{Threats to Validity}
\label{sec:threats}

\noindent\textbf{Internal threats.}
Internal threats refer to the potential defects in implementing our proposed approach and baselines. To alleviate this, we double-checked and peer-reviewed our code to ensure the fairness of the results. 
For all PCGMs, we used their publicly available models. 
For the attack baselines and method name generation baselines, we ran their open-source code directly or re-implemented them according to the original studies.

\noindent\textbf{External threats.}
External threats refer to the choice of corpora and PCGMs. 
To alleviate this, we collected two datasets based on well-maintained open-source projects with high reputations according to the relevant heuristic rules for fine-tuning code generation tasks.
\yg{
For the zero-shot code generation task, we select the Human-Eval dataset.
To ensure a fair comparison, we follow the settings from a previous study~\cite{iyer2018mapping} when dividing the dataset.
In terms of the choice of PCGMs, we select three state-of-the-art models (CodeGPT, PLBART, and CodeT5) for the fine-tuning code generation task, and three state-of-the-art models (Replit, CodeGen, and CodeT5+) for the zero-shot code generation task.
For other models, such as CodePilot, they have not made models or API interfaces publicly available, and can only be accessed through plugins, which is not suitable for large-scale empirical research. While ChatGPT does offer an API interface, its output is not deterministic, 
resulting in low reproducibility. 
As a result,  these models were not included in our selection. 
}

\noindent\textbf{Construct threats.} 
Construct threats concern the performance metrics used to evaluate  {\tool} and baselines. We use a set of metrics, which are also commonly used in similar studies. Due to the difference between natural languages and programming languages, we evaluated the quality primarily through CodeBLEU for fine-tuning code generation tasks.
CodeBLEU has been widely used in the previous studies of code generation, which can not only consider the surface match similar to the original BLEU but also the grammatical correctness and the logic correctness, leveraging the abstract syntax tree and the data flow structure. 
\yg{For the zero-shot code generation task, we choose Pass@1 as the main metric.}
\section{Conclusion}
\label{sec:conclusion}



We studied the role of method names in neural code generation from a robustness perspective.
We showed that most PCGMs using both the functional description and method signature as input, albeit demonstrating impressive performance, are fragile with respect to the input method names, meaning that an ill-formed name may degrade their performance largely. We proposed approaches to synthesize method names from the functional description which can be utilized to reinstate the performance of PCGMs.

For future work, we  plan to investigate the robustness of (now widely-adopted) deep learning models in software engineering systemically. This would shed light on, for instance, the performance and interpretability of these models in solving challenging SE tasks. 
\yg{We also want to investigate the influence of natural language descriptions and parameter lists on the performance of PCGMs, as well as identify suitable defense mechanisms to enhance their robustness.}

\begin{acks}
The authors are grateful for the valuable feedback from domain experts. This work was partially supported by the National Natural Science Foundation of China (NSFC, No.\ 61972197), the Natural Science Foundation of Jiangsu Province (No.\ BK20201292), the Collaborative Innovation Center of Novel Software Technology and Industrialization, the Postgraduate Research \& Practice Innovation Program of Jiangsu Province (No.\ KYCX23\_0396), and the Qing Lan Project. T.\ Chen is partially supported by an oversea grant from the State Key Laboratory of Novel Software Technology, Nanjing University (KFKT2022A03) and Birkbeck BEI School Project (EFFECT).
\end{acks}
 
\bibliographystyle{Reference}
\bibliography{main}


\begin{thebibliography}{104}


\ifx \showCODEN    \undefined \def \showCODEN     #1{\unskip}     \fi
\ifx \showDOI      \undefined \def \showDOI       #1{#1}\fi
\ifx \showISBNx    \undefined \def \showISBNx     #1{\unskip}     \fi
\ifx \showISBNxiii \undefined \def \showISBNxiii  #1{\unskip}     \fi
\ifx \showISSN     \undefined \def \showISSN      #1{\unskip}     \fi
\ifx \showLCCN     \undefined \def \showLCCN      #1{\unskip}     \fi
\ifx \shownote     \undefined \def \shownote      #1{#1}          \fi
\ifx \showarticletitle \undefined \def \showarticletitle #1{#1}   \fi
\ifx \showURL      \undefined \def \showURL       {\relax}        \fi
\providecommand\bibfield[2]{#2}
\providecommand\bibinfo[2]{#2}
\providecommand\natexlab[1]{#1}
\providecommand\showeprint[2][]{arXiv:#2}

\bibitem[\protect\citeauthoryear{Agashe, Iyer, and Zettlemoyer}{Agashe
  et~al\mbox{.}}{2019}]%
        {agashe2019juice}
\bibfield{author}{\bibinfo{person}{Rajas Agashe}, \bibinfo{person}{Srinivasan
  Iyer}, {and} \bibinfo{person}{Luke Zettlemoyer}.}
  \bibinfo{year}{2019}\natexlab{}.
\newblock \showarticletitle{JuICe: A Large Scale Distantly Supervised Dataset
  for Open Domain Context-based Code Generation}. In
  \bibinfo{booktitle}{\emph{Proceedings of the 2019 Conference on Empirical
  Methods in Natural Language Processing and the 9th International Joint
  Conference on Natural Language Processing (EMNLP-IJCNLP)}}.
  \bibinfo{pages}{5436--5446}.
\newblock


\bibitem[\protect\citeauthoryear{Ahmad, Chakraborty, Ray, and Chang}{Ahmad
  et~al\mbox{.}}{2021}]%
        {ahmad2021unified}
\bibfield{author}{\bibinfo{person}{Wasi Ahmad}, \bibinfo{person}{Saikat
  Chakraborty}, \bibinfo{person}{Baishakhi Ray}, {and} \bibinfo{person}{Kai-Wei
  Chang}.} \bibinfo{year}{2021}\natexlab{}.
\newblock \showarticletitle{Unified Pre-training for Program Understanding and
  Generation}. In \bibinfo{booktitle}{\emph{Proceedings of the 2021 Conference
  of the North American Chapter of the Association for Computational
  Linguistics: Human Language Technologies}}. \bibinfo{pages}{2655--2668}.
\newblock


\bibitem[\protect\citeauthoryear{Aizawa}{Aizawa}{2003}]%
        {aizawa2003information}
\bibfield{author}{\bibinfo{person}{Akiko Aizawa}.}
  \bibinfo{year}{2003}\natexlab{}.
\newblock \showarticletitle{An information-theoretic perspective of tf--idf
  measures}.
\newblock \bibinfo{journal}{\emph{Information Processing \& Management}}
  \bibinfo{volume}{39}, \bibinfo{number}{1} (\bibinfo{year}{2003}),
  \bibinfo{pages}{45--65}.
\newblock


\bibitem[\protect\citeauthoryear{Allamanis, Barr, Devanbu, and
  Sutton}{Allamanis et~al\mbox{.}}{2018}]%
        {allamanis2018survey}
\bibfield{author}{\bibinfo{person}{Miltiadis Allamanis},
  \bibinfo{person}{Earl~T Barr}, \bibinfo{person}{Premkumar Devanbu}, {and}
  \bibinfo{person}{Charles Sutton}.} \bibinfo{year}{2018}\natexlab{}.
\newblock \showarticletitle{A survey of machine learning for big code and
  naturalness}.
\newblock \bibinfo{journal}{\emph{ACM Computing Surveys (CSUR)}}
  \bibinfo{volume}{51}, \bibinfo{number}{4} (\bibinfo{year}{2018}),
  \bibinfo{pages}{1--37}.
\newblock


\bibitem[\protect\citeauthoryear{Alzantot, Sharma, Elgohary, Ho, Srivastava,
  and Chang}{Alzantot et~al\mbox{.}}{2018}]%
        {alzantot2018generating}
\bibfield{author}{\bibinfo{person}{Moustafa Alzantot}, \bibinfo{person}{Yash
  Sharma}, \bibinfo{person}{Ahmed Elgohary}, \bibinfo{person}{Bo-Jhang Ho},
  \bibinfo{person}{Mani Srivastava}, {and} \bibinfo{person}{Kai-Wei Chang}.}
  \bibinfo{year}{2018}\natexlab{}.
\newblock \showarticletitle{Generating Natural Language Adversarial Examples}.
  In \bibinfo{booktitle}{\emph{Proceedings of the 2018 Conference on Empirical
  Methods in Natural Language Processing}}. \bibinfo{pages}{2890--2896}.
\newblock


\bibitem[\protect\citeauthoryear{Applis, Panichella, and van Deursen}{Applis
  et~al\mbox{.}}{2021}]%
        {applis2021assessing}
\bibfield{author}{\bibinfo{person}{Leonhard Applis}, \bibinfo{person}{Annibale
  Panichella}, {and} \bibinfo{person}{Arie van Deursen}.}
  \bibinfo{year}{2021}\natexlab{}.
\newblock \showarticletitle{Assessing Robustness of ML-Based Program Analysis
  Tools using Metamorphic Program Transformations}. In
  \bibinfo{booktitle}{\emph{2021 36th IEEE/ACM International Conference on
  Automated Software Engineering (ASE)}}. IEEE, \bibinfo{pages}{1377--1381}.
\newblock


\bibitem[\protect\citeauthoryear{Austin, Odena, Nye, Bosma, Michalewski, Dohan,
  Jiang, Cai, Terry, Le, et~al\mbox{.}}{Austin et~al\mbox{.}}{2021}]%
        {austin2021program}
\bibfield{author}{\bibinfo{person}{Jacob Austin}, \bibinfo{person}{Augustus
  Odena}, \bibinfo{person}{Maxwell Nye}, \bibinfo{person}{Maarten Bosma},
  \bibinfo{person}{Henryk Michalewski}, \bibinfo{person}{David Dohan},
  \bibinfo{person}{Ellen Jiang}, \bibinfo{person}{Carrie Cai},
  \bibinfo{person}{Michael Terry}, \bibinfo{person}{Quoc Le}, {et~al\mbox{.}}}
  \bibinfo{year}{2021}\natexlab{}.
\newblock \showarticletitle{Program synthesis with large language models}.
\newblock \bibinfo{journal}{\emph{arXiv preprint arXiv:2108.07732}}
  (\bibinfo{year}{2021}).
\newblock


\bibitem[\protect\citeauthoryear{Bahrami, Shrikanth, Ruangwan, Liu, Mizobuchi,
  Fukuyori, Chen, Munakata, and Menzies}{Bahrami et~al\mbox{.}}{2021}]%
        {bahrami2021pytorrent}
\bibfield{author}{\bibinfo{person}{Mehdi Bahrami}, \bibinfo{person}{NC
  Shrikanth}, \bibinfo{person}{Shade Ruangwan}, \bibinfo{person}{Lei Liu},
  \bibinfo{person}{Yuji Mizobuchi}, \bibinfo{person}{Masahiro Fukuyori},
  \bibinfo{person}{Wei-Peng Chen}, \bibinfo{person}{Kazuki Munakata}, {and}
  \bibinfo{person}{Tim Menzies}.} \bibinfo{year}{2021}\natexlab{}.
\newblock \showarticletitle{Pytorrent: A python library corpus for large-scale
  language models}.
\newblock \bibinfo{journal}{\emph{arXiv preprint arXiv:2110.01710}}
  (\bibinfo{year}{2021}).
\newblock


\bibitem[\protect\citeauthoryear{Bielik and Vechev}{Bielik and Vechev}{2020}]%
        {bielik2020adversarial}
\bibfield{author}{\bibinfo{person}{Pavol Bielik} {and} \bibinfo{person}{Martin
  Vechev}.} \bibinfo{year}{2020}\natexlab{}.
\newblock \showarticletitle{Adversarial robustness for code}. In
  \bibinfo{booktitle}{\emph{International Conference on Machine Learning}}.
  PMLR, \bibinfo{pages}{896--907}.
\newblock


\bibitem[\protect\citeauthoryear{Bradbury, Frostig, Hawkins, Johnson, Leary,
  Maclaurin, Necula, Paszke, VanderPlas, Wanderman-Milne,
  et~al\mbox{.}}{Bradbury et~al\mbox{.}}{2018}]%
        {bradbury2018jax}
\bibfield{author}{\bibinfo{person}{James Bradbury}, \bibinfo{person}{Roy
  Frostig}, \bibinfo{person}{Peter Hawkins}, \bibinfo{person}{Matthew~James
  Johnson}, \bibinfo{person}{Chris Leary}, \bibinfo{person}{Dougal Maclaurin},
  \bibinfo{person}{George Necula}, \bibinfo{person}{Adam Paszke},
  \bibinfo{person}{Jake VanderPlas}, \bibinfo{person}{Skye Wanderman-Milne},
  {et~al\mbox{.}}} \bibinfo{year}{2018}\natexlab{}.
\newblock \showarticletitle{JAX: Composable Transformations of Python+ NumPy
  Programs (v0. 2.5)}.
\newblock \bibinfo{journal}{\emph{Software available from https://github.
  com/google/jax}} (\bibinfo{year}{2018}).
\newblock


\bibitem[\protect\citeauthoryear{Cai, Wang, Bi, Tu, Liu, and Shi}{Cai
  et~al\mbox{.}}{2019}]%
        {cai2019retrieval}
\bibfield{author}{\bibinfo{person}{Deng Cai}, \bibinfo{person}{Yan Wang},
  \bibinfo{person}{Wei Bi}, \bibinfo{person}{Zhaopeng Tu},
  \bibinfo{person}{Xiaojiang Liu}, {and} \bibinfo{person}{Shuming Shi}.}
  \bibinfo{year}{2019}\natexlab{}.
\newblock \showarticletitle{Retrieval-guided dialogue response generation via a
  matching-to-generation framework}. In \bibinfo{booktitle}{\emph{Proceedings
  of the 2019 Conference on Empirical Methods in Natural Language Processing
  and the 9th International Joint Conference on Natural Language Processing
  (EMNLP-IJCNLP)}}. \bibinfo{pages}{1866--1875}.
\newblock


\bibitem[\protect\citeauthoryear{Caliskan-Islam, Harang, Liu, Narayanan, Voss,
  Yamaguchi, and Greenstadt}{Caliskan-Islam et~al\mbox{.}}{2015}]%
        {caliskan2015anonymizing}
\bibfield{author}{\bibinfo{person}{Aylin Caliskan-Islam},
  \bibinfo{person}{Richard Harang}, \bibinfo{person}{Andrew Liu},
  \bibinfo{person}{Arvind Narayanan}, \bibinfo{person}{Clare Voss},
  \bibinfo{person}{Fabian Yamaguchi}, {and} \bibinfo{person}{Rachel
  Greenstadt}.} \bibinfo{year}{2015}\natexlab{}.
\newblock \showarticletitle{De-anonymizing programmers via code stylometry}. In
  \bibinfo{booktitle}{\emph{24th USENIX security symposium (USENIX Security
  15)}}. \bibinfo{pages}{255--270}.
\newblock


\bibitem[\protect\citeauthoryear{Carlini and Wagner}{Carlini and
  Wagner}{2017}]%
        {carlini2017towards}
\bibfield{author}{\bibinfo{person}{Nicholas Carlini} {and}
  \bibinfo{person}{David Wagner}.} \bibinfo{year}{2017}\natexlab{}.
\newblock \showarticletitle{Towards evaluating the robustness of neural
  networks}. In \bibinfo{booktitle}{\emph{2017 ieee symposium on security and
  privacy (sp)}}. Ieee, \bibinfo{pages}{39--57}.
\newblock


\bibitem[\protect\citeauthoryear{Chakraborty, Ahmed, Ding, Devanbu, and
  Ray}{Chakraborty et~al\mbox{.}}{2022}]%
        {chakraborty2022natgen}
\bibfield{author}{\bibinfo{person}{Saikat Chakraborty},
  \bibinfo{person}{Toufique Ahmed}, \bibinfo{person}{Yangruibo Ding},
  \bibinfo{person}{Premkumar~T. Devanbu}, {and} \bibinfo{person}{Baishakhi
  Ray}.} \bibinfo{year}{2022}\natexlab{}.
\newblock \showarticletitle{NatGen: generative pre-training by "naturalizing"
  source code}. In \bibinfo{booktitle}{\emph{Proceedings of the 30th {ACM}
  Joint European Software Engineering Conference and Symposium on the
  Foundations of Software Engineering, {ESEC/FSE} 2022, Singapore, Singapore,
  November 14-18, 2022}}, \bibfield{editor}{\bibinfo{person}{Abhik
  Roychoudhury}, \bibinfo{person}{Cristian Cadar}, {and}
  \bibinfo{person}{Miryung Kim}} (Eds.). \bibinfo{publisher}{{ACM}},
  \bibinfo{pages}{18--30}.
\newblock
\urldef\tempurl%
\url{https://doi.org/10.1145/3540250.3549162}
\showDOI{\tempurl}


\bibitem[\protect\citeauthoryear{Chen, Zhang, Nguyen, Zan, Lin, Lou, and
  Chen}{Chen et~al\mbox{.}}{2022}]%
        {chen2022codet}
\bibfield{author}{\bibinfo{person}{Bei Chen}, \bibinfo{person}{Fengji Zhang},
  \bibinfo{person}{Anh Nguyen}, \bibinfo{person}{Daoguang Zan},
  \bibinfo{person}{Zeqi Lin}, \bibinfo{person}{Jian-Guang Lou}, {and}
  \bibinfo{person}{Weizhu Chen}.} \bibinfo{year}{2022}\natexlab{}.
\newblock \showarticletitle{CodeT: Code Generation with Generated Tests}.
\newblock \bibinfo{journal}{\emph{arXiv preprint arXiv:2207.10397}}
  (\bibinfo{year}{2022}).
\newblock


\bibitem[\protect\citeauthoryear{Chen, Tworek, Jun, Yuan, Pinto, Kaplan,
  Edwards, Burda, Joseph, Brockman, et~al\mbox{.}}{Chen et~al\mbox{.}}{2021}]%
        {chen2021evaluating}
\bibfield{author}{\bibinfo{person}{Mark Chen}, \bibinfo{person}{Jerry Tworek},
  \bibinfo{person}{Heewoo Jun}, \bibinfo{person}{Qiming Yuan},
  \bibinfo{person}{Henrique Ponde de~Oliveira Pinto}, \bibinfo{person}{Jared
  Kaplan}, \bibinfo{person}{Harri Edwards}, \bibinfo{person}{Yuri Burda},
  \bibinfo{person}{Nicholas Joseph}, \bibinfo{person}{Greg Brockman},
  {et~al\mbox{.}}} \bibinfo{year}{2021}\natexlab{}.
\newblock \showarticletitle{Evaluating large language models trained on code}.
\newblock \bibinfo{journal}{\emph{arXiv preprint arXiv:2107.03374}}
  (\bibinfo{year}{2021}).
\newblock


\bibitem[\protect\citeauthoryear{Chowdhery, Narang, Devlin, Bosma, Mishra,
  Roberts, Barham, Chung, Sutton, Gehrmann, et~al\mbox{.}}{Chowdhery
  et~al\mbox{.}}{2022}]%
        {chowdhery2022palm}
\bibfield{author}{\bibinfo{person}{Aakanksha Chowdhery},
  \bibinfo{person}{Sharan Narang}, \bibinfo{person}{Jacob Devlin},
  \bibinfo{person}{Maarten Bosma}, \bibinfo{person}{Gaurav Mishra},
  \bibinfo{person}{Adam Roberts}, \bibinfo{person}{Paul Barham},
  \bibinfo{person}{Hyung~Won Chung}, \bibinfo{person}{Charles Sutton},
  \bibinfo{person}{Sebastian Gehrmann}, {et~al\mbox{.}}}
  \bibinfo{year}{2022}\natexlab{}.
\newblock \showarticletitle{Palm: Scaling language modeling with pathways}.
\newblock \bibinfo{journal}{\emph{arXiv preprint arXiv:2204.02311}}
  (\bibinfo{year}{2022}).
\newblock


\bibitem[\protect\citeauthoryear{Christopoulou, Lampouras, Gritta, Zhang, Guo,
  Li, Zhang, Xiao, Shen, Li, et~al\mbox{.}}{Christopoulou
  et~al\mbox{.}}{2022}]%
        {christopoulou2022pangu}
\bibfield{author}{\bibinfo{person}{Fenia Christopoulou},
  \bibinfo{person}{Gerasimos Lampouras}, \bibinfo{person}{Milan Gritta},
  \bibinfo{person}{Guchun Zhang}, \bibinfo{person}{Yinpeng Guo},
  \bibinfo{person}{Zhongqi Li}, \bibinfo{person}{Qi Zhang},
  \bibinfo{person}{Meng Xiao}, \bibinfo{person}{Bo Shen}, \bibinfo{person}{Lin
  Li}, {et~al\mbox{.}}} \bibinfo{year}{2022}\natexlab{}.
\newblock \showarticletitle{PanGu-Coder: Program Synthesis with Function-Level
  Language Modeling}.
\newblock \bibinfo{journal}{\emph{arXiv preprint arXiv:2207.11280}}
  (\bibinfo{year}{2022}).
\newblock


\bibitem[\protect\citeauthoryear{Clement, Drain, Timcheck, Svyatkovskiy, and
  Sundaresan}{Clement et~al\mbox{.}}{2020}]%
        {clement2020pymt5}
\bibfield{author}{\bibinfo{person}{Colin Clement}, \bibinfo{person}{Dawn
  Drain}, \bibinfo{person}{Jonathan Timcheck}, \bibinfo{person}{Alexey
  Svyatkovskiy}, {and} \bibinfo{person}{Neel Sundaresan}.}
  \bibinfo{year}{2020}\natexlab{}.
\newblock \showarticletitle{PyMT5: multi-mode translation of natural language
  and Python code with transformers}. In \bibinfo{booktitle}{\emph{Proceedings
  of the 2020 Conference on Empirical Methods in Natural Language Processing
  (EMNLP)}}. \bibinfo{pages}{9052--9065}.
\newblock


\bibitem[\protect\citeauthoryear{Committee et~al\mbox{.}}{Committee
  et~al\mbox{.}}{1990}]%
        {isc1990ieee}
\bibfield{author}{\bibinfo{person}{ISC Committee} {et~al\mbox{.}}}
  \bibinfo{year}{1990}\natexlab{}.
\newblock \showarticletitle{IEEE standard glossary of software engineering
  terminology (IEEE Std 610.12-1990). Los Alamitos}.
\newblock \bibinfo{journal}{\emph{CA IEEE Comput. Soc}} (\bibinfo{year}{1990}).
\newblock


\bibitem[\protect\citeauthoryear{Dao, Fu, Ermon, Rudra, and R{\'e}}{Dao
  et~al\mbox{.}}{2022}]%
        {dao2022flashattention}
\bibfield{author}{\bibinfo{person}{Tri Dao}, \bibinfo{person}{Dan Fu},
  \bibinfo{person}{Stefano Ermon}, \bibinfo{person}{Atri Rudra}, {and}
  \bibinfo{person}{Christopher R{\'e}}.} \bibinfo{year}{2022}\natexlab{}.
\newblock \showarticletitle{Flashattention: Fast and memory-efficient exact
  attention with io-awareness}.
\newblock \bibinfo{journal}{\emph{Advances in Neural Information Processing
  Systems}}  \bibinfo{volume}{35} (\bibinfo{year}{2022}),
  \bibinfo{pages}{16344--16359}.
\newblock


\bibitem[\protect\citeauthoryear{Deng, Zheng, Zhang, Chen, Lou, and Kim}{Deng
  et~al\mbox{.}}{2020}]%
        {deng2020analysis}
\bibfield{author}{\bibinfo{person}{Yao Deng}, \bibinfo{person}{Xi Zheng},
  \bibinfo{person}{Tianyi Zhang}, \bibinfo{person}{Chen Chen},
  \bibinfo{person}{Guannan Lou}, {and} \bibinfo{person}{Miryung Kim}.}
  \bibinfo{year}{2020}\natexlab{}.
\newblock \showarticletitle{An analysis of adversarial attacks and defenses on
  autonomous driving models}. In \bibinfo{booktitle}{\emph{2020 IEEE
  international conference on pervasive computing and communications
  (PerCom)}}. IEEE, \bibinfo{pages}{1--10}.
\newblock


\bibitem[\protect\citeauthoryear{Feng, Guo, Tang, Duan, Feng, Gong, Shou, Qin,
  Liu, Jiang, et~al\mbox{.}}{Feng et~al\mbox{.}}{2020}]%
        {feng2020codebert}
\bibfield{author}{\bibinfo{person}{Zhangyin Feng}, \bibinfo{person}{Daya Guo},
  \bibinfo{person}{Duyu Tang}, \bibinfo{person}{Nan Duan},
  \bibinfo{person}{Xiaocheng Feng}, \bibinfo{person}{Ming Gong},
  \bibinfo{person}{Linjun Shou}, \bibinfo{person}{Bing Qin},
  \bibinfo{person}{Ting Liu}, \bibinfo{person}{Daxin Jiang}, {et~al\mbox{.}}}
  \bibinfo{year}{2020}\natexlab{}.
\newblock \showarticletitle{CodeBERT: A Pre-Trained Model for Programming and
  Natural Languages}. In \bibinfo{booktitle}{\emph{Findings of the Association
  for Computational Linguistics: EMNLP 2020}}. \bibinfo{pages}{1536--1547}.
\newblock


\bibitem[\protect\citeauthoryear{Fried, Aghajanyan, Lin, Wang, Wallace, Shi,
  Zhong, Yih, Zettlemoyer, and Lewis}{Fried et~al\mbox{.}}{2022}]%
        {fried2022incoder}
\bibfield{author}{\bibinfo{person}{Daniel Fried}, \bibinfo{person}{Armen
  Aghajanyan}, \bibinfo{person}{Jessy Lin}, \bibinfo{person}{Sida Wang},
  \bibinfo{person}{Eric Wallace}, \bibinfo{person}{Freda Shi},
  \bibinfo{person}{Ruiqi Zhong}, \bibinfo{person}{Wen-tau Yih},
  \bibinfo{person}{Luke Zettlemoyer}, {and} \bibinfo{person}{Mike Lewis}.}
  \bibinfo{year}{2022}\natexlab{}.
\newblock \showarticletitle{Incoder: A generative model for code infilling and
  synthesis}.
\newblock \bibinfo{journal}{\emph{arXiv preprint arXiv:2204.05999}}
  (\bibinfo{year}{2022}).
\newblock


\bibitem[\protect\citeauthoryear{Gao, Chen, Xing, Ma, Song, and Lin}{Gao
  et~al\mbox{.}}{2019}]%
        {gao2019neural}
\bibfield{author}{\bibinfo{person}{Sa Gao}, \bibinfo{person}{Chunyang Chen},
  \bibinfo{person}{Zhenchang Xing}, \bibinfo{person}{Yukun Ma},
  \bibinfo{person}{Wen Song}, {and} \bibinfo{person}{Shang-Wei Lin}.}
  \bibinfo{year}{2019}\natexlab{}.
\newblock \showarticletitle{A neural model for method name generation from
  functional description}. In \bibinfo{booktitle}{\emph{2019 IEEE 26th
  International Conference on Software Analysis, Evolution and Reengineering
  (SANER)}}. IEEE, \bibinfo{pages}{414--421}.
\newblock


\bibitem[\protect\citeauthoryear{Garc{\'\i}a-Pedrajas, Ortiz-Boyer, and
  Herv{\'a}s-Mart{\'\i}nez}{Garc{\'\i}a-Pedrajas et~al\mbox{.}}{2006}]%
        {garcia2006alternative}
\bibfield{author}{\bibinfo{person}{Nicol{\'a}s Garc{\'\i}a-Pedrajas},
  \bibinfo{person}{Domingo Ortiz-Boyer}, {and} \bibinfo{person}{C{\'e}sar
  Herv{\'a}s-Mart{\'\i}nez}.} \bibinfo{year}{2006}\natexlab{}.
\newblock \showarticletitle{An alternative approach for neural network
  evolution with a genetic algorithm: Crossover by combinatorial optimization}.
\newblock \bibinfo{journal}{\emph{Neural Networks}} \bibinfo{volume}{19},
  \bibinfo{number}{4} (\bibinfo{year}{2006}), \bibinfo{pages}{514--528}.
\newblock


\bibitem[\protect\citeauthoryear{Ge and Kuang}{Ge and Kuang}{2021}]%
        {ge2021keywords}
\bibfield{author}{\bibinfo{person}{Fan Ge} {and} \bibinfo{person}{Li Kuang}.}
  \bibinfo{year}{2021}\natexlab{}.
\newblock \showarticletitle{Keywords guided method name generation}. In
  \bibinfo{booktitle}{\emph{2021 IEEE/ACM 29th International Conference on
  Program Comprehension (ICPC)}}. IEEE, \bibinfo{pages}{196--206}.
\newblock


\bibitem[\protect\citeauthoryear{Goodfellow, Shlens, and Szegedy}{Goodfellow
  et~al\mbox{.}}{2015}]%
        {goodfellow2014explaining}
\bibfield{author}{\bibinfo{person}{Ian~J. Goodfellow},
  \bibinfo{person}{Jonathon Shlens}, {and} \bibinfo{person}{Christian
  Szegedy}.} \bibinfo{year}{2015}\natexlab{}.
\newblock \showarticletitle{Explaining and Harnessing Adversarial Examples}. In
  \bibinfo{booktitle}{\emph{3rd International Conference on Learning
  Representations, {ICLR} 2015, San Diego, CA, USA, May 7-9, 2015, Conference
  Track Proceedings}}, \bibfield{editor}{\bibinfo{person}{Yoshua Bengio} {and}
  \bibinfo{person}{Yann LeCun}} (Eds.).
\newblock
\urldef\tempurl%
\url{http://arxiv.org/abs/1412.6572}
\showURL{%
\tempurl}


\bibitem[\protect\citeauthoryear{Goswami, Ratha, Agarwal, Singh, and
  Vatsa}{Goswami et~al\mbox{.}}{2018}]%
        {goswami2018unravelling}
\bibfield{author}{\bibinfo{person}{Gaurav Goswami}, \bibinfo{person}{Nalini
  Ratha}, \bibinfo{person}{Akshay Agarwal}, \bibinfo{person}{Richa Singh},
  {and} \bibinfo{person}{Mayank Vatsa}.} \bibinfo{year}{2018}\natexlab{}.
\newblock \showarticletitle{Unravelling robustness of deep learning based face
  recognition against adversarial attacks}. In
  \bibinfo{booktitle}{\emph{Proceedings of the AAAI Conference on Artificial
  Intelligence}}, Vol.~\bibinfo{volume}{32}.
\newblock


\bibitem[\protect\citeauthoryear{Gu, Wang, Cho, and Li}{Gu
  et~al\mbox{.}}{2018}]%
        {gu2018search}
\bibfield{author}{\bibinfo{person}{Jiatao Gu}, \bibinfo{person}{Yong Wang},
  \bibinfo{person}{Kyunghyun Cho}, {and} \bibinfo{person}{Victor~OK Li}.}
  \bibinfo{year}{2018}\natexlab{}.
\newblock \showarticletitle{Search engine guided neural machine translation}.
  In \bibinfo{booktitle}{\emph{Proceedings of the AAAI Conference on Artificial
  Intelligence}}, Vol.~\bibinfo{volume}{32}.
\newblock


\bibitem[\protect\citeauthoryear{Gunasekar, Zhang, Aneja, Mendes, Del~Giorno,
  Gopi, Javaheripi, Kauffmann, de~Rosa, Saarikivi, et~al\mbox{.}}{Gunasekar
  et~al\mbox{.}}{2023}]%
        {gunasekar2023textbooks}
\bibfield{author}{\bibinfo{person}{Suriya Gunasekar}, \bibinfo{person}{Yi
  Zhang}, \bibinfo{person}{Jyoti Aneja}, \bibinfo{person}{Caio
  C{\'e}sar~Teodoro Mendes}, \bibinfo{person}{Allie Del~Giorno},
  \bibinfo{person}{Sivakanth Gopi}, \bibinfo{person}{Mojan Javaheripi},
  \bibinfo{person}{Piero Kauffmann}, \bibinfo{person}{Gustavo de Rosa},
  \bibinfo{person}{Olli Saarikivi}, {et~al\mbox{.}}}
  \bibinfo{year}{2023}\natexlab{}.
\newblock \showarticletitle{Textbooks Are All You Need}.
\newblock \bibinfo{journal}{\emph{arXiv preprint arXiv:2306.11644}}
  (\bibinfo{year}{2023}).
\newblock


\bibitem[\protect\citeauthoryear{Guo, Lu, Duan, Wang, Zhou, and Yin}{Guo
  et~al\mbox{.}}{2022}]%
        {guo2022unixcoder}
\bibfield{author}{\bibinfo{person}{Daya Guo}, \bibinfo{person}{Shuai Lu},
  \bibinfo{person}{Nan Duan}, \bibinfo{person}{Yanlin Wang},
  \bibinfo{person}{Ming Zhou}, {and} \bibinfo{person}{Jian Yin}.}
  \bibinfo{year}{2022}\natexlab{}.
\newblock \showarticletitle{UniXcoder: Unified Cross-Modal Pre-training for
  Code Representation}. In \bibinfo{booktitle}{\emph{Proceedings of the 60th
  Annual Meeting of the Association for Computational Linguistics (Volume 1:
  Long Papers)}}. \bibinfo{pages}{7212--7225}.
\newblock


\bibitem[\protect\citeauthoryear{Guo, Ren, Lu, Feng, Tang, Liu, Zhou, Duan,
  Svyatkovskiy, Fu, et~al\mbox{.}}{Guo et~al\mbox{.}}{2021}]%
        {guo2021graphcodebert}
\bibfield{author}{\bibinfo{person}{Daya Guo}, \bibinfo{person}{Shuo Ren},
  \bibinfo{person}{Shuai Lu}, \bibinfo{person}{Zhangyin Feng},
  \bibinfo{person}{Duyu Tang}, \bibinfo{person}{Shujie Liu},
  \bibinfo{person}{Long Zhou}, \bibinfo{person}{Nan Duan},
  \bibinfo{person}{Alexey Svyatkovskiy}, \bibinfo{person}{Shengyu Fu},
  {et~al\mbox{.}}} \bibinfo{year}{2021}\natexlab{}.
\newblock \showarticletitle{GraphCodeBERT: Pre-training Code Representations
  with Data Flow}. In \bibinfo{booktitle}{\emph{ICLR}}.
\newblock


\bibitem[\protect\citeauthoryear{Hao, Li, Liu, Miao, Zong, Jiang, Liu, and
  Wei}{Hao et~al\mbox{.}}{2022}]%
        {hao2022AixBench}
\bibfield{author}{\bibinfo{person}{Yiyang Hao}, \bibinfo{person}{Ge Li},
  \bibinfo{person}{Yongqiang Liu}, \bibinfo{person}{Xiaowei Miao},
  \bibinfo{person}{He Zong}, \bibinfo{person}{Siyuan Jiang},
  \bibinfo{person}{Yang Liu}, {and} \bibinfo{person}{He Wei}.}
  \bibinfo{year}{2022}\natexlab{}.
\newblock \bibinfo{title}{AixBench: A Code Generation Benchmark Dataset}.
\newblock
\newblock
\showeprint{arXiv:2206.13179}


\bibitem[\protect\citeauthoryear{Hashimoto, Guu, Oren, and Liang}{Hashimoto
  et~al\mbox{.}}{2018}]%
        {hashimoto2018retrieve}
\bibfield{author}{\bibinfo{person}{Tatsunori~B Hashimoto},
  \bibinfo{person}{Kelvin Guu}, \bibinfo{person}{Yonatan Oren}, {and}
  \bibinfo{person}{Percy~S Liang}.} \bibinfo{year}{2018}\natexlab{}.
\newblock \showarticletitle{A retrieve-and-edit framework for predicting
  structured outputs}.
\newblock \bibinfo{journal}{\emph{Advances in Neural Information Processing
  Systems}}  \bibinfo{volume}{31} (\bibinfo{year}{2018}).
\newblock


\bibitem[\protect\citeauthoryear{Hayati, Olivier, Avvaru, Yin, Tomasic, and
  Neubig}{Hayati et~al\mbox{.}}{2018}]%
        {hayati2018retrieval}
\bibfield{author}{\bibinfo{person}{Shirley~Anugrah Hayati},
  \bibinfo{person}{Raphael Olivier}, \bibinfo{person}{Pravalika Avvaru},
  \bibinfo{person}{Pengcheng Yin}, \bibinfo{person}{Anthony Tomasic}, {and}
  \bibinfo{person}{Graham Neubig}.} \bibinfo{year}{2018}\natexlab{}.
\newblock \showarticletitle{Retrieval-Based Neural Code Generation}. In
  \bibinfo{booktitle}{\emph{Proceedings of the 2018 Conference on Empirical
  Methods in Natural Language Processing}}. \bibinfo{pages}{925--930}.
\newblock


\bibitem[\protect\citeauthoryear{Hestness, Narang, Ardalani, Diamos, Jun,
  Kianinejad, Patwary, Yang, and Zhou}{Hestness et~al\mbox{.}}{2017}]%
        {hestness2017deep}
\bibfield{author}{\bibinfo{person}{Joel Hestness}, \bibinfo{person}{Sharan
  Narang}, \bibinfo{person}{Newsha Ardalani}, \bibinfo{person}{Gregory Diamos},
  \bibinfo{person}{Heewoo Jun}, \bibinfo{person}{Hassan Kianinejad},
  \bibinfo{person}{Md~Mostofa~Ali Patwary}, \bibinfo{person}{Yang Yang}, {and}
  \bibinfo{person}{Yanqi Zhou}.} \bibinfo{year}{2017}\natexlab{}.
\newblock \showarticletitle{Deep learning scaling is predictable, empirically}.
\newblock \bibinfo{journal}{\emph{arXiv preprint arXiv:1712.00409}}
  (\bibinfo{year}{2017}).
\newblock


\bibitem[\protect\citeauthoryear{Hindle, Barr, Gabel, Su, and Devanbu}{Hindle
  et~al\mbox{.}}{2016}]%
        {hindle2016naturalness}
\bibfield{author}{\bibinfo{person}{Abram Hindle}, \bibinfo{person}{Earl~T
  Barr}, \bibinfo{person}{Mark Gabel}, \bibinfo{person}{Zhendong Su}, {and}
  \bibinfo{person}{Premkumar Devanbu}.} \bibinfo{year}{2016}\natexlab{}.
\newblock \showarticletitle{On the naturalness of software}.
\newblock \bibinfo{journal}{\emph{Commun. ACM}} \bibinfo{volume}{59},
  \bibinfo{number}{5} (\bibinfo{year}{2016}), \bibinfo{pages}{122--131}.
\newblock


\bibitem[\protect\citeauthoryear{Hofmeister, Siegmund, and Holt}{Hofmeister
  et~al\mbox{.}}{2017}]%
        {hofmeister2017shorter}
\bibfield{author}{\bibinfo{person}{Johannes Hofmeister}, \bibinfo{person}{Janet
  Siegmund}, {and} \bibinfo{person}{Daniel~V Holt}.}
  \bibinfo{year}{2017}\natexlab{}.
\newblock \showarticletitle{Shorter identifier names take longer to
  comprehend}. In \bibinfo{booktitle}{\emph{2017 IEEE 24th International
  conference on software analysis, evolution and reengineering (SANER)}}. IEEE,
  \bibinfo{pages}{217--227}.
\newblock


\bibitem[\protect\citeauthoryear{Huang, Chen, Wu, Zhao, Xie, and Sun}{Huang
  et~al\mbox{.}}{2021}]%
        {huang2021named}
\bibfield{author}{\bibinfo{person}{Xiusheng Huang}, \bibinfo{person}{Yubo
  Chen}, \bibinfo{person}{Shun Wu}, \bibinfo{person}{Jun Zhao},
  \bibinfo{person}{Yuantao Xie}, {and} \bibinfo{person}{Weijian Sun}.}
  \bibinfo{year}{2021}\natexlab{}.
\newblock \showarticletitle{Named Entity Recognition via Noise Aware Training
  Mechanism with Data Filter}. In \bibinfo{booktitle}{\emph{Findings of the
  Association for Computational Linguistics: ACL-IJCNLP 2021}}.
  \bibinfo{pages}{4791--4803}.
\newblock


\bibitem[\protect\citeauthoryear{Husain, Wu, Gazit, Allamanis, and
  Brockschmidt}{Husain et~al\mbox{.}}{2019}]%
        {husain2019codesearchnet}
\bibfield{author}{\bibinfo{person}{Hamel Husain}, \bibinfo{person}{Ho-Hsiang
  Wu}, \bibinfo{person}{Tiferet Gazit}, \bibinfo{person}{Miltiadis Allamanis},
  {and} \bibinfo{person}{Marc Brockschmidt}.} \bibinfo{year}{2019}\natexlab{}.
\newblock \showarticletitle{Codesearchnet challenge: Evaluating the state of
  semantic code search}.
\newblock \bibinfo{journal}{\emph{arXiv preprint arXiv:1909.09436}}
  (\bibinfo{year}{2019}).
\newblock


\bibitem[\protect\citeauthoryear{Iyer, Konstas, Cheung, and Zettlemoyer}{Iyer
  et~al\mbox{.}}{2018}]%
        {iyer2018mapping}
\bibfield{author}{\bibinfo{person}{Srinivasan Iyer}, \bibinfo{person}{Ioannis
  Konstas}, \bibinfo{person}{Alvin Cheung}, {and} \bibinfo{person}{Luke
  Zettlemoyer}.} \bibinfo{year}{2018}\natexlab{}.
\newblock \showarticletitle{Mapping Language to Code in Programmatic Context}.
  In \bibinfo{booktitle}{\emph{Proceedings of the 2018 Conference on Empirical
  Methods in Natural Language Processing}}. \bibinfo{pages}{1643--1652}.
\newblock


\bibitem[\protect\citeauthoryear{Kaplan, McCandlish, Henighan, Brown, Chess,
  Child, Gray, Radford, Wu, and Amodei}{Kaplan et~al\mbox{.}}{2020}]%
        {kaplan2020scaling}
\bibfield{author}{\bibinfo{person}{Jared Kaplan}, \bibinfo{person}{Sam
  McCandlish}, \bibinfo{person}{Tom Henighan}, \bibinfo{person}{Tom~B Brown},
  \bibinfo{person}{Benjamin Chess}, \bibinfo{person}{Rewon Child},
  \bibinfo{person}{Scott Gray}, \bibinfo{person}{Alec Radford},
  \bibinfo{person}{Jeffrey Wu}, {and} \bibinfo{person}{Dario Amodei}.}
  \bibinfo{year}{2020}\natexlab{}.
\newblock \showarticletitle{Scaling laws for neural language models}.
\newblock \bibinfo{journal}{\emph{arXiv preprint arXiv:2001.08361}}
  (\bibinfo{year}{2020}).
\newblock


\bibitem[\protect\citeauthoryear{Lewis, Liu, Goyal, Ghazvininejad, Mohamed,
  Levy, Stoyanov, and Zettlemoyer}{Lewis et~al\mbox{.}}{2020}]%
        {lewis2020bart}
\bibfield{author}{\bibinfo{person}{Mike Lewis}, \bibinfo{person}{Yinhan Liu},
  \bibinfo{person}{Naman Goyal}, \bibinfo{person}{Marjan Ghazvininejad},
  \bibinfo{person}{Abdelrahman Mohamed}, \bibinfo{person}{Omer Levy},
  \bibinfo{person}{Veselin Stoyanov}, {and} \bibinfo{person}{Luke
  Zettlemoyer}.} \bibinfo{year}{2020}\natexlab{}.
\newblock \showarticletitle{BART: Denoising Sequence-to-Sequence Pre-training
  for Natural Language Generation, Translation, and Comprehension}. In
  \bibinfo{booktitle}{\emph{Proceedings of the 58th Annual Meeting of the
  Association for Computational Linguistics}}. \bibinfo{pages}{7871--7880}.
\newblock


\bibitem[\protect\citeauthoryear{Li, Su, Cai, Wang, and Liu}{Li
  et~al\mbox{.}}{2022b}]%
        {li2022survey}
\bibfield{author}{\bibinfo{person}{Huayang Li}, \bibinfo{person}{Yixuan Su},
  \bibinfo{person}{Deng Cai}, \bibinfo{person}{Yan Wang}, {and}
  \bibinfo{person}{Lemao Liu}.} \bibinfo{year}{2022}\natexlab{b}.
\newblock \showarticletitle{A Survey on Retrieval-Augmented Text Generation}.
\newblock \bibinfo{journal}{\emph{arXiv preprint arXiv:2202.01110}}
  (\bibinfo{year}{2022}).
\newblock


\bibitem[\protect\citeauthoryear{Li, Ji, Du, Li, and Wang}{Li
  et~al\mbox{.}}{2019}]%
        {li2019textbugger}
\bibfield{author}{\bibinfo{person}{J Li}, \bibinfo{person}{S Ji},
  \bibinfo{person}{T Du}, \bibinfo{person}{B Li}, {and} \bibinfo{person}{T
  Wang}.} \bibinfo{year}{2019}\natexlab{}.
\newblock \showarticletitle{TextBugger: Generating Adversarial Text Against
  Real-world Applications}. In \bibinfo{booktitle}{\emph{26th Annual Network
  and Distributed System Security Symposium}}.
\newblock


\bibitem[\protect\citeauthoryear{Li, Allal, Zi, Muennighoff, Kocetkov, Mou,
  Marone, Akiki, Li, Chim, et~al\mbox{.}}{Li et~al\mbox{.}}{2023}]%
        {li2023starcoder}
\bibfield{author}{\bibinfo{person}{Raymond Li}, \bibinfo{person}{Loubna~Ben
  Allal}, \bibinfo{person}{Yangtian Zi}, \bibinfo{person}{Niklas Muennighoff},
  \bibinfo{person}{Denis Kocetkov}, \bibinfo{person}{Chenghao Mou},
  \bibinfo{person}{Marc Marone}, \bibinfo{person}{Christopher Akiki},
  \bibinfo{person}{Jia Li}, \bibinfo{person}{Jenny Chim}, {et~al\mbox{.}}}
  \bibinfo{year}{2023}\natexlab{}.
\newblock \showarticletitle{StarCoder: may the source be with you!}
\newblock \bibinfo{journal}{\emph{arXiv preprint arXiv:2305.06161}}
  (\bibinfo{year}{2023}).
\newblock


\bibitem[\protect\citeauthoryear{Li, Choi, Chung, Kushman, Schrittwieser,
  Leblond, Eccles, Keeling, Gimeno, Dal~Lago, et~al\mbox{.}}{Li
  et~al\mbox{.}}{2022a}]%
        {li2022competition}
\bibfield{author}{\bibinfo{person}{Yujia Li}, \bibinfo{person}{David Choi},
  \bibinfo{person}{Junyoung Chung}, \bibinfo{person}{Nate Kushman},
  \bibinfo{person}{Julian Schrittwieser}, \bibinfo{person}{R{\'e}mi Leblond},
  \bibinfo{person}{Tom Eccles}, \bibinfo{person}{James Keeling},
  \bibinfo{person}{Felix Gimeno}, \bibinfo{person}{Agustin Dal~Lago},
  {et~al\mbox{.}}} \bibinfo{year}{2022}\natexlab{a}.
\newblock \showarticletitle{Competition-level code generation with alphacode}.
\newblock \bibinfo{journal}{\emph{Science}} \bibinfo{volume}{378},
  \bibinfo{number}{6624} (\bibinfo{year}{2022}), \bibinfo{pages}{1092--1097}.
\newblock


\bibitem[\protect\citeauthoryear{Liguori, Al-Hossami, Cotroneo, Natella, Cukic,
  and Shaikh}{Liguori et~al\mbox{.}}{2022}]%
        {liguori2022can}
\bibfield{author}{\bibinfo{person}{Pietro Liguori}, \bibinfo{person}{Erfan
  Al-Hossami}, \bibinfo{person}{Domenico Cotroneo}, \bibinfo{person}{Roberto
  Natella}, \bibinfo{person}{Bojan Cukic}, {and} \bibinfo{person}{Samira
  Shaikh}.} \bibinfo{year}{2022}\natexlab{}.
\newblock \showarticletitle{Can we generate shellcodes via natural language? An
  empirical study}.
\newblock \bibinfo{journal}{\emph{Automated Software Engineering}}
  \bibinfo{volume}{29}, \bibinfo{number}{1} (\bibinfo{year}{2022}),
  \bibinfo{pages}{1--34}.
\newblock


\bibitem[\protect\citeauthoryear{Liguori, Al-Hossami, Orbinato, Natella,
  Shaikh, Cotroneo, and Cukic}{Liguori et~al\mbox{.}}{2021}]%
        {liguori2021evil}
\bibfield{author}{\bibinfo{person}{Pietro Liguori}, \bibinfo{person}{Erfan
  Al-Hossami}, \bibinfo{person}{Vittorio Orbinato}, \bibinfo{person}{Roberto
  Natella}, \bibinfo{person}{Samira Shaikh}, \bibinfo{person}{Domenico
  Cotroneo}, {and} \bibinfo{person}{Bojan Cukic}.}
  \bibinfo{year}{2021}\natexlab{}.
\newblock \showarticletitle{EVIL: exploiting software via natural language}. In
  \bibinfo{booktitle}{\emph{2021 IEEE 32nd International Symposium on Software
  Reliability Engineering (ISSRE)}}. IEEE, \bibinfo{pages}{321--332}.
\newblock


\bibitem[\protect\citeauthoryear{Lin}{Lin}{2004}]%
        {lin2004rouge}
\bibfield{author}{\bibinfo{person}{Chin-Yew Lin}.}
  \bibinfo{year}{2004}\natexlab{}.
\newblock \showarticletitle{Rouge: A package for automatic evaluation of
  summaries}. In \bibinfo{booktitle}{\emph{Text summarization branches out}}.
  \bibinfo{pages}{74--81}.
\newblock


\bibitem[\protect\citeauthoryear{Lin, Wang, Zettlemoyer, and Ernst}{Lin
  et~al\mbox{.}}{2018}]%
        {lin2018nl2bash}
\bibfield{author}{\bibinfo{person}{Xi~Victoria Lin}, \bibinfo{person}{Chenglong
  Wang}, \bibinfo{person}{Luke Zettlemoyer}, {and} \bibinfo{person}{Michael~D
  Ernst}.} \bibinfo{year}{2018}\natexlab{}.
\newblock \showarticletitle{NL2Bash: A Corpus and Semantic Parser for Natural
  Language Interface to the Linux Operating System}. In
  \bibinfo{booktitle}{\emph{Proceedings of the Eleventh International
  Conference on Language Resources and Evaluation (LREC 2018)}}.
\newblock


\bibitem[\protect\citeauthoryear{Ling, Blunsom, Grefenstette, Hermann,
  Ko{\v{c}}isk{\`y}, Wang, and Senior}{Ling et~al\mbox{.}}{2016}]%
        {ling2016latent}
\bibfield{author}{\bibinfo{person}{Wang Ling}, \bibinfo{person}{Phil Blunsom},
  \bibinfo{person}{Edward Grefenstette}, \bibinfo{person}{Karl~Moritz Hermann},
  \bibinfo{person}{Tom{\'a}{\v{s}} Ko{\v{c}}isk{\`y}}, \bibinfo{person}{Fumin
  Wang}, {and} \bibinfo{person}{Andrew Senior}.}
  \bibinfo{year}{2016}\natexlab{}.
\newblock \showarticletitle{Latent Predictor Networks for Code Generation}. In
  \bibinfo{booktitle}{\emph{Proceedings of the 54th Annual Meeting of the
  Association for Computational Linguistics (Volume 1: Long Papers)}}.
  \bibinfo{pages}{599--609}.
\newblock


\bibitem[\protect\citeauthoryear{Liu, Kim, Bissyand{\'e}, Kim, Kim, Koyuncu,
  Kim, and Le~Traon}{Liu et~al\mbox{.}}{2019}]%
        {liu2019learning}
\bibfield{author}{\bibinfo{person}{Kui Liu}, \bibinfo{person}{Dongsun Kim},
  \bibinfo{person}{Tegawend{\'e}~F Bissyand{\'e}}, \bibinfo{person}{Taeyoung
  Kim}, \bibinfo{person}{Kisub Kim}, \bibinfo{person}{Anil Koyuncu},
  \bibinfo{person}{Suntae Kim}, {and} \bibinfo{person}{Yves Le~Traon}.}
  \bibinfo{year}{2019}\natexlab{}.
\newblock \showarticletitle{Learning to spot and refactor inconsistent method
  names}. In \bibinfo{booktitle}{\emph{2019 IEEE/ACM 41st International
  Conference on Software Engineering (ICSE)}}. IEEE, \bibinfo{pages}{1--12}.
\newblock


\bibitem[\protect\citeauthoryear{Liu, Xia, Hassan, Lo, Xing, and Wang}{Liu
  et~al\mbox{.}}{2018}]%
        {liu2018neural}
\bibfield{author}{\bibinfo{person}{Zhongxin Liu}, \bibinfo{person}{Xin Xia},
  \bibinfo{person}{Ahmed~E Hassan}, \bibinfo{person}{David Lo},
  \bibinfo{person}{Zhenchang Xing}, {and} \bibinfo{person}{Xinyu Wang}.}
  \bibinfo{year}{2018}\natexlab{}.
\newblock \showarticletitle{Neural-machine-translation-based commit message
  generation: how far are we?}. In \bibinfo{booktitle}{\emph{Proceedings of the
  33rd ACM/IEEE International Conference on Automated Software Engineering}}.
  \bibinfo{pages}{373--384}.
\newblock


\bibitem[\protect\citeauthoryear{Locascio, Narasimhan, DeLeon, Kushman, and
  Barzilay}{Locascio et~al\mbox{.}}{2016}]%
        {locascio2016neural}
\bibfield{author}{\bibinfo{person}{Nicholas Locascio}, \bibinfo{person}{Karthik
  Narasimhan}, \bibinfo{person}{Eduardo DeLeon}, \bibinfo{person}{Nate
  Kushman}, {and} \bibinfo{person}{Regina Barzilay}.}
  \bibinfo{year}{2016}\natexlab{}.
\newblock \showarticletitle{Neural Generation of Regular Expressions from
  Natural Language with Minimal Domain Knowledge}. In
  \bibinfo{booktitle}{\emph{Proceedings of the 2016 Conference on Empirical
  Methods in Natural Language Processing}}. \bibinfo{pages}{1918--1923}.
\newblock


\bibitem[\protect\citeauthoryear{Lu, Guo, Ren, Huang, Svyatkovskiy, Blanco,
  Clement, Drain, Jiang, Tang, Li, Zhou, Shou, Zhou, Tufano, Gong, Zhou, Duan,
  Sundaresan, Deng, Fu, and Liu}{Lu et~al\mbox{.}}{2021}]%
        {lu2021codexglue}
\bibfield{author}{\bibinfo{person}{Shuai Lu}, \bibinfo{person}{Daya Guo},
  \bibinfo{person}{Shuo Ren}, \bibinfo{person}{Junjie Huang},
  \bibinfo{person}{Alexey Svyatkovskiy}, \bibinfo{person}{Ambrosio Blanco},
  \bibinfo{person}{Colin~B. Clement}, \bibinfo{person}{Dawn Drain},
  \bibinfo{person}{Daxin Jiang}, \bibinfo{person}{Duyu Tang},
  \bibinfo{person}{Ge Li}, \bibinfo{person}{Lidong Zhou},
  \bibinfo{person}{Linjun Shou}, \bibinfo{person}{Long Zhou},
  \bibinfo{person}{Michele Tufano}, \bibinfo{person}{Ming Gong},
  \bibinfo{person}{Ming Zhou}, \bibinfo{person}{Nan Duan},
  \bibinfo{person}{Neel Sundaresan}, \bibinfo{person}{Shao~Kun Deng},
  \bibinfo{person}{Shengyu Fu}, {and} \bibinfo{person}{Shujie Liu}.}
  \bibinfo{year}{2021}\natexlab{}.
\newblock \showarticletitle{CodeXGLUE: {A} Machine Learning Benchmark Dataset
  for Code Understanding and Generation}. In
  \bibinfo{booktitle}{\emph{Proceedings of the Neural Information Processing
  Systems Track on Datasets and Benchmarks 1, NeurIPS Datasets and Benchmarks
  2021, December 2021, virtual}}, \bibfield{editor}{\bibinfo{person}{Joaquin
  Vanschoren} {and} \bibinfo{person}{Sai{-}Kit Yeung}} (Eds.).
\newblock
\urldef\tempurl%
\url{https://datasets-benchmarks-proceedings.neurips.cc/paper/2021/hash/c16a5320fa475530d9583c34fd356ef5-Abstract-round1.html}
\showURL{%
\tempurl}


\bibitem[\protect\citeauthoryear{Lundberg and Lee}{Lundberg and Lee}{2017}]%
        {NIPS2017_8a20a862}
\bibfield{author}{\bibinfo{person}{Scott~M Lundberg} {and}
  \bibinfo{person}{Su-In Lee}.} \bibinfo{year}{2017}\natexlab{}.
\newblock \showarticletitle{A Unified Approach to Interpreting Model
  Predictions}. In \bibinfo{booktitle}{\emph{Advances in Neural Information
  Processing Systems}}, \bibfield{editor}{\bibinfo{person}{I.~Guyon},
  \bibinfo{person}{U.~Von Luxburg}, \bibinfo{person}{S.~Bengio},
  \bibinfo{person}{H.~Wallach}, \bibinfo{person}{R.~Fergus},
  \bibinfo{person}{S.~Vishwanathan}, {and} \bibinfo{person}{R.~Garnett}}
  (Eds.), Vol.~\bibinfo{volume}{30}. \bibinfo{publisher}{Curran Associates,
  Inc.}
\newblock
\urldef\tempurl%
\url{https://proceedings.neurips.cc/paper/2017/file/8a20a8621978632d76c43dfd28b67767-Paper.pdf}
\showURL{%
\tempurl}


\bibitem[\protect\citeauthoryear{Mikolov, Sutskever, Chen, Corrado, and
  Dean}{Mikolov et~al\mbox{.}}{2013}]%
        {mikolov2013distributed}
\bibfield{author}{\bibinfo{person}{Tomas Mikolov}, \bibinfo{person}{Ilya
  Sutskever}, \bibinfo{person}{Kai Chen}, \bibinfo{person}{Greg~S Corrado},
  {and} \bibinfo{person}{Jeff Dean}.} \bibinfo{year}{2013}\natexlab{}.
\newblock \showarticletitle{Distributed representations of words and phrases
  and their compositionality}.
\newblock \bibinfo{journal}{\emph{Advances in neural information processing
  systems}}  \bibinfo{volume}{26} (\bibinfo{year}{2013}).
\newblock


\bibitem[\protect\citeauthoryear{Mou, Men, Li, Zhang, and Jin}{Mou
  et~al\mbox{.}}{2015}]%
        {mou2015end}
\bibfield{author}{\bibinfo{person}{Lili Mou}, \bibinfo{person}{Rui Men},
  \bibinfo{person}{Ge Li}, \bibinfo{person}{Lu Zhang}, {and}
  \bibinfo{person}{Zhi Jin}.} \bibinfo{year}{2015}\natexlab{}.
\newblock \showarticletitle{On end-to-end program generation from user
  intention by deep neural networks}.
\newblock \bibinfo{journal}{\emph{arXiv preprint arXiv:1510.07211}}
  (\bibinfo{year}{2015}).
\newblock


\bibitem[\protect\citeauthoryear{Murphy-Hill, Parnin, and Black}{Murphy-Hill
  et~al\mbox{.}}{2011}]%
        {murphy2011we}
\bibfield{author}{\bibinfo{person}{Emerson Murphy-Hill}, \bibinfo{person}{Chris
  Parnin}, {and} \bibinfo{person}{Andrew~P Black}.}
  \bibinfo{year}{2011}\natexlab{}.
\newblock \showarticletitle{How we refactor, and how we know it}.
\newblock \bibinfo{journal}{\emph{IEEE Transactions on Software Engineering}}
  \bibinfo{volume}{38}, \bibinfo{number}{1} (\bibinfo{year}{2011}),
  \bibinfo{pages}{5--18}.
\newblock


\bibitem[\protect\citeauthoryear{Nijkamp, Hayashi, Xiong, Savarese, and
  Zhou}{Nijkamp et~al\mbox{.}}{2023}]%
        {nijkamp2023codegen2}
\bibfield{author}{\bibinfo{person}{Erik Nijkamp}, \bibinfo{person}{Hiroaki
  Hayashi}, \bibinfo{person}{Caiming Xiong}, \bibinfo{person}{Silvio Savarese},
  {and} \bibinfo{person}{Yingbo Zhou}.} \bibinfo{year}{2023}\natexlab{}.
\newblock \showarticletitle{Codegen2: Lessons for training llms on programming
  and natural languages}.
\newblock \bibinfo{journal}{\emph{arXiv preprint arXiv:2305.02309}}
  (\bibinfo{year}{2023}).
\newblock


\bibitem[\protect\citeauthoryear{Nijkamp, Pang, Hayashi, Tu, Wang, Zhou,
  Savarese, and Xiong}{Nijkamp et~al\mbox{.}}{2022}]%
        {nijkamp2022codegen}
\bibfield{author}{\bibinfo{person}{Erik Nijkamp}, \bibinfo{person}{Bo Pang},
  \bibinfo{person}{Hiroaki Hayashi}, \bibinfo{person}{Lifu Tu},
  \bibinfo{person}{Huan Wang}, \bibinfo{person}{Yingbo Zhou},
  \bibinfo{person}{Silvio Savarese}, {and} \bibinfo{person}{Caiming Xiong}.}
  \bibinfo{year}{2022}\natexlab{}.
\newblock \showarticletitle{Codegen: An open large language model for code with
  multi-turn program synthesis}.
\newblock \bibinfo{journal}{\emph{arXiv preprint arXiv:2203.13474}}
  (\bibinfo{year}{2022}).
\newblock


\bibitem[\protect\citeauthoryear{Oda, Fudaba, Neubig, Hata, Sakti, Toda, and
  Nakamura}{Oda et~al\mbox{.}}{2015}]%
        {oda2015learning}
\bibfield{author}{\bibinfo{person}{Yusuke Oda}, \bibinfo{person}{Hiroyuki
  Fudaba}, \bibinfo{person}{Graham Neubig}, \bibinfo{person}{Hideaki Hata},
  \bibinfo{person}{Sakriani Sakti}, \bibinfo{person}{Tomoki Toda}, {and}
  \bibinfo{person}{Satoshi Nakamura}.} \bibinfo{year}{2015}\natexlab{}.
\newblock \showarticletitle{Learning to generate pseudo-code from source code
  using statistical machine translation}. In \bibinfo{booktitle}{\emph{2015
  30th IEEE/ACM International Conference on Automated Software Engineering
  (ASE)}}. IEEE, \bibinfo{pages}{574--584}.
\newblock


\bibitem[\protect\citeauthoryear{Papineni, Roukos, Ward, and Zhu}{Papineni
  et~al\mbox{.}}{2002}]%
        {papineni2002bleu}
\bibfield{author}{\bibinfo{person}{Kishore Papineni}, \bibinfo{person}{Salim
  Roukos}, \bibinfo{person}{Todd Ward}, {and} \bibinfo{person}{Wei-Jing Zhu}.}
  \bibinfo{year}{2002}\natexlab{}.
\newblock \showarticletitle{BLEU: a method for automatic evaluation of machine
  translation}. In \bibinfo{booktitle}{\emph{Proceedings of the 40th annual
  meeting of the Association for Computational Linguistics}}.
  \bibinfo{pages}{311--318}.
\newblock


\bibitem[\protect\citeauthoryear{Parvez, Ahmad, Chakraborty, Ray, and
  Chang}{Parvez et~al\mbox{.}}{2021}]%
        {parvez2021retrieval}
\bibfield{author}{\bibinfo{person}{Md~Rizwan Parvez}, \bibinfo{person}{Wasi
  Ahmad}, \bibinfo{person}{Saikat Chakraborty}, \bibinfo{person}{Baishakhi
  Ray}, {and} \bibinfo{person}{Kai-Wei Chang}.}
  \bibinfo{year}{2021}\natexlab{}.
\newblock \showarticletitle{Retrieval Augmented Code Generation and
  Summarization}. In \bibinfo{booktitle}{\emph{Findings of the Association for
  Computational Linguistics: EMNLP 2021}}. \bibinfo{pages}{2719--2734}.
\newblock


\bibitem[\protect\citeauthoryear{Phan, Tran, Le, Nguyen, Annibal, Peltekian,
  and Ye}{Phan et~al\mbox{.}}{2021}]%
        {phan2021cotext}
\bibfield{author}{\bibinfo{person}{Long Phan}, \bibinfo{person}{Hieu Tran},
  \bibinfo{person}{Daniel Le}, \bibinfo{person}{Hieu Nguyen},
  \bibinfo{person}{James Annibal}, \bibinfo{person}{Alec Peltekian}, {and}
  \bibinfo{person}{Yanfang Ye}.} \bibinfo{year}{2021}\natexlab{}.
\newblock \showarticletitle{CoTexT: Multi-task Learning with Code-Text
  Transformer}. In \bibinfo{booktitle}{\emph{Proceedings of the 1st Workshop on
  Natural Language Processing for Programming (NLP4Prog 2021)}}.
  \bibinfo{pages}{40--47}.
\newblock


\bibitem[\protect\citeauthoryear{Press, Smith, and Lewis}{Press
  et~al\mbox{.}}{2021}]%
        {press2021train}
\bibfield{author}{\bibinfo{person}{Ofir Press}, \bibinfo{person}{Noah Smith},
  {and} \bibinfo{person}{Mike Lewis}.} \bibinfo{year}{2021}\natexlab{}.
\newblock \showarticletitle{Train Short, Test Long: Attention with Linear
  Biases Enables Input Length Extrapolation}. In
  \bibinfo{booktitle}{\emph{International Conference on Learning
  Representations}}.
\newblock


\bibitem[\protect\citeauthoryear{Qu, Hu, Zeng, Cai, and Yang}{Qu
  et~al\mbox{.}}{2022}]%
        {qu2022method}
\bibfield{author}{\bibinfo{person}{Zhiheng Qu}, \bibinfo{person}{Yi Hu},
  \bibinfo{person}{Jianhui Zeng}, \bibinfo{person}{Bo Cai}, {and}
  \bibinfo{person}{Shun Yang}.} \bibinfo{year}{2022}\natexlab{}.
\newblock \showarticletitle{Method Name Generation Based on Code Structure
  Guidance}. In \bibinfo{booktitle}{\emph{2022 IEEE International Conference on
  Software Analysis, Evolution and Reengineering (SANER)}}. IEEE,
  \bibinfo{pages}{1101--1110}.
\newblock


\bibitem[\protect\citeauthoryear{Rabin, Bui, Wang, Yu, Jiang, and
  Alipour}{Rabin et~al\mbox{.}}{2021}]%
        {rabin2021generalizability}
\bibfield{author}{\bibinfo{person}{Md~Rafiqul~Islam Rabin},
  \bibinfo{person}{Nghi~DQ Bui}, \bibinfo{person}{Ke Wang},
  \bibinfo{person}{Yijun Yu}, \bibinfo{person}{Lingxiao Jiang}, {and}
  \bibinfo{person}{Mohammad~Amin Alipour}.} \bibinfo{year}{2021}\natexlab{}.
\newblock \showarticletitle{On the generalizability of Neural Program Models
  with respect to semantic-preserving program transformations}.
\newblock \bibinfo{journal}{\emph{Information and Software Technology}}
  \bibinfo{volume}{135} (\bibinfo{year}{2021}), \bibinfo{pages}{106552}.
\newblock


\bibitem[\protect\citeauthoryear{Radford, Wu, Child, Luan, Amodei, Sutskever,
  et~al\mbox{.}}{Radford et~al\mbox{.}}{2019}]%
        {radford2019language}
\bibfield{author}{\bibinfo{person}{Alec Radford}, \bibinfo{person}{Jeffrey Wu},
  \bibinfo{person}{Rewon Child}, \bibinfo{person}{David Luan},
  \bibinfo{person}{Dario Amodei}, \bibinfo{person}{Ilya Sutskever},
  {et~al\mbox{.}}} \bibinfo{year}{2019}\natexlab{}.
\newblock \showarticletitle{Language models are unsupervised multitask
  learners}.
\newblock \bibinfo{journal}{\emph{OpenAI blog}} \bibinfo{volume}{1},
  \bibinfo{number}{8} (\bibinfo{year}{2019}), \bibinfo{pages}{9}.
\newblock


\bibitem[\protect\citeauthoryear{Raffel, Shazeer, Roberts, Lee, Narang, Matena,
  Zhou, Li, Liu, et~al\mbox{.}}{Raffel et~al\mbox{.}}{2020}]%
        {raffel2020exploring}
\bibfield{author}{\bibinfo{person}{Colin Raffel}, \bibinfo{person}{Noam
  Shazeer}, \bibinfo{person}{Adam Roberts}, \bibinfo{person}{Katherine Lee},
  \bibinfo{person}{Sharan Narang}, \bibinfo{person}{Michael Matena},
  \bibinfo{person}{Yanqi Zhou}, \bibinfo{person}{Wei Li},
  \bibinfo{person}{Peter~J Liu}, {et~al\mbox{.}}}
  \bibinfo{year}{2020}\natexlab{}.
\newblock \showarticletitle{Exploring the limits of transfer learning with a
  unified text-to-text transformer.}
\newblock \bibinfo{journal}{\emph{J. Mach. Learn. Res.}} \bibinfo{volume}{21},
  \bibinfo{number}{140} (\bibinfo{year}{2020}), \bibinfo{pages}{1--67}.
\newblock


\bibitem[\protect\citeauthoryear{Rawlinson}{Rawlinson}{2007}]%
        {rawlinson2007significance}
\bibfield{author}{\bibinfo{person}{Graham Rawlinson}.}
  \bibinfo{year}{2007}\natexlab{}.
\newblock \showarticletitle{The significance of letter position in word
  recognition}.
\newblock \bibinfo{journal}{\emph{IEEE Aerospace and Electronic Systems
  Magazine}} \bibinfo{volume}{22}, \bibinfo{number}{1} (\bibinfo{year}{2007}),
  \bibinfo{pages}{26--27}.
\newblock


\bibitem[\protect\citeauthoryear{Ren, Guo, Lu, Zhou, Liu, Tang, Sundaresan,
  Zhou, Blanco, and Ma}{Ren et~al\mbox{.}}{2020}]%
        {ren2020codebleu}
\bibfield{author}{\bibinfo{person}{Shuo Ren}, \bibinfo{person}{Daya Guo},
  \bibinfo{person}{Shuai Lu}, \bibinfo{person}{Long Zhou},
  \bibinfo{person}{Shujie Liu}, \bibinfo{person}{Duyu Tang},
  \bibinfo{person}{Neel Sundaresan}, \bibinfo{person}{Ming Zhou},
  \bibinfo{person}{Ambrosio Blanco}, {and} \bibinfo{person}{Shuai Ma}.}
  \bibinfo{year}{2020}\natexlab{}.
\newblock \showarticletitle{Codebleu: a method for automatic evaluation of code
  synthesis}.
\newblock \bibinfo{journal}{\emph{arXiv preprint arXiv:2009.10297}}
  (\bibinfo{year}{2020}).
\newblock


\bibitem[\protect\citeauthoryear{replit}{replit}{2023}]%
        {replit}
\bibfield{author}{\bibinfo{person}{replit}.} \bibinfo{year}{2023}\natexlab{}.
\newblock \bibinfo{title}{replit-code-v1-3b}.
\newblock
\newblock
\urldef\tempurl%
\url{https://huggingface.co/replit/replit-code-v1-3b}
\showURL{%
\tempurl}


\bibitem[\protect\citeauthoryear{Robertson and Zaragoza}{Robertson and
  Zaragoza}{2009}]%
        {robertson2009probabilistic}
\bibfield{author}{\bibinfo{person}{Stephen Robertson} {and}
  \bibinfo{person}{Hugo Zaragoza}.} \bibinfo{year}{2009}\natexlab{}.
\newblock \bibinfo{booktitle}{\emph{The probabilistic relevance framework: BM25
  and beyond}}.
\newblock \bibinfo{publisher}{Now Publishers Inc}.
\newblock


\bibitem[\protect\citeauthoryear{See, Liu, and Manning}{See
  et~al\mbox{.}}{2017}]%
        {see2017get}
\bibfield{author}{\bibinfo{person}{Abigail See}, \bibinfo{person}{Peter~J Liu},
  {and} \bibinfo{person}{Christopher~D Manning}.}
  \bibinfo{year}{2017}\natexlab{}.
\newblock \showarticletitle{Get To The Point: Summarization with
  Pointer-Generator Networks}. In \bibinfo{booktitle}{\emph{Proceedings of the
  55th Annual Meeting of the Association for Computational Linguistics (Volume
  1: Long Papers)}}. \bibinfo{pages}{1073--1083}.
\newblock


\bibitem[\protect\citeauthoryear{Stapleton, Gambhir, LeClair, Eberhart, Weimer,
  Leach, and Huang}{Stapleton et~al\mbox{.}}{2020}]%
        {stapleton2020human}
\bibfield{author}{\bibinfo{person}{Sean Stapleton}, \bibinfo{person}{Yashmeet
  Gambhir}, \bibinfo{person}{Alexander LeClair}, \bibinfo{person}{Zachary
  Eberhart}, \bibinfo{person}{Westley Weimer}, \bibinfo{person}{Kevin Leach},
  {and} \bibinfo{person}{Yu Huang}.} \bibinfo{year}{2020}\natexlab{}.
\newblock \showarticletitle{A human study of comprehension and code
  summarization}. In \bibinfo{booktitle}{\emph{Proceedings of the 28th
  International Conference on Program Comprehension}}. \bibinfo{pages}{2--13}.
\newblock


\bibitem[\protect\citeauthoryear{Sun, Zhu, Xiong, Sun, Mou, and Zhang}{Sun
  et~al\mbox{.}}{2020}]%
        {sun2020treegen}
\bibfield{author}{\bibinfo{person}{Zeyu Sun}, \bibinfo{person}{Qihao Zhu},
  \bibinfo{person}{Yingfei Xiong}, \bibinfo{person}{Yican Sun},
  \bibinfo{person}{Lili Mou}, {and} \bibinfo{person}{Lu Zhang}.}
  \bibinfo{year}{2020}\natexlab{}.
\newblock \showarticletitle{Treegen: A tree-based transformer architecture for
  code generation}. In \bibinfo{booktitle}{\emph{Proceedings of the AAAI
  Conference on Artificial Intelligence}}, Vol.~\bibinfo{volume}{34}.
  \bibinfo{pages}{8984--8991}.
\newblock


\bibitem[\protect\citeauthoryear{Svyatkovskiy, Deng, Fu, and
  Sundaresan}{Svyatkovskiy et~al\mbox{.}}{2020}]%
        {svyatkovskiy2020intellicode}
\bibfield{author}{\bibinfo{person}{Alexey Svyatkovskiy},
  \bibinfo{person}{Shao~Kun Deng}, \bibinfo{person}{Shengyu Fu}, {and}
  \bibinfo{person}{Neel Sundaresan}.} \bibinfo{year}{2020}\natexlab{}.
\newblock \showarticletitle{Intellicode compose: Code generation using
  transformer}. In \bibinfo{booktitle}{\emph{Proceedings of the 28th ACM Joint
  Meeting on European Software Engineering Conference and Symposium on the
  Foundations of Software Engineering}}. \bibinfo{pages}{1433--1443}.
\newblock


\bibitem[\protect\citeauthoryear{Tian, Wang, Li, and Wen}{Tian
  et~al\mbox{.}}{2021}]%
        {tian2021generating}
\bibfield{author}{\bibinfo{person}{Junfeng Tian}, \bibinfo{person}{Chenxin
  Wang}, \bibinfo{person}{Zhen Li}, {and} \bibinfo{person}{Yu Wen}.}
  \bibinfo{year}{2021}\natexlab{}.
\newblock \showarticletitle{Generating Adversarial Examples of Source Code
  Classification Models via Q-Learning-Based Markov Decision Process}. In
  \bibinfo{booktitle}{\emph{2021 IEEE 21st International Conference on Software
  Quality, Reliability and Security (QRS)}}. IEEE, \bibinfo{pages}{807--818}.
\newblock


\bibitem[\protect\citeauthoryear{Vaswani, Shazeer, Parmar, Uszkoreit, Jones,
  Gomez, Kaiser, and Polosukhin}{Vaswani et~al\mbox{.}}{2017}]%
        {vaswani2017attention}
\bibfield{author}{\bibinfo{person}{Ashish Vaswani}, \bibinfo{person}{Noam
  Shazeer}, \bibinfo{person}{Niki Parmar}, \bibinfo{person}{Jakob Uszkoreit},
  \bibinfo{person}{Llion Jones}, \bibinfo{person}{Aidan~N Gomez},
  \bibinfo{person}{{\L}ukasz Kaiser}, {and} \bibinfo{person}{Illia
  Polosukhin}.} \bibinfo{year}{2017}\natexlab{}.
\newblock \showarticletitle{Attention is all you need}.
\newblock \bibinfo{journal}{\emph{Advances in neural information processing
  systems}}  \bibinfo{volume}{30} (\bibinfo{year}{2017}).
\newblock


\bibitem[\protect\citeauthoryear{Wang, Chen, Sun, Ma, Wang, Sun, and
  Cheng}{Wang et~al\mbox{.}}{2021a}]%
        {wang2021robot}
\bibfield{author}{\bibinfo{person}{Jingyi Wang}, \bibinfo{person}{Jialuo Chen},
  \bibinfo{person}{Youcheng Sun}, \bibinfo{person}{Xingjun Ma},
  \bibinfo{person}{Dongxia Wang}, \bibinfo{person}{Jun Sun}, {and}
  \bibinfo{person}{Peng Cheng}.} \bibinfo{year}{2021}\natexlab{a}.
\newblock \showarticletitle{Robot: Robustness-oriented testing for deep
  learning systems}. In \bibinfo{booktitle}{\emph{2021 IEEE/ACM 43rd
  International Conference on Software Engineering (ICSE)}}. IEEE,
  \bibinfo{pages}{300--311}.
\newblock


\bibitem[\protect\citeauthoryear{Wang, Li, Qian, Yang, Wang, Shang, Kumar, Tan,
  Ray, Bhatia, et~al\mbox{.}}{Wang et~al\mbox{.}}{2022a}]%
        {wang2022recode}
\bibfield{author}{\bibinfo{person}{Shiqi Wang}, \bibinfo{person}{Zheng Li},
  \bibinfo{person}{Haifeng Qian}, \bibinfo{person}{Chenghao Yang},
  \bibinfo{person}{Zijian Wang}, \bibinfo{person}{Mingyue Shang},
  \bibinfo{person}{Varun Kumar}, \bibinfo{person}{Samson Tan},
  \bibinfo{person}{Baishakhi Ray}, \bibinfo{person}{Parminder Bhatia},
  {et~al\mbox{.}}} \bibinfo{year}{2022}\natexlab{a}.
\newblock \showarticletitle{ReCode: Robustness Evaluation of Code Generation
  Models}.
\newblock \bibinfo{journal}{\emph{arXiv preprint arXiv:2212.10264}}
  (\bibinfo{year}{2022}).
\newblock


\bibitem[\protect\citeauthoryear{Wang, Wen, Lin, and Mao}{Wang
  et~al\mbox{.}}{2021c}]%
        {wang2021lightweight}
\bibfield{author}{\bibinfo{person}{Shangwen Wang}, \bibinfo{person}{Ming Wen},
  \bibinfo{person}{Bo Lin}, {and} \bibinfo{person}{Xiaoguang Mao}.}
  \bibinfo{year}{2021}\natexlab{c}.
\newblock \showarticletitle{Lightweight global and local contexts guided method
  name recommendation with prior knowledge}. In
  \bibinfo{booktitle}{\emph{Proceedings of the 29th ACM Joint Meeting on
  European Software Engineering Conference and Symposium on the Foundations of
  Software Engineering}}. \bibinfo{pages}{741--753}.
\newblock


\bibitem[\protect\citeauthoryear{Wang, Xu, Fang, Liu, Sun, Xu, Zhu, and
  Zeng}{Wang et~al\mbox{.}}{2022b}]%
        {wang2022training}
\bibfield{author}{\bibinfo{person}{Shuohang Wang}, \bibinfo{person}{Yichong
  Xu}, \bibinfo{person}{Yuwei Fang}, \bibinfo{person}{Yang Liu},
  \bibinfo{person}{Siqi Sun}, \bibinfo{person}{Ruochen Xu},
  \bibinfo{person}{Chenguang Zhu}, {and} \bibinfo{person}{Michael Zeng}.}
  \bibinfo{year}{2022}\natexlab{b}.
\newblock \showarticletitle{Training Data is More Valuable than You Think: A
  Simple and Effective Method by Retrieving from Training Data}. In
  \bibinfo{booktitle}{\emph{Proceedings of the 60th Annual Meeting of the
  Association for Computational Linguistics (Volume 1: Long Papers)}}.
  \bibinfo{pages}{3170--3179}.
\newblock


\bibitem[\protect\citeauthoryear{Wang, Le, Gotmare, Bui, Li, and Hoi}{Wang
  et~al\mbox{.}}{2023}]%
        {wang2023codet5+}
\bibfield{author}{\bibinfo{person}{Yue Wang}, \bibinfo{person}{Hung Le},
  \bibinfo{person}{Akhilesh~Deepak Gotmare}, \bibinfo{person}{Nghi~DQ Bui},
  \bibinfo{person}{Junnan Li}, {and} \bibinfo{person}{Steven~CH Hoi}.}
  \bibinfo{year}{2023}\natexlab{}.
\newblock \showarticletitle{Codet5+: Open code large language models for code
  understanding and generation}.
\newblock \bibinfo{journal}{\emph{arXiv preprint arXiv:2305.07922}}
  (\bibinfo{year}{2023}).
\newblock


\bibitem[\protect\citeauthoryear{Wang, Wang, Joty, and Hoi}{Wang
  et~al\mbox{.}}{2021b}]%
        {wang2021codet5}
\bibfield{author}{\bibinfo{person}{Yue Wang}, \bibinfo{person}{Weishi Wang},
  \bibinfo{person}{Shafiq Joty}, {and} \bibinfo{person}{Steven~CH Hoi}.}
  \bibinfo{year}{2021}\natexlab{b}.
\newblock \showarticletitle{CodeT5: Identifier-aware Unified Pre-trained
  Encoder-Decoder Models for Code Understanding and Generation}. In
  \bibinfo{booktitle}{\emph{Proceedings of the 2021 Conference on Empirical
  Methods in Natural Language Processing}}. \bibinfo{pages}{8696--8708}.
\newblock


\bibitem[\protect\citeauthoryear{Wei, Tay, Bommasani, Raffel, Zoph, Borgeaud,
  Yogatama, Bosma, Zhou, Metzler, et~al\mbox{.}}{Wei et~al\mbox{.}}{[n.d.]}]%
        {weiemergent}
\bibfield{author}{\bibinfo{person}{Jason Wei}, \bibinfo{person}{Yi Tay},
  \bibinfo{person}{Rishi Bommasani}, \bibinfo{person}{Colin Raffel},
  \bibinfo{person}{Barret Zoph}, \bibinfo{person}{Sebastian Borgeaud},
  \bibinfo{person}{Dani Yogatama}, \bibinfo{person}{Maarten Bosma},
  \bibinfo{person}{Denny Zhou}, \bibinfo{person}{Donald Metzler},
  {et~al\mbox{.}}} \bibinfo{year}{[n.d.]}\natexlab{}.
\newblock \showarticletitle{Emergent Abilities of Large Language Models}.
\newblock \bibinfo{journal}{\emph{Transactions on Machine Learning Research}}
  (\bibinfo{year}{[n.\,d.]}).
\newblock


\bibitem[\protect\citeauthoryear{Xu, Zhang, Wang, Cao, Guo, and Xu}{Xu
  et~al\mbox{.}}{2019}]%
        {xu2019method}
\bibfield{author}{\bibinfo{person}{Sihan Xu}, \bibinfo{person}{Sen Zhang},
  \bibinfo{person}{Weijing Wang}, \bibinfo{person}{Xinya Cao},
  \bibinfo{person}{Chenkai Guo}, {and} \bibinfo{person}{Jing Xu}.}
  \bibinfo{year}{2019}\natexlab{}.
\newblock \showarticletitle{Method name suggestion with hierarchical attention
  networks}. In \bibinfo{booktitle}{\emph{Proceedings of the 2019 ACM SIGPLAN
  workshop on partial evaluation and program manipulation}}.
  \bibinfo{pages}{10--21}.
\newblock


\bibitem[\protect\citeauthoryear{Yang, Liu, Chen, Zhou, Yu, and Lin}{Yang
  et~al\mbox{.}}{2022a}]%
        {yang2022ccgir}
\bibfield{author}{\bibinfo{person}{Guang Yang}, \bibinfo{person}{Ke Liu},
  \bibinfo{person}{Xiang Chen}, \bibinfo{person}{Yanlin Zhou},
  \bibinfo{person}{Chi Yu}, {and} \bibinfo{person}{Hao Lin}.}
  \bibinfo{year}{2022}\natexlab{a}.
\newblock \showarticletitle{CCGIR: Information retrieval-based code comment
  generation method for smart contracts}.
\newblock \bibinfo{journal}{\emph{Knowledge-Based Systems}}
  \bibinfo{volume}{237} (\bibinfo{year}{2022}), \bibinfo{pages}{107858}.
\newblock


\bibitem[\protect\citeauthoryear{Yang, Chen, Hsieh, Wang, and Jordan}{Yang
  et~al\mbox{.}}{2020}]%
        {yang2020greedy}
\bibfield{author}{\bibinfo{person}{Puyudi Yang}, \bibinfo{person}{Jianbo Chen},
  \bibinfo{person}{Cho-Jui Hsieh}, \bibinfo{person}{Jane-Ling Wang}, {and}
  \bibinfo{person}{Michael~I Jordan}.} \bibinfo{year}{2020}\natexlab{}.
\newblock \showarticletitle{Greedy attack and gumbel attack: Generating
  adversarial examples for discrete data}.
\newblock \bibinfo{journal}{\emph{The Journal of Machine Learning Research}}
  \bibinfo{volume}{21}, \bibinfo{number}{1} (\bibinfo{year}{2020}),
  \bibinfo{pages}{1613--1648}.
\newblock


\bibitem[\protect\citeauthoryear{Yang, Shi, He, and Lo}{Yang
  et~al\mbox{.}}{2022b}]%
        {yang2022natural}
\bibfield{author}{\bibinfo{person}{Zhou Yang}, \bibinfo{person}{Jieke Shi},
  \bibinfo{person}{Junda He}, {and} \bibinfo{person}{David Lo}.}
  \bibinfo{year}{2022}\natexlab{b}.
\newblock \showarticletitle{Natural Attack for Pre-trained Models of Code}. In
  \bibinfo{booktitle}{\emph{44th {IEEE/ACM} 44th International Conference on
  Software Engineering, {ICSE} 2022, Pittsburgh, PA, USA, May 25-27, 2022}}.
  \bibinfo{publisher}{{ACM}}, \bibinfo{pages}{1482--1493}.
\newblock
\urldef\tempurl%
\url{https://doi.org/10.1145/3510003.3510146}
\showDOI{\tempurl}


\bibitem[\protect\citeauthoryear{Yefet, Alon, and Yahav}{Yefet
  et~al\mbox{.}}{2020}]%
        {yefet2020adversarial}
\bibfield{author}{\bibinfo{person}{Noam Yefet}, \bibinfo{person}{Uri Alon},
  {and} \bibinfo{person}{Eran Yahav}.} \bibinfo{year}{2020}\natexlab{}.
\newblock \showarticletitle{Adversarial examples for models of code}.
\newblock \bibinfo{journal}{\emph{Proceedings of the ACM on Programming
  Languages}} \bibinfo{volume}{4}, \bibinfo{number}{OOPSLA}
  (\bibinfo{year}{2020}), \bibinfo{pages}{1--30}.
\newblock


\bibitem[\protect\citeauthoryear{Yin, Deng, Chen, Vasilescu, and Neubig}{Yin
  et~al\mbox{.}}{2018}]%
        {yin2018learning}
\bibfield{author}{\bibinfo{person}{Pengcheng Yin}, \bibinfo{person}{Bowen
  Deng}, \bibinfo{person}{Edgar Chen}, \bibinfo{person}{Bogdan Vasilescu},
  {and} \bibinfo{person}{Graham Neubig}.} \bibinfo{year}{2018}\natexlab{}.
\newblock \showarticletitle{Learning to mine aligned code and natural language
  pairs from stack overflow}. In \bibinfo{booktitle}{\emph{Proceedings of the
  15th International Conference on Mining Software Repositories}}.
  \bibinfo{pages}{476--486}.
\newblock


\bibitem[\protect\citeauthoryear{Yin and Neubig}{Yin and Neubig}{2017}]%
        {yin2017syntactic}
\bibfield{author}{\bibinfo{person}{Pengcheng Yin} {and} \bibinfo{person}{Graham
  Neubig}.} \bibinfo{year}{2017}\natexlab{}.
\newblock \showarticletitle{A Syntactic Neural Model for General-Purpose Code
  Generation}. In \bibinfo{booktitle}{\emph{Proceedings of the 55th Annual
  Meeting of the Association for Computational Linguistics (Volume 1: Long
  Papers)}}. \bibinfo{pages}{440--450}.
\newblock


\bibitem[\protect\citeauthoryear{Zang, Qi, Yang, Liu, Zhang, Liu, and Sun}{Zang
  et~al\mbox{.}}{2020}]%
        {zang2020word}
\bibfield{author}{\bibinfo{person}{Yuan Zang}, \bibinfo{person}{Fanchao Qi},
  \bibinfo{person}{Chenghao Yang}, \bibinfo{person}{Zhiyuan Liu},
  \bibinfo{person}{Meng Zhang}, \bibinfo{person}{Qun Liu}, {and}
  \bibinfo{person}{Maosong Sun}.} \bibinfo{year}{2020}\natexlab{}.
\newblock \showarticletitle{Word-level Textual Adversarial Attacking as
  Combinatorial Optimization}. In \bibinfo{booktitle}{\emph{Proceedings of the
  58th Annual Meeting of the Association for Computational Linguistics}}.
  \bibinfo{pages}{6066--6080}.
\newblock


\bibitem[\protect\citeauthoryear{Zeng, Tan, Zhang, Li, Zhang, and Zhang}{Zeng
  et~al\mbox{.}}{2022}]%
        {zeng2022extensive}
\bibfield{author}{\bibinfo{person}{Zhengran Zeng}, \bibinfo{person}{Hanzhuo
  Tan}, \bibinfo{person}{Haotian Zhang}, \bibinfo{person}{Jing Li},
  \bibinfo{person}{Yuqun Zhang}, {and} \bibinfo{person}{Lingming Zhang}.}
  \bibinfo{year}{2022}\natexlab{}.
\newblock \showarticletitle{An extensive study on pre-trained models for
  program understanding and generation}. In
  \bibinfo{booktitle}{\emph{Proceedings of the 31st ACM SIGSOFT international
  symposium on software testing and analysis}}. \bibinfo{pages}{39--51}.
\newblock


\bibitem[\protect\citeauthoryear{Zhang, Fu, Li, Ma, Zhao, Yang, Sun, Liu, and
  Jin}{Zhang et~al\mbox{.}}{2022}]%
        {zhang2022towards}
\bibfield{author}{\bibinfo{person}{Huangzhao Zhang}, \bibinfo{person}{Zhiyi
  Fu}, \bibinfo{person}{Ge Li}, \bibinfo{person}{Lei Ma},
  \bibinfo{person}{Zhehao Zhao}, \bibinfo{person}{Hua’an Yang},
  \bibinfo{person}{Yizhe Sun}, \bibinfo{person}{Yang Liu}, {and}
  \bibinfo{person}{Zhi Jin}.} \bibinfo{year}{2022}\natexlab{}.
\newblock \showarticletitle{Towards robustness of deep program processing
  models—detection, estimation, and enhancement}.
\newblock \bibinfo{journal}{\emph{ACM Transactions on Software Engineering and
  Methodology (TOSEM)}} \bibinfo{volume}{31}, \bibinfo{number}{3}
  (\bibinfo{year}{2022}), \bibinfo{pages}{1--40}.
\newblock


\bibitem[\protect\citeauthoryear{Zhang, Li, Li, Ma, Liu, and Jin}{Zhang
  et~al\mbox{.}}{2020a}]%
        {zhang2020generating}
\bibfield{author}{\bibinfo{person}{Huangzhao Zhang}, \bibinfo{person}{Zhuo Li},
  \bibinfo{person}{Ge Li}, \bibinfo{person}{Lei Ma}, \bibinfo{person}{Yang
  Liu}, {and} \bibinfo{person}{Zhi Jin}.} \bibinfo{year}{2020}\natexlab{a}.
\newblock \showarticletitle{Generating adversarial examples for holding
  robustness of source code processing models}. In
  \bibinfo{booktitle}{\emph{Proceedings of the AAAI Conference on Artificial
  Intelligence}}, Vol.~\bibinfo{volume}{34}. \bibinfo{pages}{1169--1176}.
\newblock


\bibitem[\protect\citeauthoryear{Zhang, Wang, Zhang, Sun, and Liu}{Zhang
  et~al\mbox{.}}{2020b}]%
        {zhang2020retrieval}
\bibfield{author}{\bibinfo{person}{Jian Zhang}, \bibinfo{person}{Xu Wang},
  \bibinfo{person}{Hongyu Zhang}, \bibinfo{person}{Hailong Sun}, {and}
  \bibinfo{person}{Xudong Liu}.} \bibinfo{year}{2020}\natexlab{b}.
\newblock \showarticletitle{Retrieval-based neural source code summarization}.
  In \bibinfo{booktitle}{\emph{Proceedings of the ACM/IEEE 42nd International
  Conference on Software Engineering}}. \bibinfo{pages}{1385--1397}.
\newblock


\bibitem[\protect\citeauthoryear{Zhang, Zhou, Han, and Chen}{Zhang
  et~al\mbox{.}}{2020c}]%
        {zhang2020training}
\bibfield{author}{\bibinfo{person}{Xiaoqing Zhang}, \bibinfo{person}{Yu Zhou},
  \bibinfo{person}{Tingting Han}, {and} \bibinfo{person}{Taolue Chen}.}
  \bibinfo{year}{2020}\natexlab{c}.
\newblock \showarticletitle{Training deep code comment generation models via
  data augmentation}. In \bibinfo{booktitle}{\emph{12th Asia-Pacific Symposium
  on Internetware}}. \bibinfo{pages}{185--188}.
\newblock


\bibitem[\protect\citeauthoryear{Zheng, Xia, Zou, Dong, Wang, Xue, Wang, Shen,
  Wang, Li, et~al\mbox{.}}{Zheng et~al\mbox{.}}{2023}]%
        {zheng2023codegeex}
\bibfield{author}{\bibinfo{person}{Qinkai Zheng}, \bibinfo{person}{Xiao Xia},
  \bibinfo{person}{Xu Zou}, \bibinfo{person}{Yuxiao Dong},
  \bibinfo{person}{Shan Wang}, \bibinfo{person}{Yufei Xue},
  \bibinfo{person}{Zihan Wang}, \bibinfo{person}{Lei Shen},
  \bibinfo{person}{Andi Wang}, \bibinfo{person}{Yang Li}, {et~al\mbox{.}}}
  \bibinfo{year}{2023}\natexlab{}.
\newblock \showarticletitle{Codegeex: A pre-trained model for code generation
  with multilingual evaluations on humaneval-x}.
\newblock \bibinfo{journal}{\emph{arXiv preprint arXiv:2303.17568}}
  (\bibinfo{year}{2023}).
\newblock


\bibitem[\protect\citeauthoryear{Zhou, Zhang, Shen, Han, Chen, and Gall}{Zhou
  et~al\mbox{.}}{2021}]%
        {zhou2021adversarial}
\bibfield{author}{\bibinfo{person}{Yu Zhou}, \bibinfo{person}{Xiaoqing Zhang},
  \bibinfo{person}{Juanjuan Shen}, \bibinfo{person}{Tingting Han},
  \bibinfo{person}{Taolue Chen}, {and} \bibinfo{person}{Harald Gall}.}
  \bibinfo{year}{2021}\natexlab{}.
\newblock \showarticletitle{Adversarial robustness of deep code comment
  generation}.
\newblock \bibinfo{journal}{\emph{ACM Transactions on Software Engineering and
  Methodology}} (\bibinfo{year}{2021}).
\newblock


\end{thebibliography}
 
\end{document}